\documentclass[twocolumn]{aastex61}
\newcommand\refbold[1]{#1}

\shorttitle{Modelling Solar Energetic Particle transport near a wavy Heliospheric Current Sheet}
\shortauthors{Battarbee, Dalla, \& Marsh}

\usepackage{graphicx}
\usepackage{amsmath}
\usepackage[percent]{overpic}

\begin{document}

\title{Modelling Solar Energetic Particle transport near a wavy Heliospheric Current Sheet}

\correspondingauthor{Markus Battarbee}
\email{markus.battarbee@gmail.com}

\author[0000-0001-7055-551X]{Markus Battarbee}
\altaffiliation{Currently at the Department of Physics, University of Helsinki, Finland}
\affiliation{Jeremiah Horrocks Institute, University of Central Lancashire, PR1 2HE, UK}

\author{Silvia Dalla}
\affiliation{Jeremiah Horrocks Institute, University of Central Lancashire, PR1 2HE, UK}

\author{Mike S. Marsh}
\affiliation{Met Office, Exeter, EX1 3PB, UK}



\begin{abstract}
Understanding the transport of Solar Energetic Particles (SEPs) from acceleration sites at the Sun into interplanetary space and to the Earth is an important question for forecasting space weather. The Interplanetary Magnetic Field (IMF), with two distinct polarities and a complex structure, governs energetic particle transport and drifts.

We analyse for the first time the effect of a wavy Heliospheric Current Sheet (HCS) on the propagation of SEPs. We inject protons close to the Sun and propagate them by integrating fully 3D trajectories within the inner heliosphere in the presence of weak scattering. \refbold{We model the HCS position using fits based on neutral lines of} magnetic field source surface maps (SSMs).

\refbold{We map} \mbox{1 au} proton crossings, which show efficient transport in longitude via HCS, depending on the location of the injection region with respect to the HCS. For HCS tilt angles around $30^\circ-40^\circ$, we find significant qualitative differences between A+ and A-- configurations of the IMF, with stronger fluences along the HCS in the former case but with a distribution of particles across a wider range of longitudes and latitudes in the latter. 
  
We show how a wavy current sheet leads to longitudinally periodic enhancements in particle fluence. We show that for an A+ IMF configuration, a wavy HCS allows for more proton deceleration than a flat HCS. We find that A-- IMF configurations result in larger average fluences than A+ IMF configurations, due to a radial drift component at the current sheet.

\end{abstract}


%

\section{Introduction} \label{sec:intro}
Solar activity and energy release events are capable of accelerating Solar Energetic Particles (SEPs) to high energies, launching them into the heliosphere where they propagate, influencing space weather. The dangers SEPs pose to spacecraft and satellite operations, astronaut safety, communication systems, and even aircraft personnel, have been recognised as significant, warranting studies to improve our understanding and forecasting capabilities of solar activity events \citep{Turner2000}.

The propagation of SEPs from an acceleration site at or near the Sun to observers at the Earth or in interplanetary space is a topic of continuing research, utilizing transport equations (see, e.g., \citealt{Roelof1969}) and test particle simulations (see, e.g., \citealt{Marsh2013}). The physics of particle propagation has been integrated into space weather models (see, e.g., \citealt{Aran2005}, \citealt{Luhmann2007}, \refbold{\citealt{Schwadron2014}}, and \citealt{Marsh2015}). The motion of charged particles propagating through interplanetary space is strongly influenced by the solar wind's magnetic field and by its spatial and temporal variations. Particles experience scattering, as well as drift due to the Interplanetary Magnetic Field (IMF) and the motional electric field associated with the outward-flowing solar wind. Particle drifts, which are currently not modeled by SEP transport equation approaches, play an important role, especially with increasing IMF complexity (see, e.g., \citealt{Marsh2013}, \citealt{Dalla2013,Dalla2015}, and \citealt{Battarbee2017}). Analytical expressions for drift velocities associated with the gradient and curvature of the Parker spiral can be obtained \citep{Dalla2013}, showing that they increase with heliographic latitude and depend on radial distance from the Sun. \refbold{Cosmic ray propagation effects due to drifts, diffusion, and turbulence have been investigated in, e.g., \cite{Kota1983}, \cite{Zhang2003}, and \cite{Zhao2017}. Though results derived from cosmic ray studies are interesting and can provide insight to SEP studies, the dynamics of cosmic ray influx are distinct from that of localized injection close to the Sun.}

The Heliospheric Current Sheet (HCS) is the boundary between the two hemispheres of opposite magnetic polarity in interplanetary space. During solar minimum it can appear mostly planar, whereas during solar maximum it experiences a great deal of deformation. Each polarity region exhibits distinct particle drifts due to the reversal of mean magnetic field direction, with the reversal zone at the HCS providing additional current sheet drifts \citep{Burger1985}.

The solar magnetic field can, during certain phases of solar activity, be described as a magnetic dipole tilted with respect to the Sun's axis of rotation. This produces a wavy current sheet, also known as the ballerina skirt model. Previous studies of wavy current sheet effects on energetic particle propagation have been limited to galactic cosmic rays (GCRs), entering the solar system from outside the termination shock (see \citealt{Jokipii1977}, \citealt{Kota2001}, \citealt{Burger2012}, \citealt{Pei2012}, \citealt{Strauss2012}, \citealt{Guo2014}, and references therein). We have recently introduced a flat HCS to our SEP test particle model \citep{Battarbee2017}, and will be referring to this publication as Paper I. 

In the current paper, we present novel results, showing the influence of a wavy HCS configuration on SEP propagation, as derived from a 3D test particle model. We report on periodic fluence enhancements found at longitudes of increased HCS inclination and discuss how the mean IMF polarity configuration affects total fluences due to the directionality of HCS drift. We show the energy-dependence of particle access to the HCS when the injection region is not intersected by the sheet. We report on asymmetric geometry-dependent drifts in the vicinity of a small injection region.

In section \ref{sec:HCS} we describe our wavy HCS model and how it is parametrised. In section \ref{sec:simulations}, we discuss our simulation set-ups, and in section \ref{sec:results}, we present fluence maps and energy spectrograms of protons at \mbox{1 au}. We discuss the results and implications of our model in section \ref{sec:conclusions}. Additionally, in \mbox{Appendix \ref{appendix:fitting}}, we discuss how our HCS model can be fitted to coronal source surface modelling maps.


\section{A Wavy Heliospheric Current Sheet Model} \label{sec:HCS}
We investigate SEP propagation through numerical integration of equations of motion for a large number of test particles. For this task, our model (described in detail in \citealt{Marsh2013}) requires knowledge of the electric and magnetic fields throughout the heliosphere. In Paper I, we presented a simple flat formulation for a HCS, which was a 3D Parker spiral, modified as a function of latitude to account for two magnetic field directions. Additionally, we scaled the strength of the magnetic field in a small region close to the HCS. This allowed for the two hemispheres to have opposite magnetic polarity, with a smooth transition in between. Our model is constructed using the fixed heliographic inertial (HGI) frame of reference. 

If a flat current sheet, delimiting the northern and southern hemispheres of magnetic polarity, as presented in Paper I, is the zeroth order approximation of a true IMF, the first order approximation would be a wavy current sheet produced by a dipole, tilted with respect to the rotation axis of the Sun. We now extend the model presented in Paper I, using this formulation of a tilted dipole. In our model, we assume the frozen-in theory to hold and maintain a constant radial solar wind throughout the heliosphere. Thus, we continue to use the simplification of neglecting reconnection effects at the HCS, which is supported by \cite{Gosling2012}, where reconnection at the HCS was suggested to be infrequent. \refbold{Analysis and observations of reconnection in the solar wind is a current and complex research topic (see, e.g., \citealt{Xu2015}, \citealt{Zharkova2015}, and \citealt{Khabarova2017}), made especially challenging due to the point-source observations and the challenge of acquiring reliable electric field measurements. Disturbances and discontinuities are certain to cause some degree of reconnection in the solar wind, but for the purpose of this study, we assume such effects to cause negligible changes to the effects our magnetic and electric fields have on particle propagation.}

We construct our model of an IMF with a wavy HCS by first defining a solar source surface, which we set at $r_\mathrm{s}=2.5R_\odot$ to match available source modelling data. At this source surface, we define a great circle as the neutral line delimiting the two hemispheres associated with the tilted dipole. One hemisphere is associated with outward-pointing magnetic field lines, the other with inward-pointing magnetic field lines. We define the dipole tilt angle as $\alpha_\mathrm{nl}$, and the longitudinal anchor point for the neutral line on the solar equator at time $t=0$ as $\phi_\mathrm{nl}$. In a Carrington map, this great circle draws what appears like a sinusoidal neutral line. As the neutral line parametrisation is inferred from magnetic fields on the rotating solar surface, we let the longitudinal anchor point rotate with the Sun as $\Phi_\mathrm{nl}(t)=\phi_\mathrm{nl}+\Omega_\odot t$, using an average solar rotation rate of $\Omega_\odot = 2.87\times10^{-6}\,\mathrm{rad}\,\mathrm{s}^{-1}$. As an extension to fit a wider variety of solar source conditions, we additionally allow for a longitudinal coordinate multiplier $n_\mathrm{nl}$, which in effect allows for the apparent sinusoidal shape of the neutral line to have one, two, or more full periods over 360 degrees of solar longitude. We discuss fitting of this neutral line model to solar observations in \mbox{Appendix \ref{appendix:fitting}}.

Throughout interplanetary space, we assume the shape of field lines for the IMF to be those of a Parker spiral, using a constant solar wind speed $u_\mathrm{sw}$ everywhere. For each position $(r, \theta, \phi)$ in space, we can trace the field line back to the source surface at $r_\mathrm{s}$ and find the intersection point $(r_\mathrm{s}, \theta_\mathrm{s}, \phi_\mathrm{s})$ of the field line and the source surface. We note that by following Parker spiral field lines, $\theta_\mathrm{s}=\theta$ always holds true. The value for $\phi_\mathrm{s}$ is found as
\begin{align}
\phi_\mathrm{s} &= \phi - \frac{\Omega_\odot}{u_\mathrm{sw}} (r-r_\mathrm{s}) \,.
\end{align}

Next, we find the smallest angular distance between the intersection point and the neutral line along a great circle. For this, we use a spherical coordinate transformation. For a point on the source surface with HGI coordinates $(\theta_\mathrm{s},\phi_\mathrm{s})$, the angular distance from the neutral line can be given as $\delta'=90^\circ-\theta'$ where $\theta'$ is the colatitude of the point in a spherical coordinate system rotated so that its z-axis aligns with the tilted dipole axis. Using a standard coordinate transformation (extended with the oscillation multiplier $n_\mathrm{nl}$), we solve $\theta'$, and by extension, $\delta'$, from
\begin{align}
\cos \theta' = & \cos \theta_\mathrm{s} \cos \alpha_\mathrm{nl} \nonumber \\
 & + \sin \theta_\mathrm{s} \sin \alpha_\mathrm{nl} \sin \left( n_\mathrm{nl}(\phi_\mathrm{s}-\Phi_\mathrm{nl}(t))\right)\,. \label{eq:tiltheta1}
\end{align}

The presented method allows finding the smallest angular distance from the HCS, for any position within the inner heliosphere, using a simple analytic equation. For a numerical full-orbit particle propagation model, it is important to have a fast method for finding the electric and magnetic field values at any position. We can then use this angular distance from the HCS to find the required field values including the correct magnetic polarity.

In Paper I, the magnetic field incorporating a flat HCS model, in spherical heliocentric coordinates, was a scaled Parker spiral magnetic field
\begin{align}
  {\bf B} &= S(\delta') {\bf B}_\mathrm{Parker} \,, \label{eq:SParker}
\end{align}
where $S$ is a shape function, ${\bf B}_\mathrm{Parker}$ is the classical Parker field
\begin{align}
  B_{r,\mathrm{Parker}} &= B_0 \frac{r_0^2}{r^2}  \label{eq:Br}\\
  B_{\theta,\mathrm{Parker}} &= 0 \label{eq:Btheta}\\
  B_{\phi,\mathrm{Parker}} &= -\frac{B_0 r_0^2 \Omega_\odot}{u_\mathrm{sw}} \frac{\sin \theta}{r} \,, \label{eq:Bphi}
\end{align}
and $\delta'$ is the smallest great circle angular distance to the HCS. The field strength is normalized through $B_0$ to provide a field strength at \mbox{1 au} of $B(1\,\mathrm{au})=3.85\,\mathrm{nT}$, in agreement with observations. The average solar rotation rate $\Omega_\odot$ and a constant radial solar wind speed $u_\mathrm{sw}=500 \,\mathrm{km}\,\mathrm{s}^{-1}$ parametrize the Parker spiral winding. For these values, the longitudinal winding angle at \mbox{1 au} is $49^\circ$. 

Both in Paper I and in the current work, we use the shape function $S$, parametrizing the direction and strength of the magnetic field at given HGI coordinates in relation to the position of the current sheet. Our shape function is defined as
\begin{align}
S(\delta') = A \left(-1 + 2\,\mathcal{S}\,(\tfrac{1}{2} + \frac{2\delta'}{l_\mathrm{HCS}}) \right) \, , \label{eq:shapefunction}
\end{align}
where $A$ takes values of $+1$ or $-1$ to define the dipole field polarity. As is standard in cosmic ray physics, we refer to an $A+$ ($A-$) configuration as defining a northern hemisphere field pointing outward (inward), with an opposite field direction in the southern hemisphere. The HCS thickness is defined by the parameter $l_\mathrm{HCS}$, normalised to provide a thickness of \mbox{5000 km} at \mbox{1 au} (see, e.g., \citealt{Eastwood2002} and \citealt{Winterhalter1994}). $\mathcal{S}$ is the \emph{Smootherstep} function (see \citealt{ebert2003texturing} and Paper I). As the shape function's controlling parameter $\delta'$ is angular, this parametrization results in a current sheet thickness which increases with distance from the Sun.

In order to solve SEP propagation, we also need to consider the motional electric field, which is given by
\begin{align}
{\bf E} = -\frac{{\bf u}_\mathrm{sw}}{c}\times{\bf B} \,, \label{eq:motionalfield}
\end{align}
where $c$ is the speed of light. Thus, equations \ref{eq:SParker} and \ref{eq:motionalfield} together with equations \ref{eq:Br}--\ref{eq:Bphi} describe the electric and magnetic fields, in which our particles propagate. Particle motion is then solved using equations (4) and (5) from \cite{Dalla2005}.


\section{Simulations}\label{sec:simulations}
To study the effects of a wavy HCS on SEP propagation, we performed numerical particle transport simulations in the fixed frame using the test particle model originally introduced in \cite{Dalla2005} and applied to heliospheric propagation by \cite{Kelly2012}, \cite{Marsh2013}, \cite{Marsh2015}, and Paper I. We solve the full three-dimensional differential equations of motion for each particle. During interplanetary travel, we scatter particles in the solar wind frame at Poisson-distributed time intervals, in effect modeling a parametrized mean free path. The interactions with magnetic and electric fields and scattering within the full-orbit calculations cause drifts and deceleration effects to arise naturally.

In order to model the effects a wavy HCS has on SEP propagation, we chose to launch protons from various different injection regions. We chose a region either overlapping the HCS (and at the solar equator), or with the closest corner within a $<2^\circ$ distance to it, or distinctly further out. The injection regions were placed at $r=2.5R_\odot$, and had an angular extent of $6^\circ \times 6^\circ$. We distributed the initial energies of our protons according to a $\gamma=-1.1$ power law, between 10 and \mbox{800 MeV}. Particles were initialised with a random pitch-angle pointing outward from the Sun along the Parker spiral. We solve the relativistic differential equations of particle motion using a self-optimizing Bulirsch-Stoer method \citep{press1996numerical}. In order to verify the accuracy of our simulations, we performed some duplicates using a Boris-Push method \citep{Boris1970}, and found the results to agree. For all simulations, we used the mean free path \mbox{$\lambda_\mathrm{mfp}=1\,\mathrm{au}$}, independent of energy. For this study, we assumed a constant solar wind speed of $u_\mathrm{sw}=500\,\mathrm{km}\,\mathrm{s}^{-1}$.

Each simulation consisted of $N=10^5$ particles and lasted for 100 hours. We recorded each particle crossing over the \mbox{1 au} sphere, so that these events can be used to construct fluence maps and energy spectrograms at \mbox{1 au}. 

We selected six heliospheric configurations to parametrize the fields in our simulations, designated A through F. The first four configurations were based on the results of applying a neutral line fitting algorithm to an actual solar source surface map, as described in Appendix \ref{appendix:fitting}. The final two configurations were manually selected to represent a strongly deformed current sheet. Using cosmic ray notation, a configuration of $A+$ (simulations A, C, and E) indicates outward-pointing fields in the northern hemisphere and inward-pointing fields in the southern hemisphere, and vice versa for a configuration of $A-$ (simulations B, D, and F).

As presented in Paper I, the IMF configuration has a significant role on how particle drifts and the HCS together alter particle propagation. As our previous study results showed little dependence on HCS thickness, we used a single thickness parameter for all simulations, resulting in a HCS with a thickness of \mbox{5000 km} at \mbox{1 au}. The simulation parameters for all 24 of our simulations are presented in Table \ref{tab:simparams}, grouped according to simulation background configuration.

\begin{table}
\caption{Simulation parameters for 24 runs, labelled A1 through F4. For each simulation, we list the IMF polarity configuration, the wavy neutral line fit parameters $n_\mathrm{nl}$, $\alpha_\mathrm{nl}$, and $\phi_\mathrm{nl}$, and the centre point of the \mbox{$6^\circ \times 6^\circ$} injection region as Carrington latitude $\delta_{0,\mathrm{inj}}$ and longitude $\phi_{0,\mathrm{inj}}$.}
\label{tab:simparams}
\centering
\begin{tabular}{l l r r r r r}
\hline
Run & IMF  & $n_\mathrm{nl}$ & $\alpha_\mathrm{nl}$ & $\phi_\mathrm{nl}$ & $\delta_{0,\mathrm{inj}}$ & $\phi_{0,\mathrm{inj}}$ \\ 
\hline
\hline
A1 & A+  & 1 & 29 & 210 & -6 & 210 \\
A2 & A+  & 1 & 29 & 210 & 0 & 210 \\
A3 & A+  & 1 & 29 & 210 & 6 & 210 \\
\hline
B1 & A-- & 1 & 29 & 210 & -6 & 218 \\ 
B2 & A-- & 1 & 29 & 210 & 0 & 218 \\ 
B3 & A-- & 1 & 29 & 210 & 6 & 218 \\ 
\hline
C1 & A+  & 2 & 37 & 77  & -9 & 167 \\
C2 & A+  & 2 & 37 & 77  & 0 & 167 \\
C3 & A+  & 2 & 37 & 77  & 9 & 167 \\
C4 & A+  & 2 & 37 & 77  & -20 & 122 \\
C5 & A+  & 2 & 37 & 77  & 20 & 212 \\
\hline
D1 & A-- & 2 & 37 & 132 & -9 & 222 \\
D2 & A-- & 2 & 37 & 132 & 0 & 222 \\
D3 & A-- & 2 & 37 & 132 & 9 & 222 \\
D4 & A-- & 2 & 37 & 132 & -20 & 177 \\
D5 & A-- & 2 & 37 & 132 & 20 & 87 \\
\hline
E1 & A+ & 2 & 85 & 0 & -30 & -10 \\
E2 & A+ & 2 & 85 & 0 & 30 & -10 \\
E3 & A+ & 2 & 85 & 0 & -30 & 10 \\
E4 & A+ & 2 & 85 & 0 & 30 & 10 \\
\hline
F1 & A-- & 2 & 85 & 0 & -30 & -10 \\
F2 & A-- & 2 & 85 & 0 & 30 & -10 \\
F3 & A-- & 2 & 85 & 0 & -30 & 10 \\
F4 & A-- & 2 & 85 & 0 & 30 & 10 \\
\hline
\end{tabular}
\end{table}


\section{Results} \label{sec:results}
We present the results of our particle transport simulations through fluence maps and energy spectrograms, counting particle crossings over the \mbox{1 au} sphere. We present simulations using six different IMF configurations, and for each one we consider several injection locations at different latitudes (see Table \ref{tab:simparams}), performing comparative analysis between the different IMF configurations. Our novel results are also compared with results published in Paper I.

In analyzing particle drifts, we consider longitudinal motion of particles relative to the field line which was best connected to the centre of the injection region at the time of injection. Due to solar rotation, the region from which injection occurred, and thus, the field lines connected to it, co-rotate westward at a set angular rate. If considering particle motion in the fixed frame, all particles and the IMF configuration itself experience the same time-dependent but energy-independent corotation, analogous with the ${\bf E}\times{\bf B}$ drift.

In our fluence maps and energy spectrograms, we either plot longitudes in relation to a stationary \mbox{1 au} observer (with corotation), or in relation to the centrally connected field line (with corotation removed). Removing the effect of corotation allows us to better analyze the remaining curvature, gradient, and HCS particle drifts, whereas leaving it in more closely represents how a true observer would detect particle flux in interplanetary space. 

If corotation is removed and the effects of a current sheet ignored, the curvature and gradient drifts result in quasi-symmetric longitudinal and unidirectional latitudinal drifts, shaped like a fan, as shown in the top two panels of Figure 4 of Paper I. Protons drift in latitude according to the magnetic field direction (southward for outward-pointing field lines and northward for inward-pointing field lines), and approximately equally both east and west in longitude. As shown in Figure 14 of Paper I, an A+ IMF configuration truncates curvature and gradient drift patterns at the location of the HCS.

\subsection{Fluence maps}
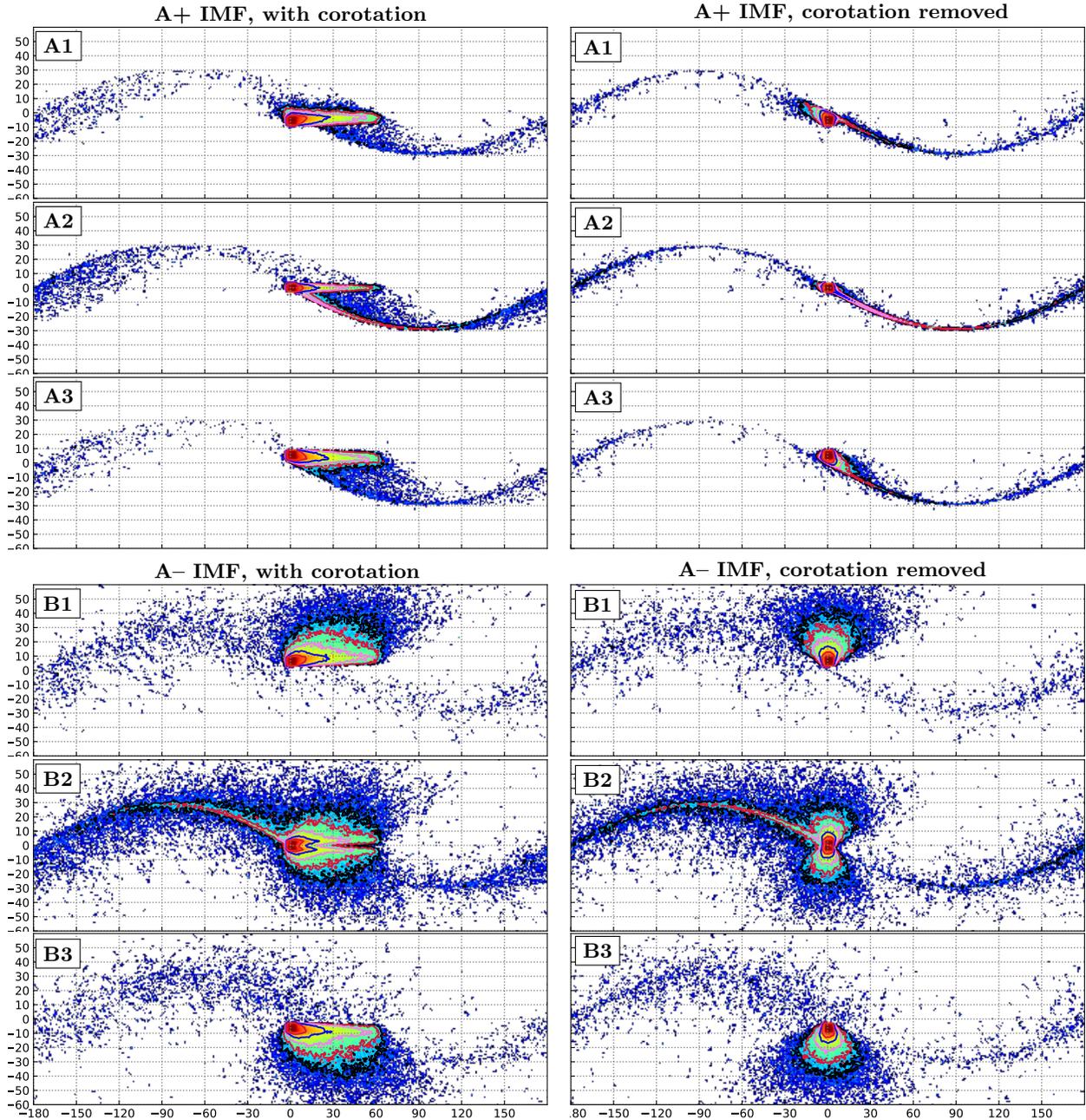
\begin{figure*}[!ht]
\centering 
\begin{overpic}[trim=10 20 8 -20, clip, scale=0.5]{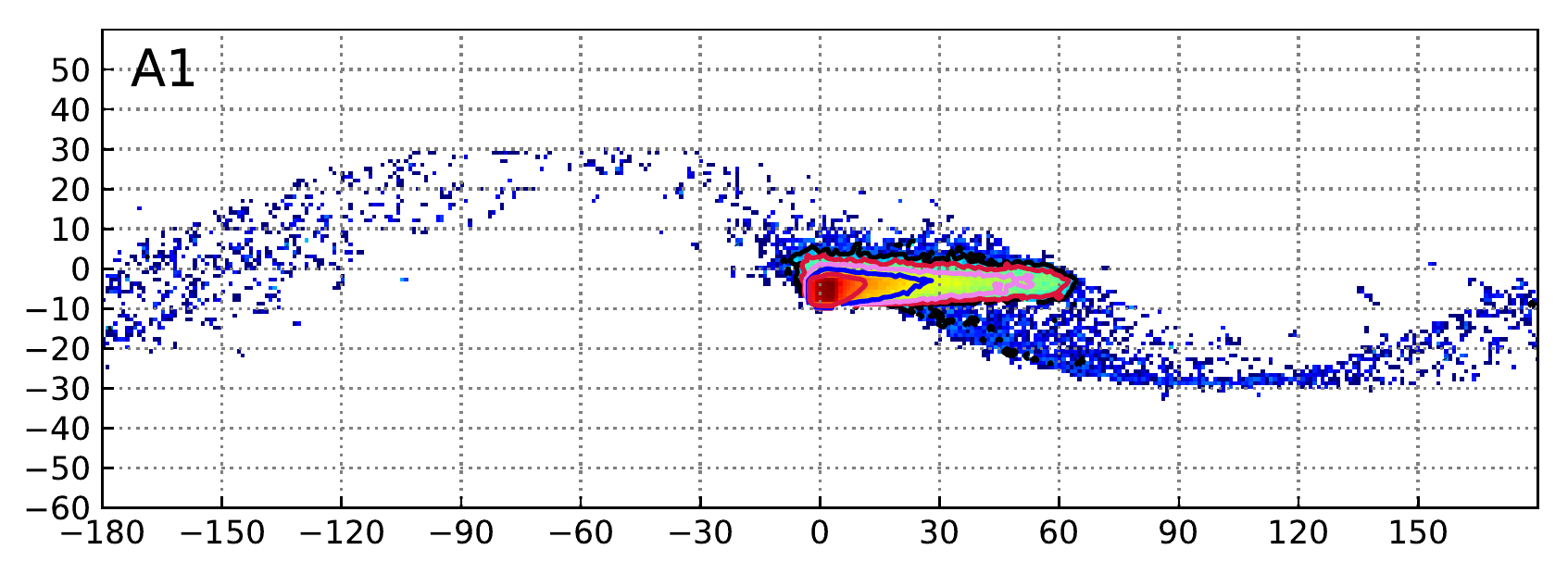}
  \put(27,33){\bf A+ IMF, with corotation}
  \put(5,27){\fcolorbox{black}{white}{\bf A1}}
\end{overpic}
\begin{overpic}[trim=31 20 8 -20, clip, scale=0.5]{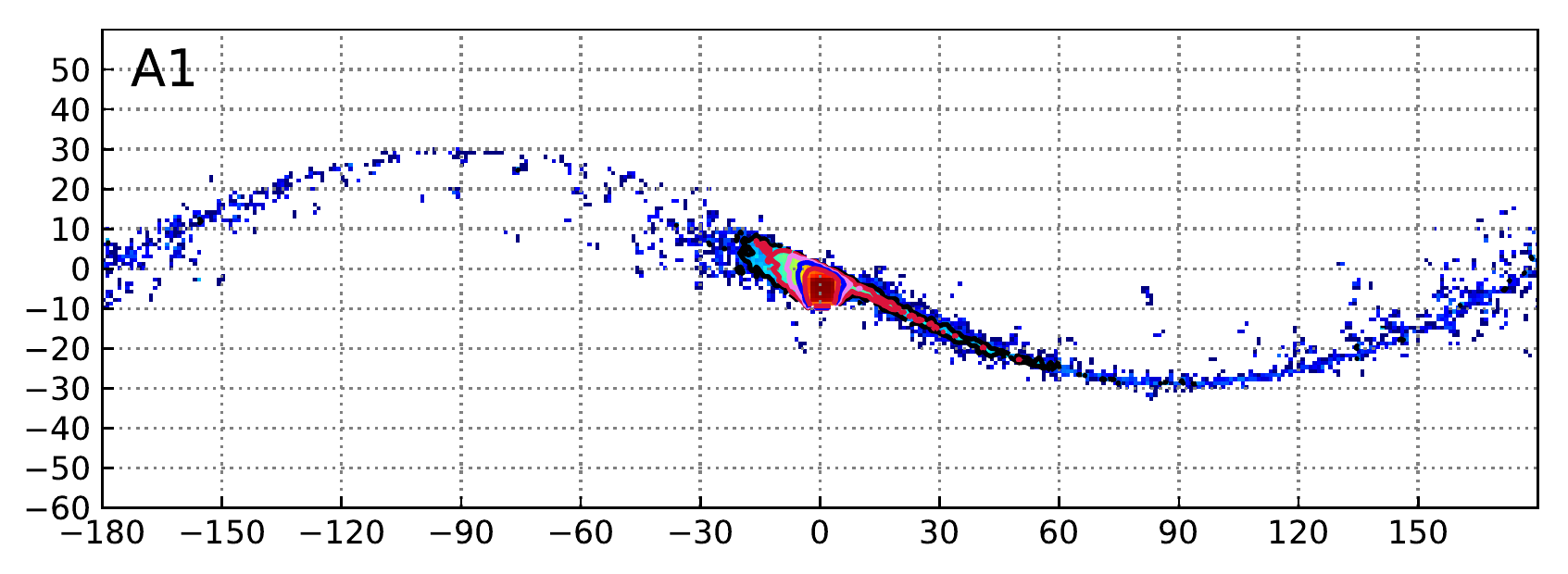}
  \put(24,35){\bf A+ IMF, corotation removed}
  \put(1,28){\fcolorbox{black}{white}{\bf A1}}
\end{overpic} \\
\begin{overpic}[trim=10 20 8 8, clip, scale=0.5]{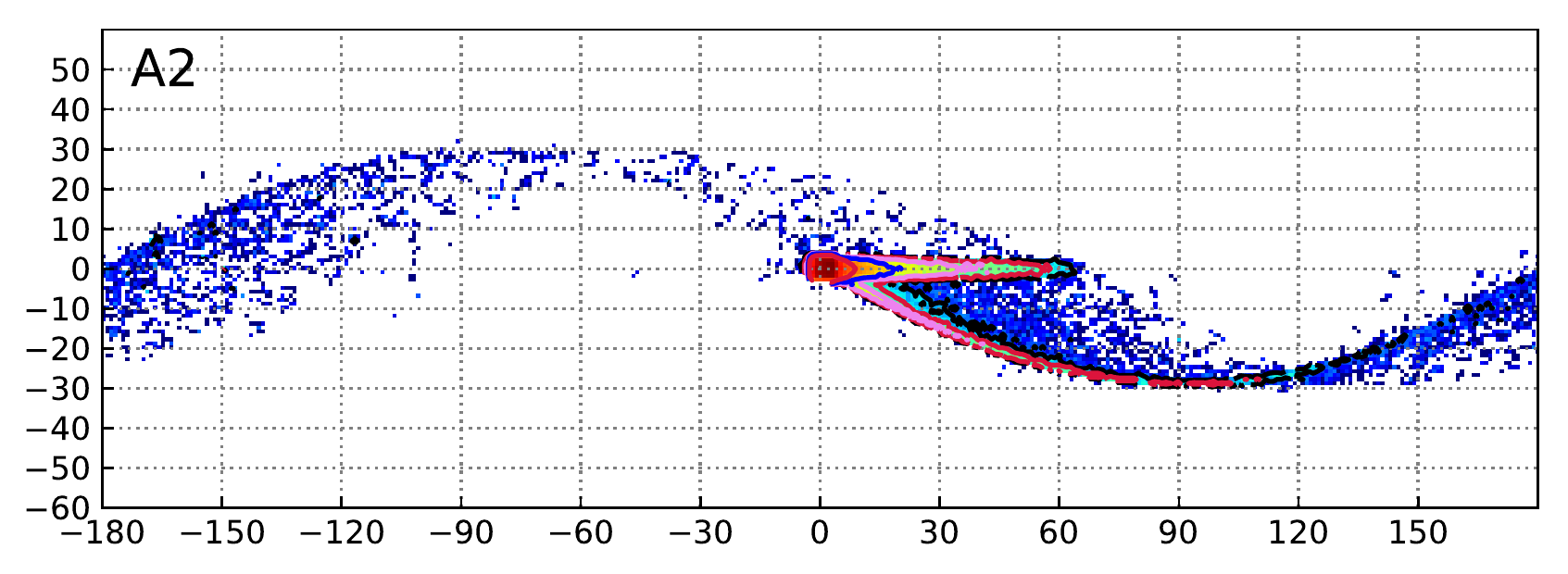}
  \put(5,27){\fcolorbox{black}{white}{\bf A2}}
\end{overpic}
\begin{overpic}[trim=31 20 8 8, clip, scale=0.5]{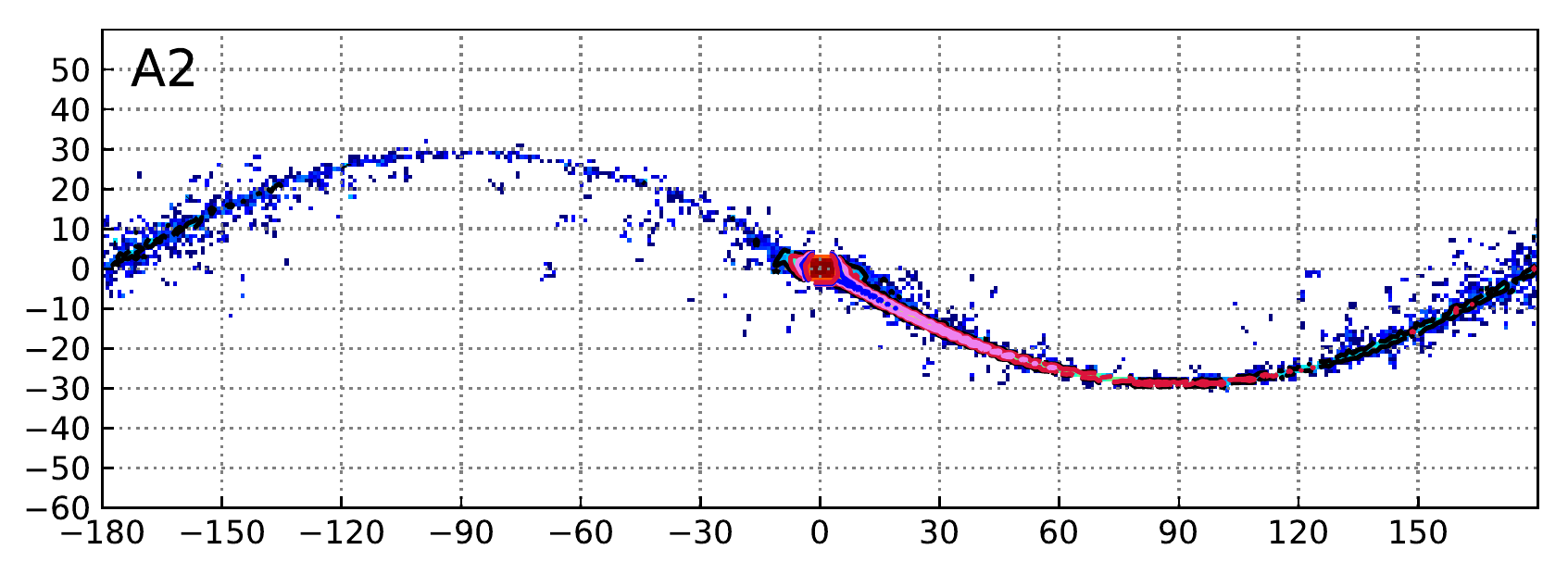}
  \put(1,28){\fcolorbox{black}{white}{\bf A2}}
\end{overpic} \\
\begin{overpic}[trim=10 20 8 8, clip, scale=0.5]{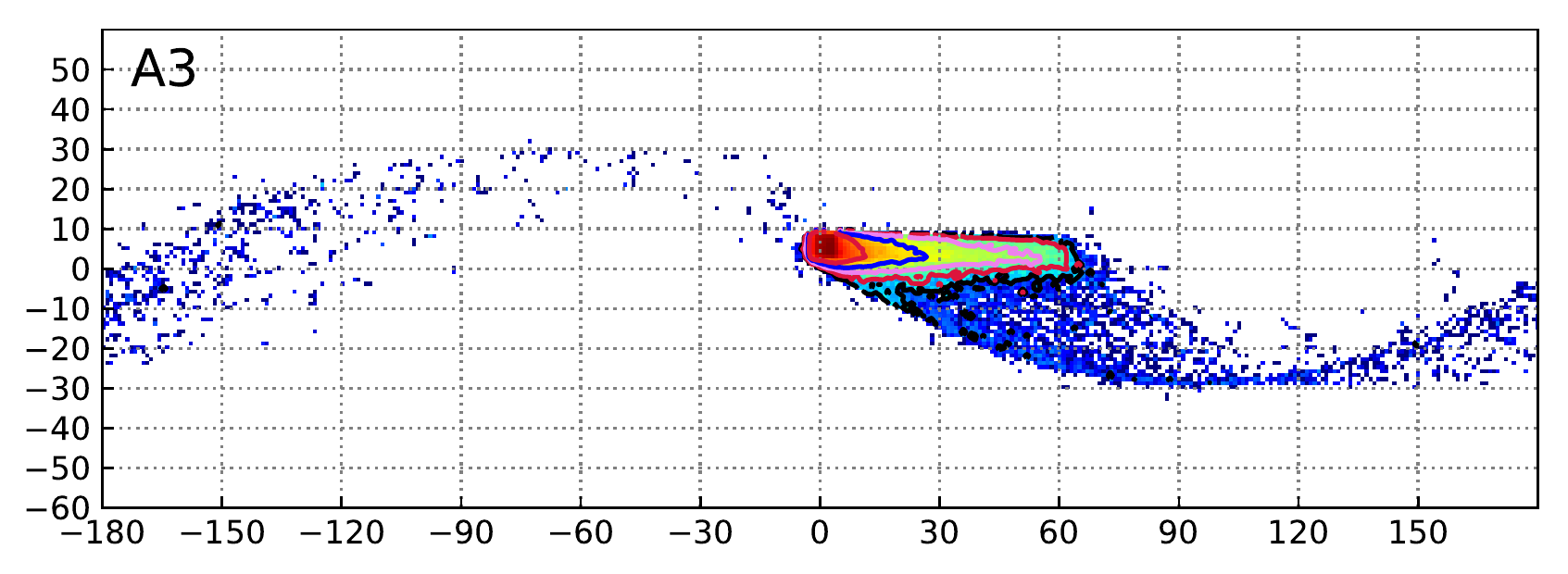}
  \put(5,27){\fcolorbox{black}{white}{\bf A3}}
\end{overpic}
\begin{overpic}[trim=31 20 8 8, clip, scale=0.5]{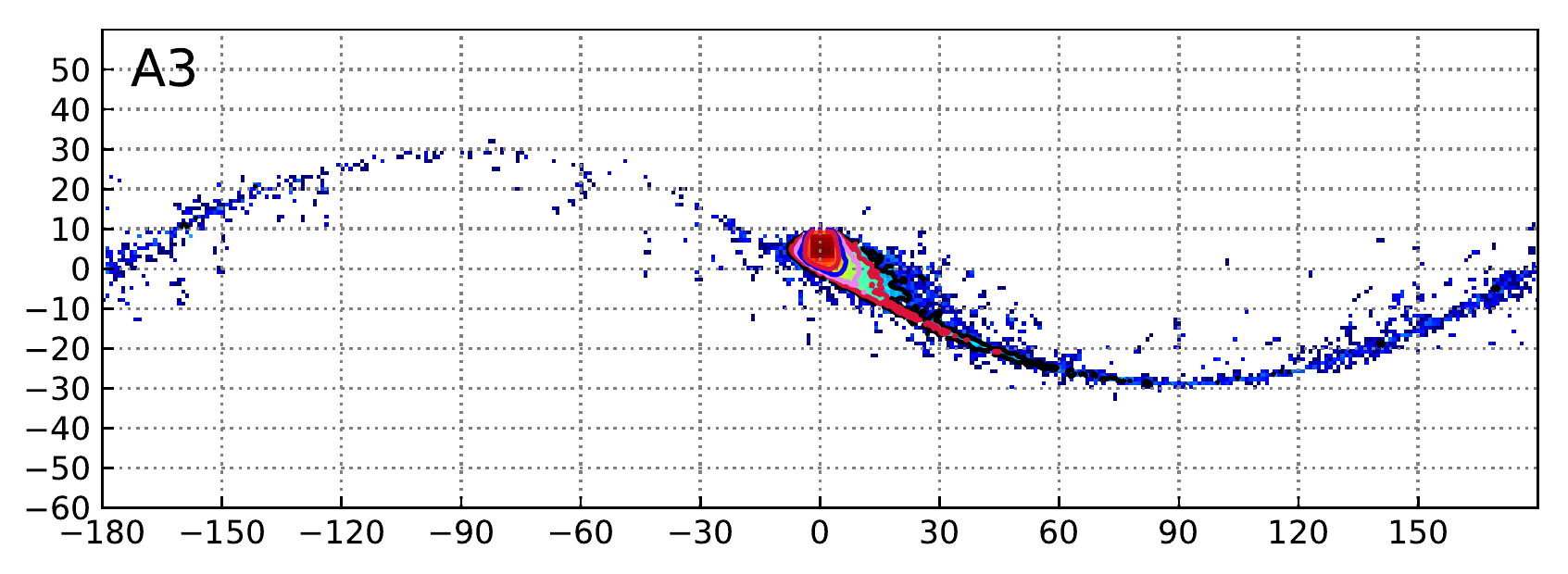}
  \put(1,28){\fcolorbox{black}{white}{\bf A3}}
\end{overpic} \\
\begin{overpic}[trim=10 20 8 -20, clip, scale=0.5]{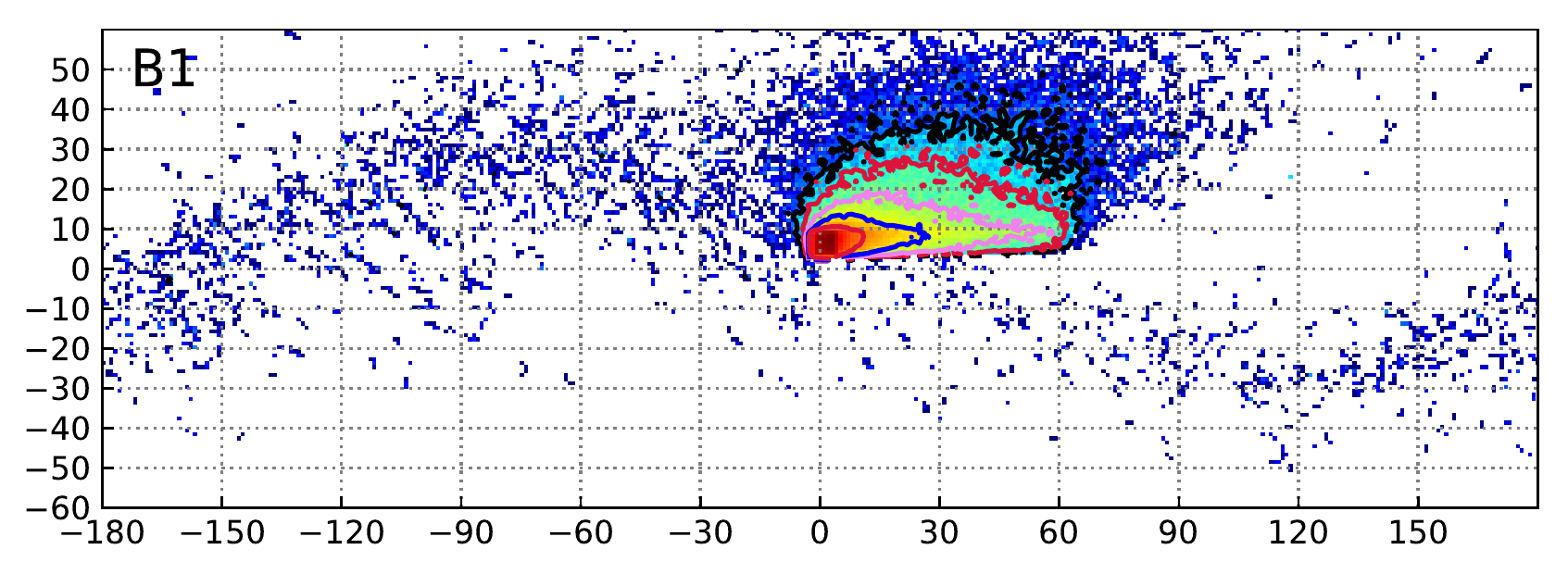}
  \put(27,33){\bf A-- IMF, with corotation}
  \put(5,27){\fcolorbox{black}{white}{\bf B1}}
\end{overpic}
\begin{overpic}[trim=31 20 8 -20, clip, scale=0.5]{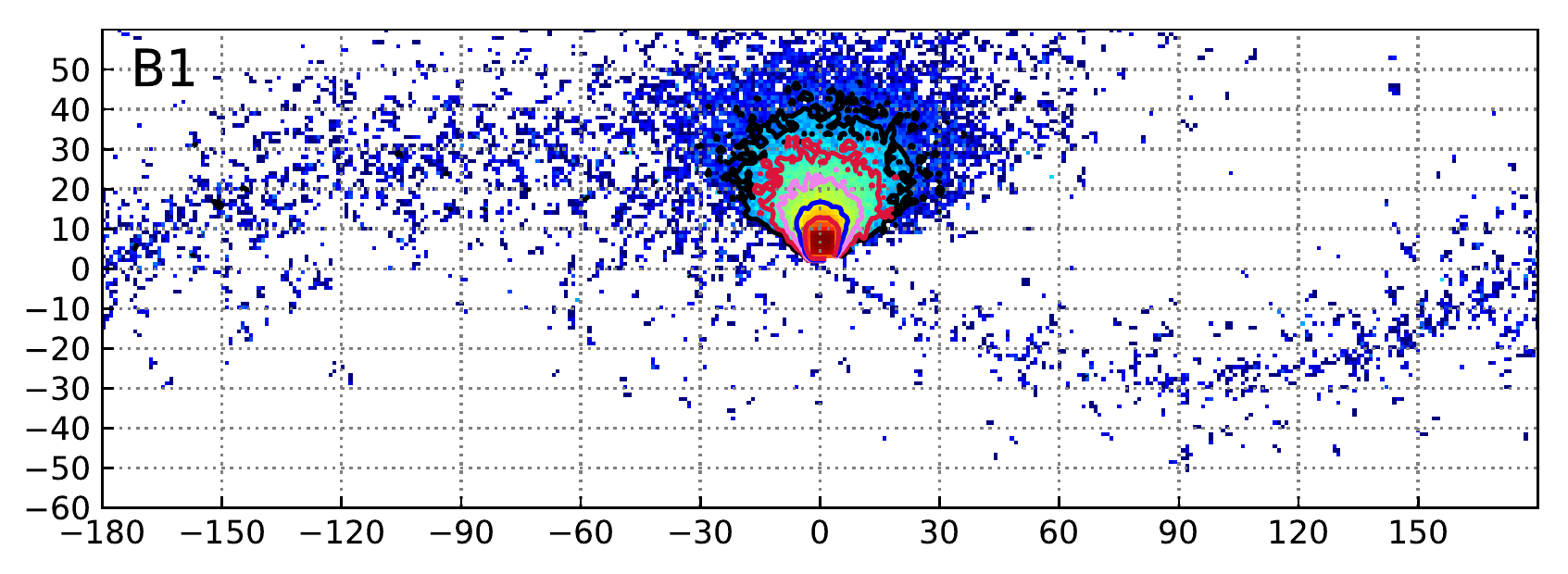}
  \put(21,35){\bf A-- IMF, corotation removed}
  \put(1,28){\fcolorbox{black}{white}{\bf B1}}
\end{overpic} \\
\begin{overpic}[trim=10 20 8 8, clip, scale=0.5]{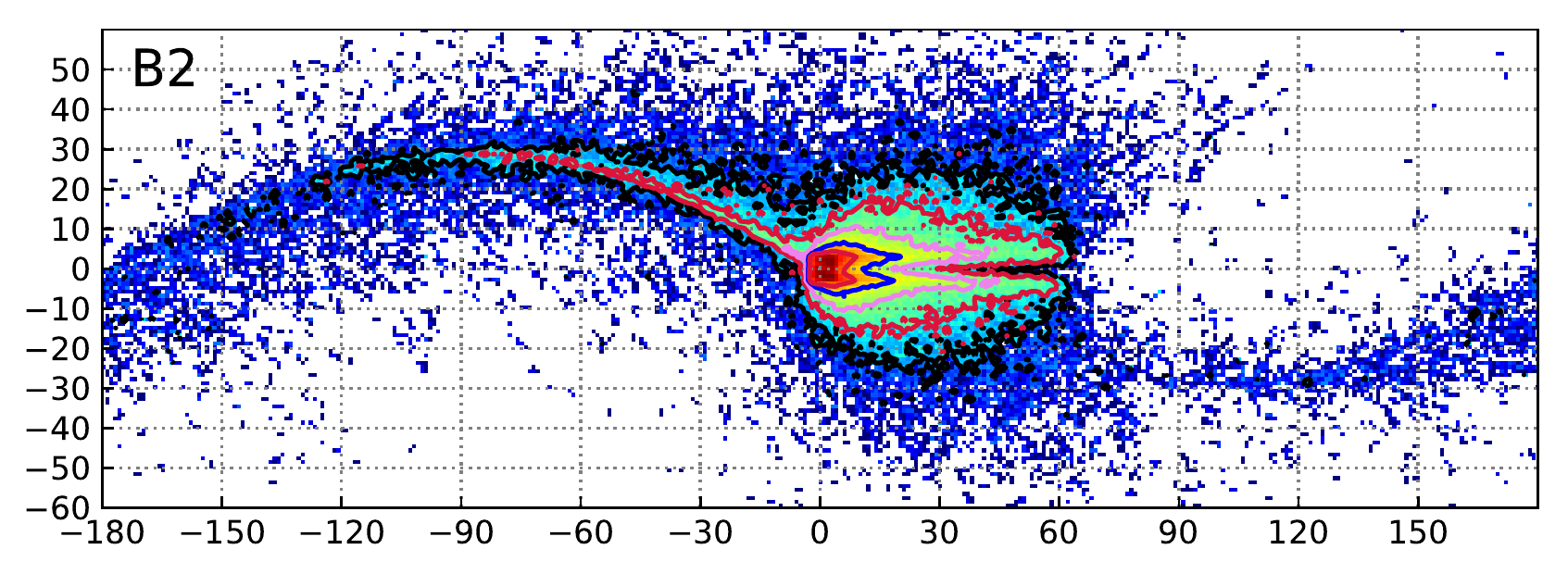}
  \put(5,27){\fcolorbox{black}{white}{\bf B2}}
\end{overpic}
\begin{overpic}[trim=31 20 8 8, clip, scale=0.5]{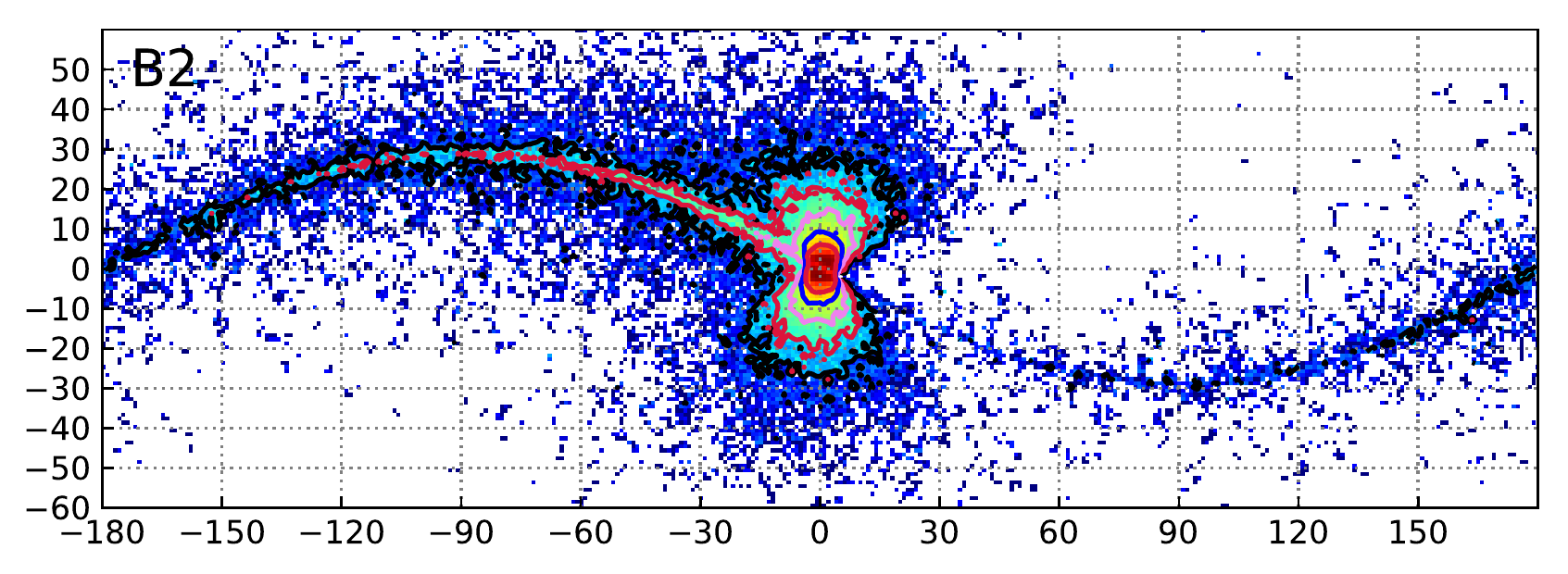}
  \put(1,28){\fcolorbox{black}{white}{\bf B2}}
\end{overpic} \\
\begin{overpic}[trim=10 0 8 8, clip, scale=0.5]{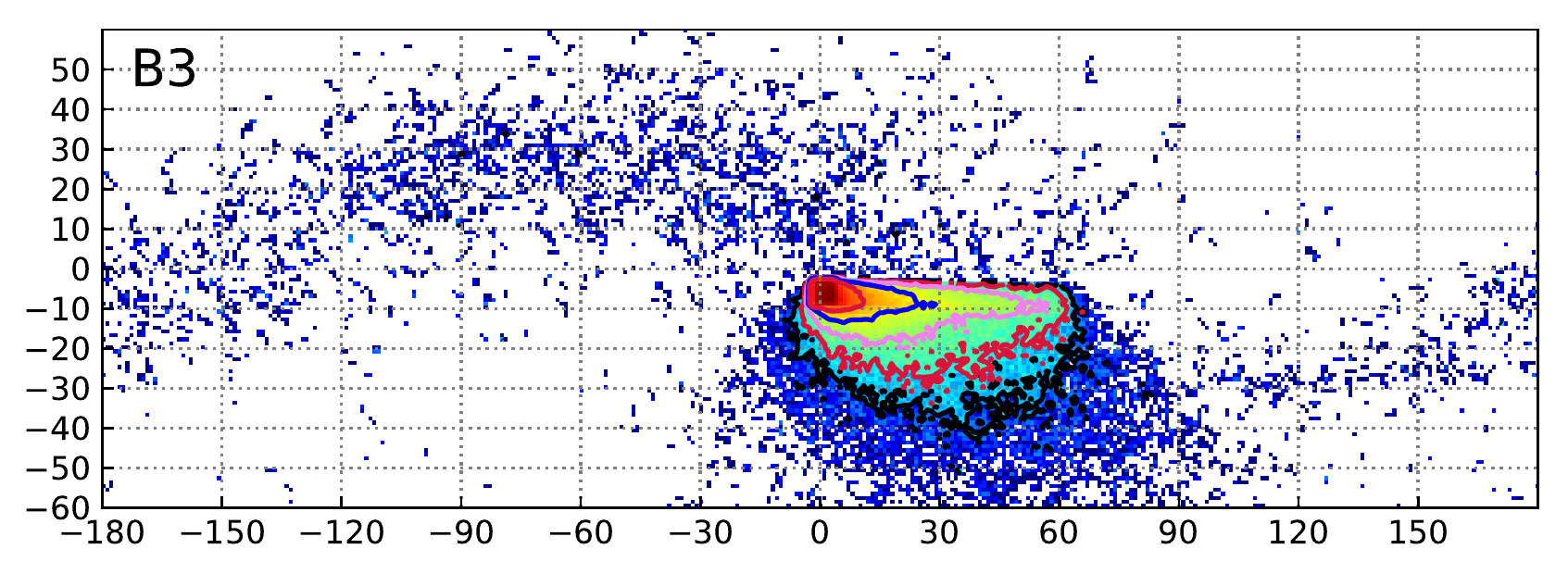}
  \put(5,31){\fcolorbox{black}{white}{\bf B3}}
\end{overpic}
\begin{overpic}[trim=31 0 8 8, clip, scale=0.5]{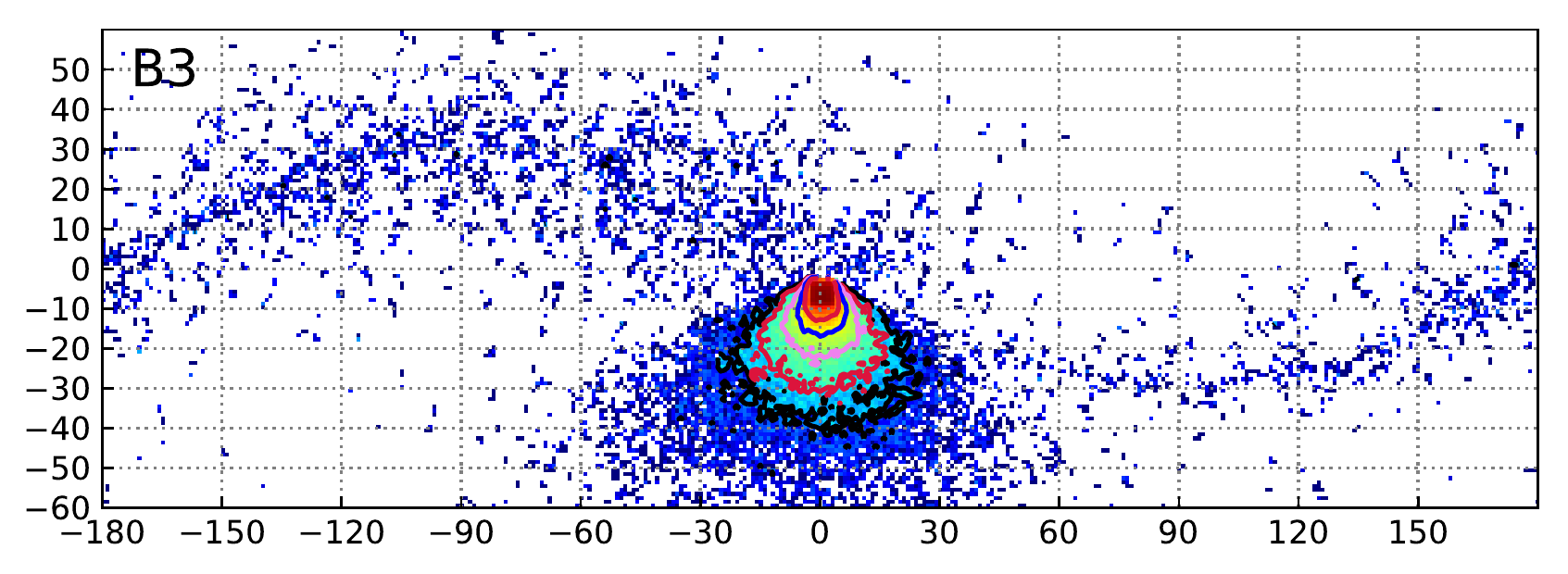}
  \put(1,33){\fcolorbox{black}{white}{\bf B3}}
\end{overpic} \\
\caption{
Fluence maps of protons, injected from a power-law with $\gamma=-1.1$, spanning the energy range from 10 to 800 MeV, crossing the \mbox{1 au} sphere, over a time of \mbox{100 hr}. Fluence colours are on a logarithmic scale, overlaid by contours, two per decade. Injection regions were $6^\circ \times 6^\circ$. Panels are labelled A1 through B3 according to simulation setup. The left column shows crossings in relation to the location of the centrally connected field line at the time of injection, and the right column in coordinates corotating with the field lines connected to the injection region, in effect removing effects of corotation.}\label{fig:corot}
\end{figure*}

In Figure \ref{fig:corot}, we plot fluence maps of protons crossing over the \mbox{1 au} sphere for simulations in groups A (rows 1 through 3) and B (rows 4 through 6). Crossings over the whole 100 hour simulation length and both outward and Sunward are included. The shape of the HCS is easily distinguished from the fluence maps. 
As is well known from GCR patterns and as shown in Paper I, the A+ IMF configuration causes southward latitudinal proton drifts in the northern hemisphere and northward latitudinal proton drifts in the southern hemisphere. For a flat HCS configuration, this would results in drifts towards the HCS. The A-- IMF configuration, in turn, causes protons to drift toward the poles.

The left and right columns in Figure \ref{fig:corot} show the effect of corotation on the apparent spread of particles. All charged particles experience drift due to the \mbox{$\bf E \times B$} force, which is identical to all particles and analogous with corotation of field lines. Over the 100 hours of simulation presented here, corotation causes field lines to move westward by $59^\circ$. In the left column, we display proton crossings in HGI coordinates, and in the right column, crossings where this effect of corotation has been removed, i.e. in the corotating frame. In effect, the right column shows fluence maps in relation to those field lines which were connected to the injection region at the start of the simulation. Although the left column provides a more accurate depiction of proton fluences from an observational point of view, the right column with corotation removed allows for more detailed analysis of less predictable drift effects.

As can be seen from Figure \ref{fig:corot}, protons which have access to the HCS start to travel along it due to current sheet drift \citep{Burger1985}. We note that as in the flat HCS model presented in Paper I, particles travel along the current sheet for great distances, up to 360 degrees in longitude. For the A+ configuration the current sheet drift is westward and for A-- it is eastward. For simulation sets 1 and 3, the injection region is not intersected by the HCS, being located below the HCS in sets 1 and above it in sets 3. Both A+ and A-- IMF configurations allow a small amount of particles to drift or scatter close enough to the sheet in order to experience HCS drift. Due to the direction of latitudinal gradient and curvature drifts, we expect the A+ IMF configuration to efficiently trap particles to the HCS, with the A-- IMF configuration allowing particles to drift away from the sheet more readily. For the most part, judging from the right-hand column in Figure \ref{fig:corot}, this expectation is fulfilled.

The general shapes of particle drifts seen in the right column of Figure \ref{fig:corot} display a scenario comparable with Figure 12 in Paper I (hereafter f12). Simulation set A, with the A+ IMF configuration resembles the bottom panel of f12, with simulation set B, with an A-- IMF configuration resembling the middle panel of f12. For B2, the injection region appears depleted at the centre of the HCS, as particles drift efficiently along the current sheet. Drifts toward the poles in each hemisphere show an hourglass shape in the fluence map surrounding the initial injection region. Both these effects are similar to that seen in the middle panel of f12.

Proton fluences at the HCS are strongest in simulations A2 and B2, as those injection regions are intersected by the current sheet, and thus, a larger portion of particles have easy access to the HCS and the drift it provides. In the left column of Figure \ref{fig:corot}, panel A2 shows two strong enhancements extending from the injection region, one to the west due to corotation and one to the southwest due to HCS drift. In a more complicated picture, panel B2 in the left column shows one extension to the northeast due to HCS drift, and a two-pronged extension to the west due to corotation. This split of the westward extension is due to the injection region allowing good acces to the HCS, with HCS drift transporting particles effeciently along the sheet, resulting in some of the fieldlines connected to the injection site being effectively depleted later in the simulation.

Best seen for simulation set A, a difference between the left and right columns of Figure \ref{fig:corot} is that corotation causes an apparent spread of particles in longitude in regions of large HCS inclination, as the HCS corotates westward (to the right). The right column  shows that with the effects of corotation removed, only little longitudinal escape of particles from the HCS is seen.

In panels A1--A3 we are also able to see how longitudinal and latitudinal drifts experienced by protons only extend until the protons reach the HCS. Curvature and gradient drifts, resulting in a fan shape, are in effect truncated by the HCS. Within regions of small HCS inclination, the HCS truncates the whole fan-shaped drift distribution, and most drifting particles end up impacting the HCS. In regions of large HCS inclination, this truncation involves a smaller portion of the fan-shape, allowing for a diagonal drift in the vicinity of the HCS to take place. We see this diagonal drift close to the injection region, in the right column, at longitudes \mbox{10--40} east in panel A1, and at longitudes \mbox{10--40} west in panel A3, as only a diagonal combination of gradient and curvature drifts either eastward (A1) or westward (A3) survives the truncation. In both cases, the westward A+ HCS drift is visible in addition to the fan-shaped drifts. It is important to note that this effect, providing effective asymmetry in drifts, depends both on the direction of inclination of the HCS, and the location of the injection region being either below or above the HCS. In panels B1--B3, gradient and curvature drifts are towards the poles, and the fan-shape of drifts remains untruncated.

\begin{figure*}[!ht]
\centering 
\begin{overpic}[trim=10 20 8 -20, clip, scale=0.5]{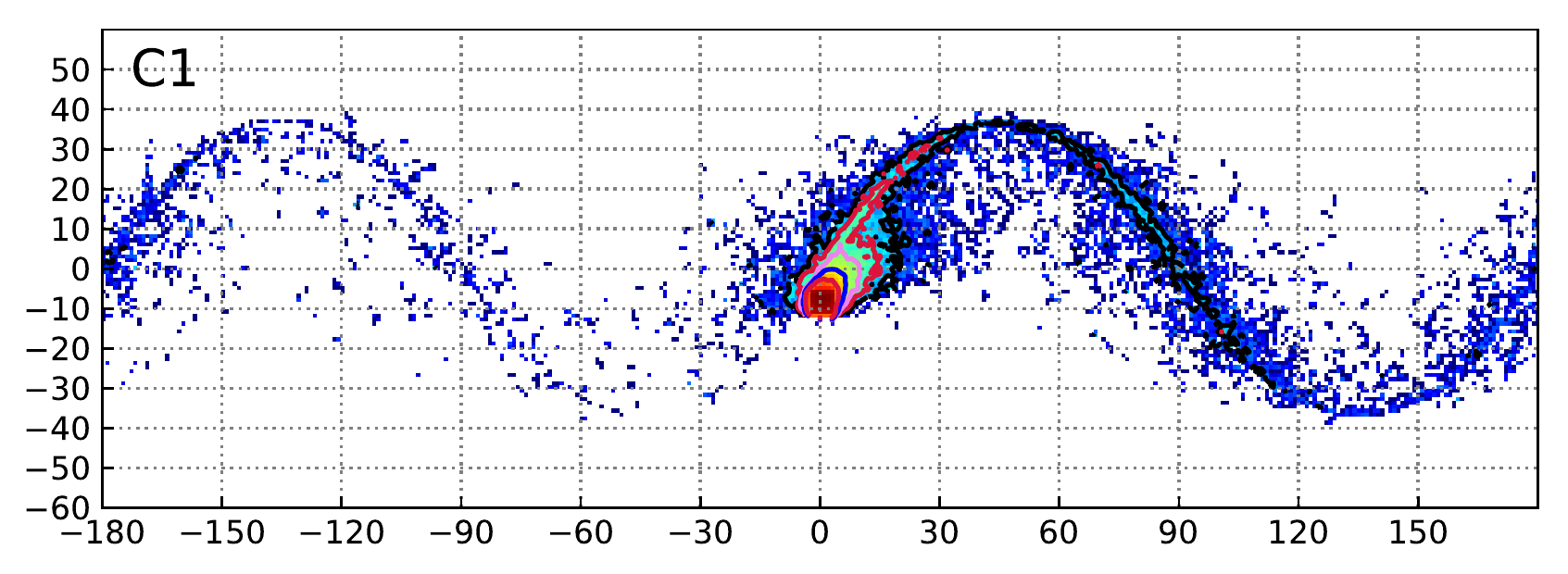}
  \put(27,33){\bf A+ IMF, corotation removed}
  \put(5,27){\fcolorbox{black}{white}{\bf C1}}
\end{overpic}
\begin{overpic}[trim=31 20 8 -20, clip, scale=0.5]{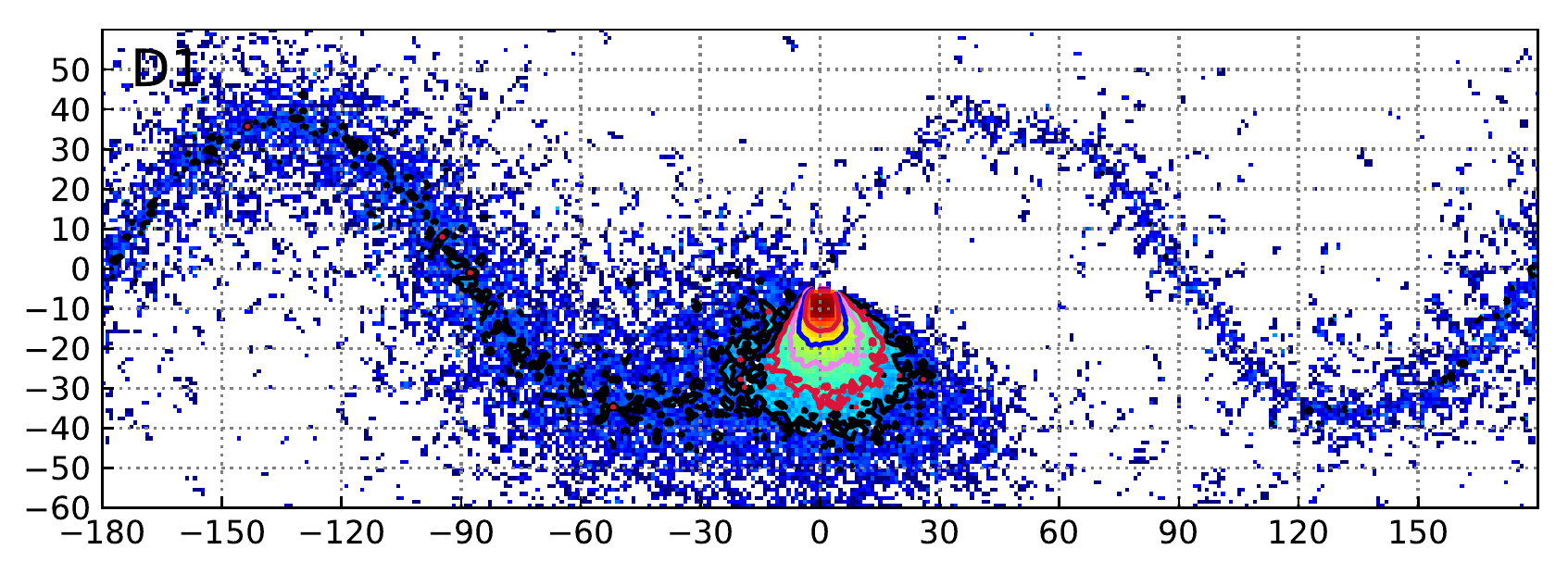}
  \put(23,35){\bf A-- IMF, corotation removed}
  \put(1,28){\fcolorbox{black}{white}{\bf D1}}
\end{overpic} \\
\begin{overpic}[trim=10 20 8 8, clip, scale=0.5]{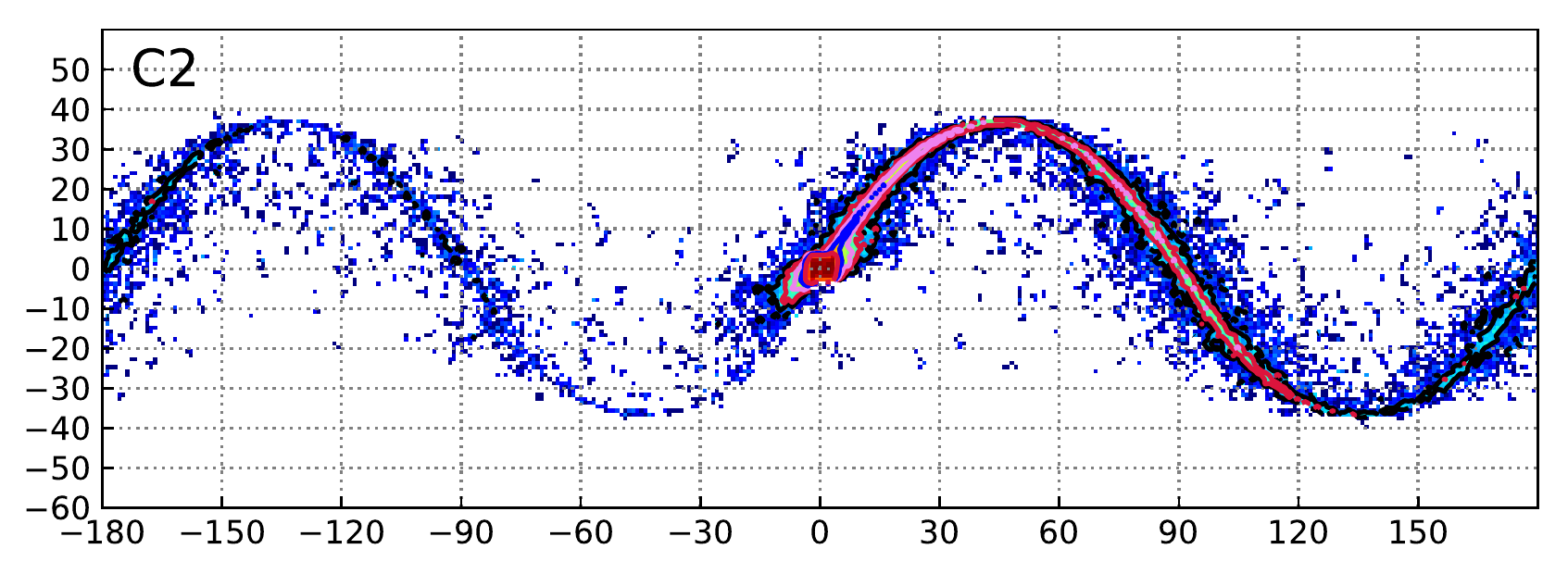}
  \put(5,27){\fcolorbox{black}{white}{\bf C2}}
\end{overpic}
\begin{overpic}[trim=31 20 8 8, clip, scale=0.5]{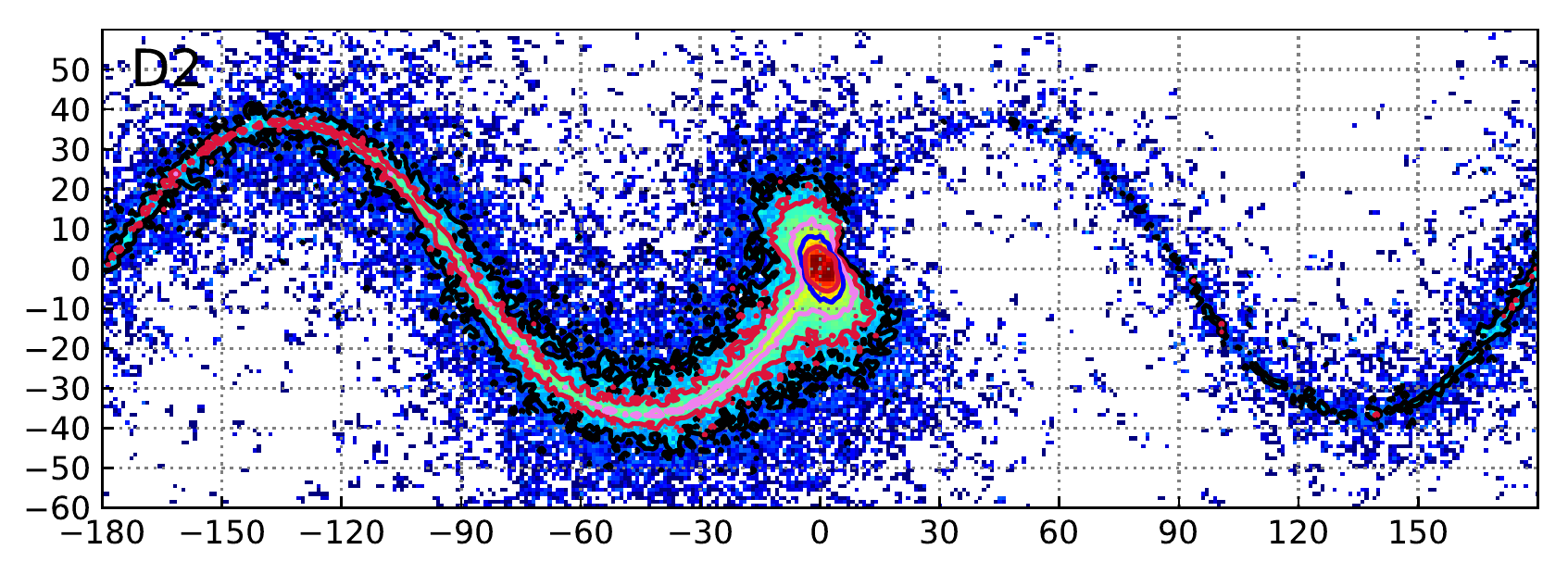}
  \put(1,28){\fcolorbox{black}{white}{\bf D2}}
\end{overpic} \\
\begin{overpic}[trim=10 20 8 8, clip, scale=0.5]{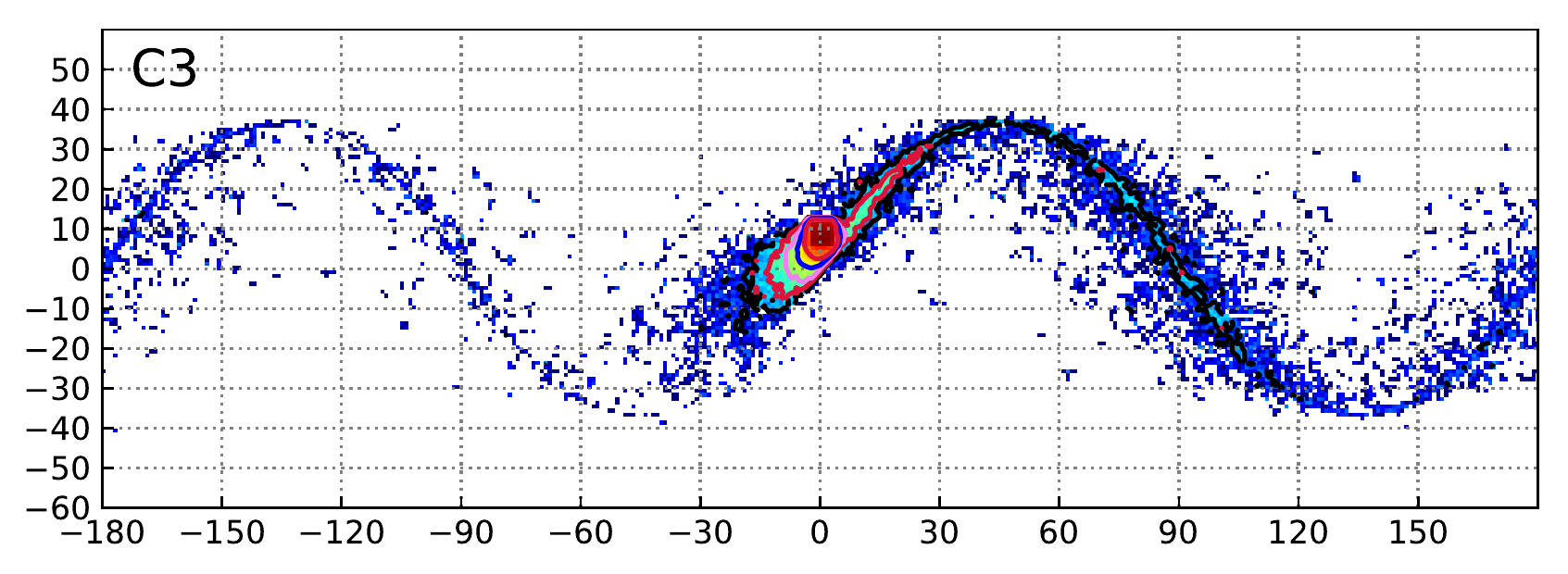}
  \put(5,27){\fcolorbox{black}{white}{\bf C3}}
\end{overpic}
\begin{overpic}[trim=31 20 8 8, clip, scale=0.5]{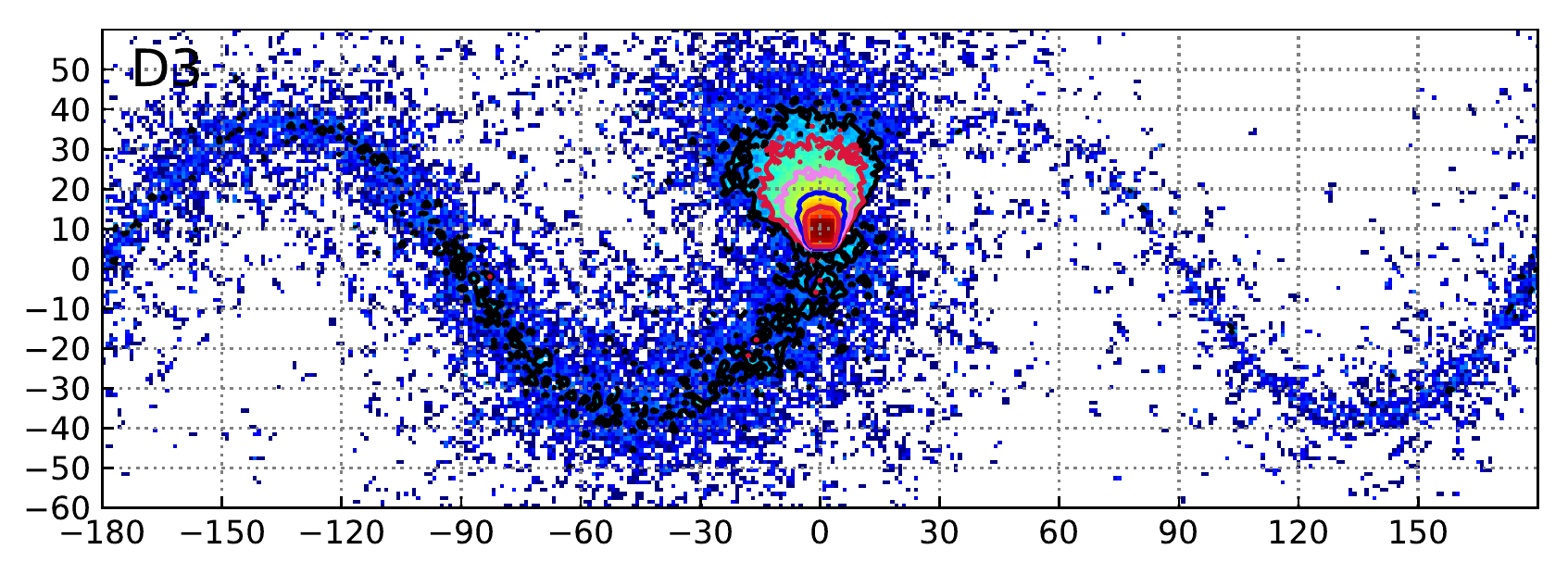}
  \put(1,28){\fcolorbox{black}{white}{\bf D3}}
\end{overpic} \\
\begin{overpic}[trim=10 20 8 8, clip, scale=0.5]{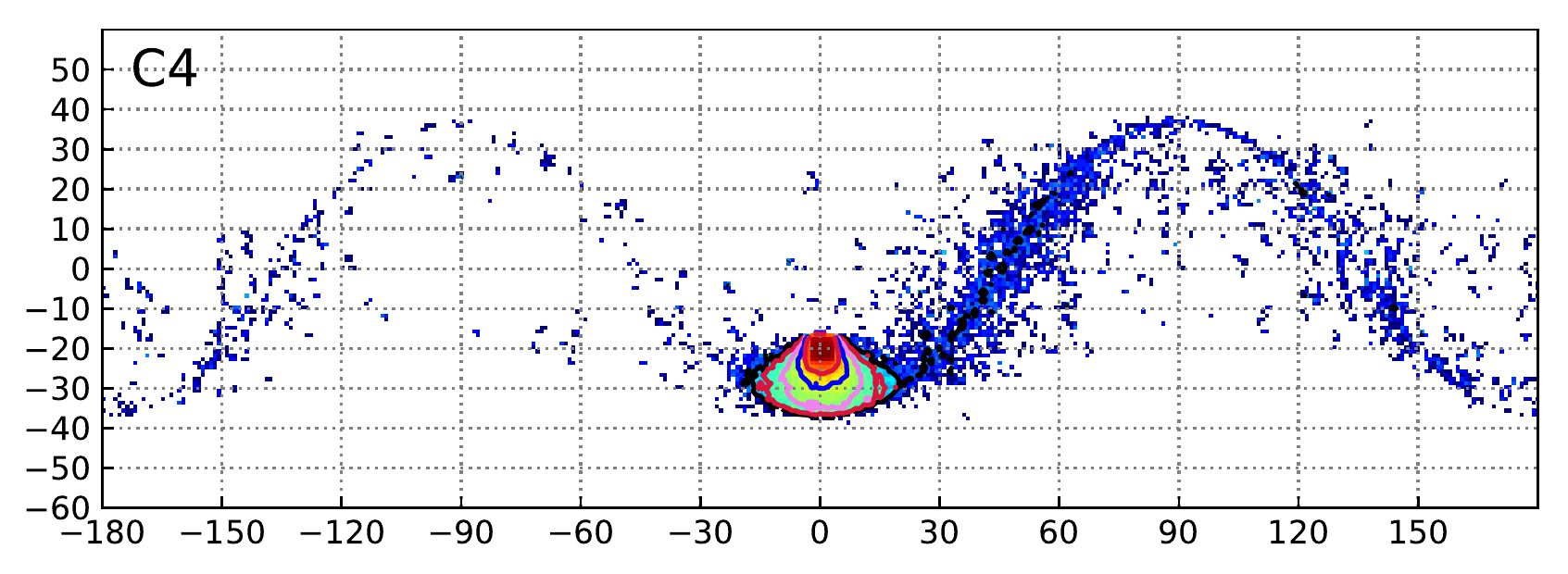}
  \put(5,27){\fcolorbox{black}{white}{\bf C4}}
\end{overpic}
\begin{overpic}[trim=31 20 8 8, clip, scale=0.5]{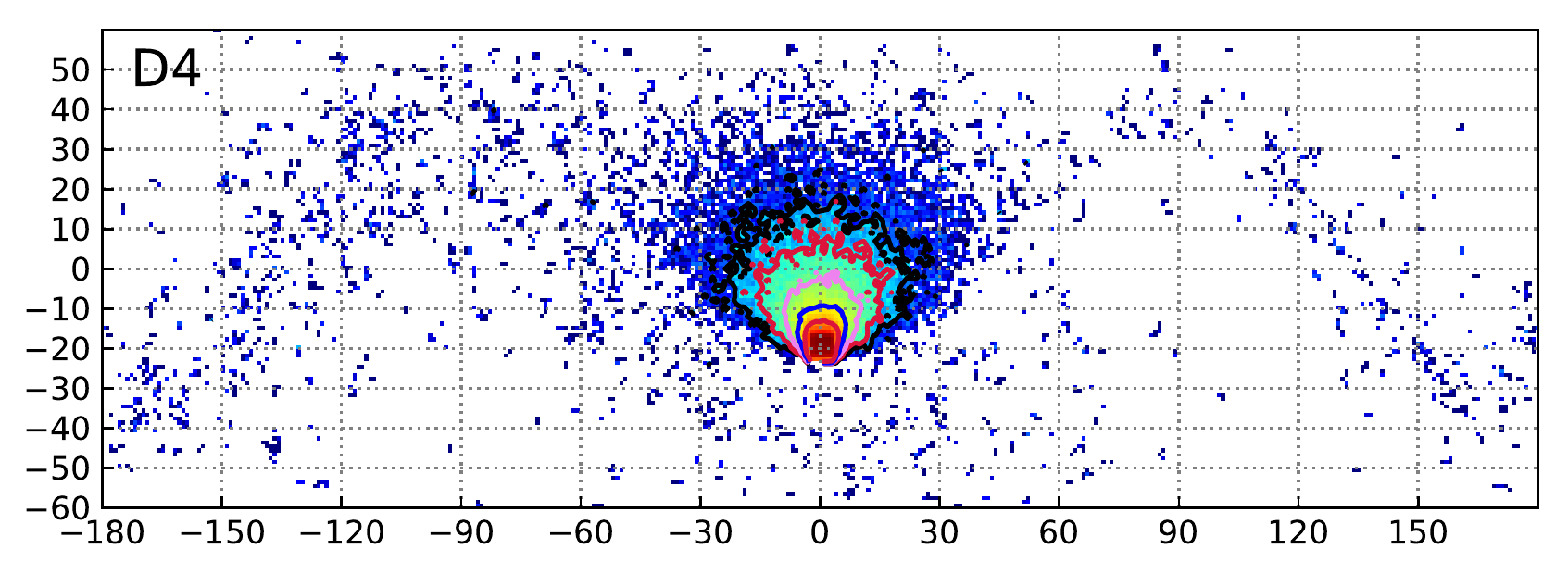}
  \put(1,28){\fcolorbox{black}{white}{\bf D4}}
\end{overpic} \\
\begin{overpic}[trim=10 0 8 8, clip, scale=0.5]{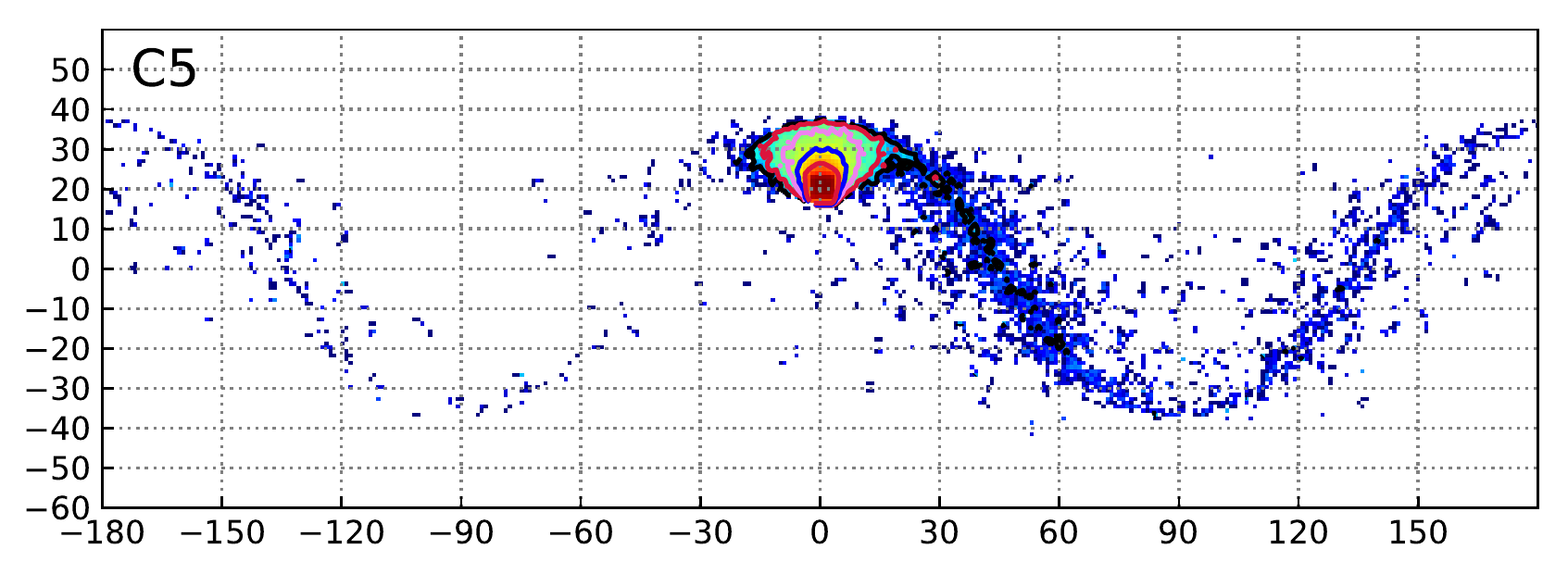}
  \put(5,31){\fcolorbox{black}{white}{\bf C5}}
\end{overpic}
\begin{overpic}[trim=31 0 8 8, clip, scale=0.5]{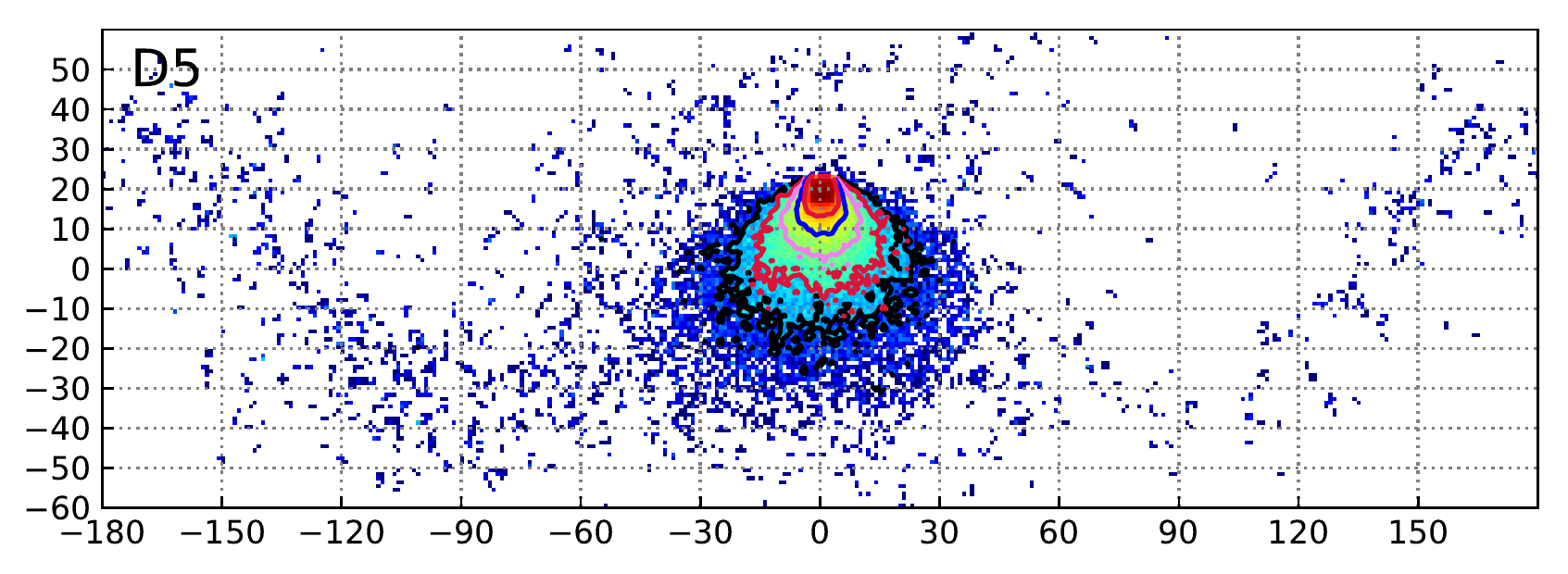}
  \put(1,33){\fcolorbox{black}{white}{\bf D5}}
\end{overpic} \\
\caption{
Fluence maps of protons, injected from a power-law with $\gamma=-1.1$, spanning the energy range from 10 to 800 MeV, crossing the \mbox{1 au} sphere, over a time of \mbox{100 hr}. Effects of corotation are removed. Fluence colours are on a logarithmic scale, overlaid by contours, two per decade. Injection regions were $6^\circ \times 6^\circ$. Panels are labelled C1 through D5 according to simulation setup.
}\label{fig:maps_CD}
\end{figure*}

In Figure \ref{fig:maps_CD}, we plot fluence maps for protons, with corotation removed, for simulations C1--C5 and D1--D5. Setups C1--C3 are much like setups A1--A3, except with a larger HCS tilt angle ($\alpha_\mathrm{nl}=37^\circ$ instead of $29^\circ$) and a waviness multiplier of $n_\mathrm{nl}=2$. Setups D1--D3 differ from setups B1--B3 in a similar manner. Additionally, whilst setups A and B injected particles at a region of negative HCS inclination, setups C1--C3 and D1--D3 injected particles at a region of positive HCS inclination. 

Panels C1--C3, with an A+ IMF configuration, show enhanced spread of particles off the current sheet in the regions of large HCS inclination. This suggests that the effect is indeed linked to the steepness of the HCS inclination. We suggest that in regions of large HCS inclination, particles are able to scatter to regions adjacent to the HCS and remain in them, as the HCS truncates a smaller portion of the fan-shape of gradient and curvature drifts. In regions of small HCS inclination, particles drift efficiently back to the HCS, and experience rapid HCS drift transporting them away in longitude. The net result is a statistical enhancement in counts in regions from which particles are not transported efficiently away, namely regions at longitudes of large HCS inclination. For A-- IMF configurations, this dependence on HCS inclination is not seen, as the HCS does not truncate curvature or gradient drifts, and there are no preferential regions of longitudinal transport.

We note that the HCS drift is indeed a fast drift if compared with gradient and curvature drifts, capable of efficient longitudinal transport. The mean propagation velocity due to HCS drift was, for a step-mode field transition and a particle travelling at speed $v$, calculated to be $\langle v_\mathrm{S}\rangle = 0.463v$ \citep{Burger1985}, well above other drifts (as can be inferred from \citealt{Dalla2013}).

We also note that in simulation sets C and D, the asymmetry of particle drifts near the injection region matches our analysis for simulations A1 and A3, with C1 providing only westward drifts, and C3 providing westward current sheet drift and eastward gradient and curvature drifts. The increased HCS inclination in panels D1 and D3 compared with B1 and B3 allows easier access for particles to the HCS, as drift over a smaller longitudinal range is enough to bring particles to the HCS. This increase in access results in an increase in fluence of particles which have experienced HCS drift.

Simulations C4, C5, D4, and D5 show an injection region placed at $\pm 20^\circ$ in latitude, well away from the current sheet. In simulations C4 and C5, the gradient and curvature drifts allow for a number of protons to drift all the way to the current sheet, although only very few particles are seen to drift along the HCS for more than $180^\circ$ in longitude. For D4 and D5, we see only very few particles reaching the HCS. We will discuss the connection between particle energy and these drifts in section \ref{subsec:energylongitude}.

\begin{figure*}[!ht]
\centering 
\begin{overpic}[trim=10 20 8 -20, clip, scale=0.66]{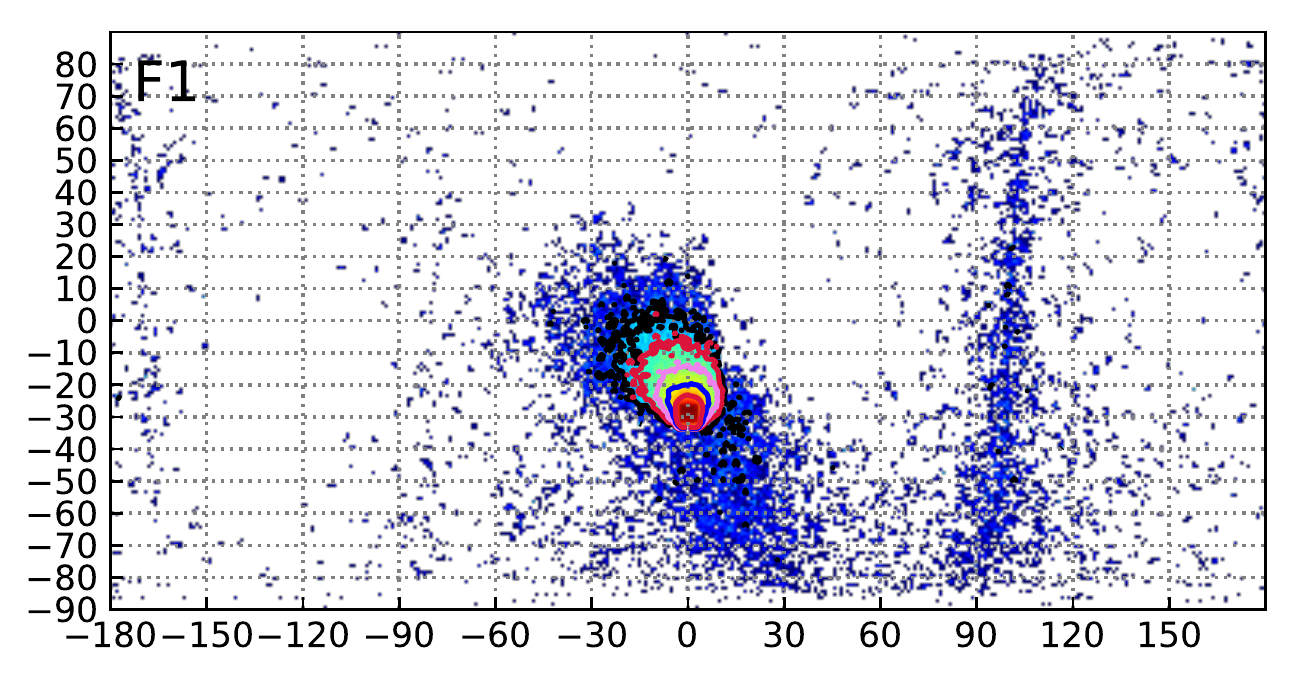}
  \put(15,50){\bf A+ IMF, corotation removed}
  \put(6,41.5){\fcolorbox{black}{white}{\bf E1}}
\end{overpic}
\begin{overpic}[trim=31 20 8 -20, clip, scale=0.66]{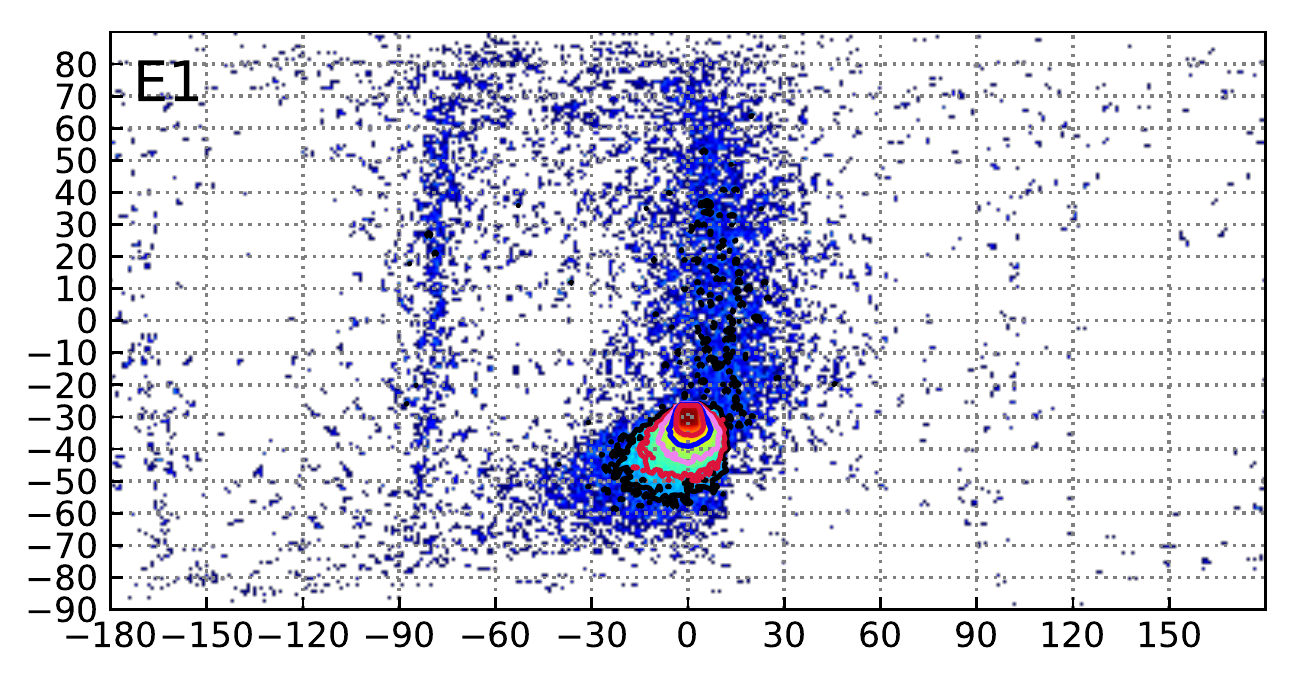}
  \put(11,53){\bf A-- IMF, corotation removed}
  \put(0.5,44){\fcolorbox{black}{white}{\bf F1}}
\end{overpic} \\
\begin{overpic}[trim=10 20 8 8, clip, scale=0.66]{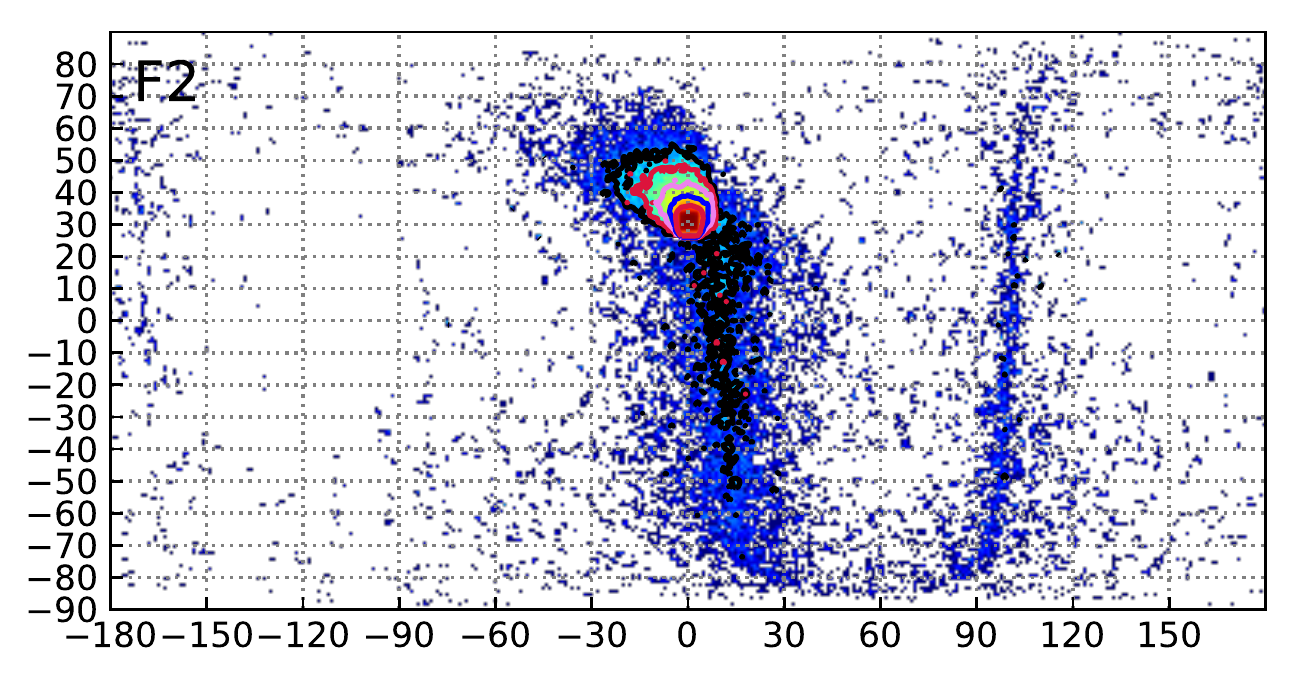}
  \put(6,41.5){\fcolorbox{black}{white}{\bf E2}}
\end{overpic}
\begin{overpic}[trim=31 20 8 8, clip, scale=0.66]{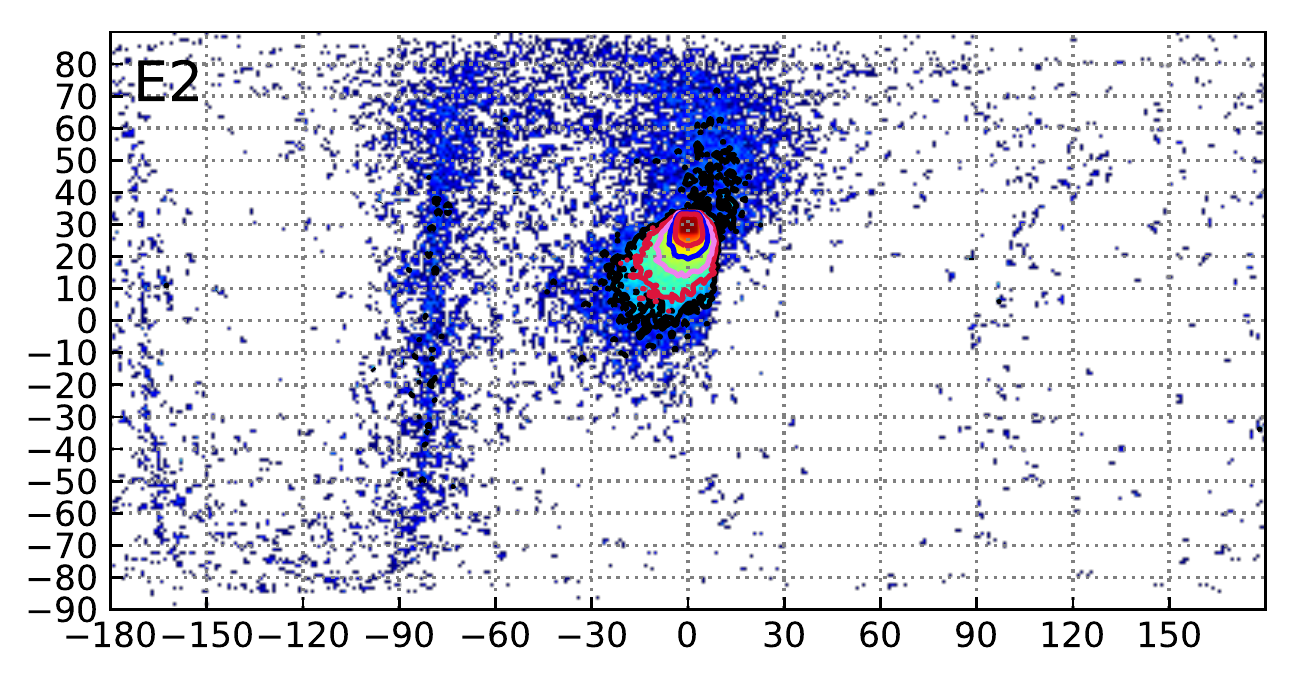}
  \put(0.5,44){\fcolorbox{black}{white}{\bf F2}}
\end{overpic} \\
\begin{overpic}[trim=10 20 8 8, clip, scale=0.66]{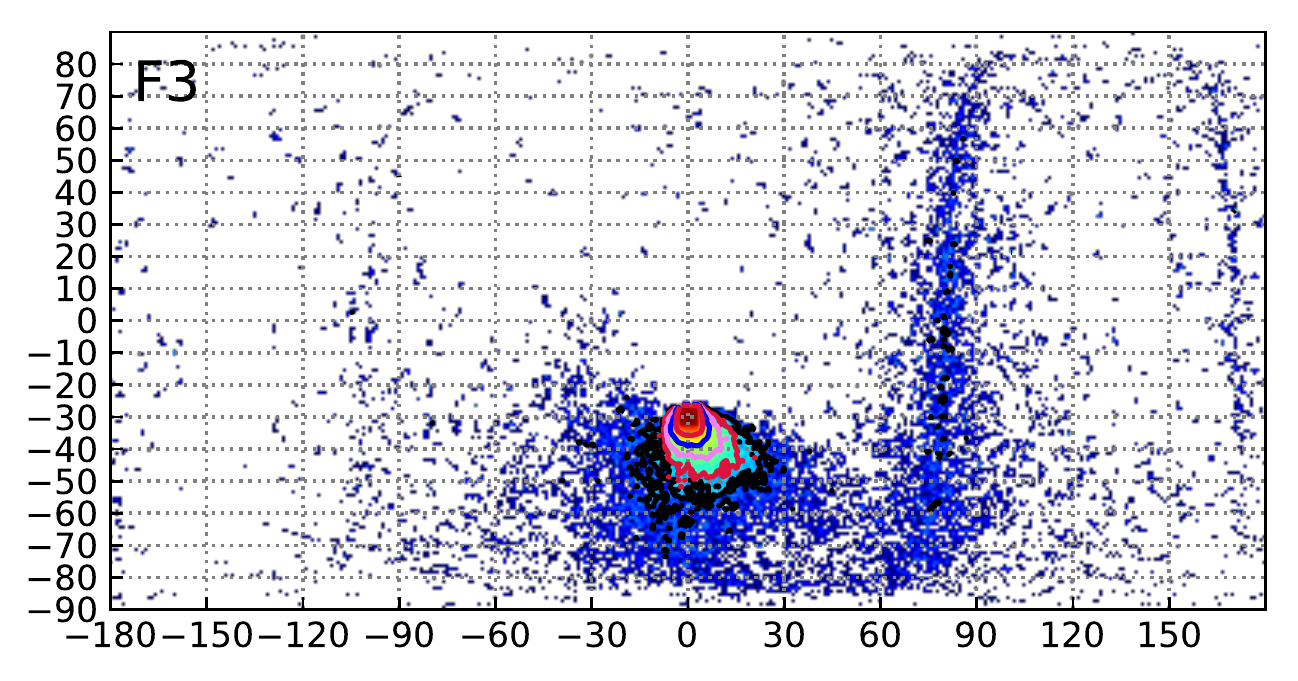}
  \put(6,41.5){\fcolorbox{black}{white}{\bf E3}}
\end{overpic}
\begin{overpic}[trim=31 20 8 8, clip, scale=0.66]{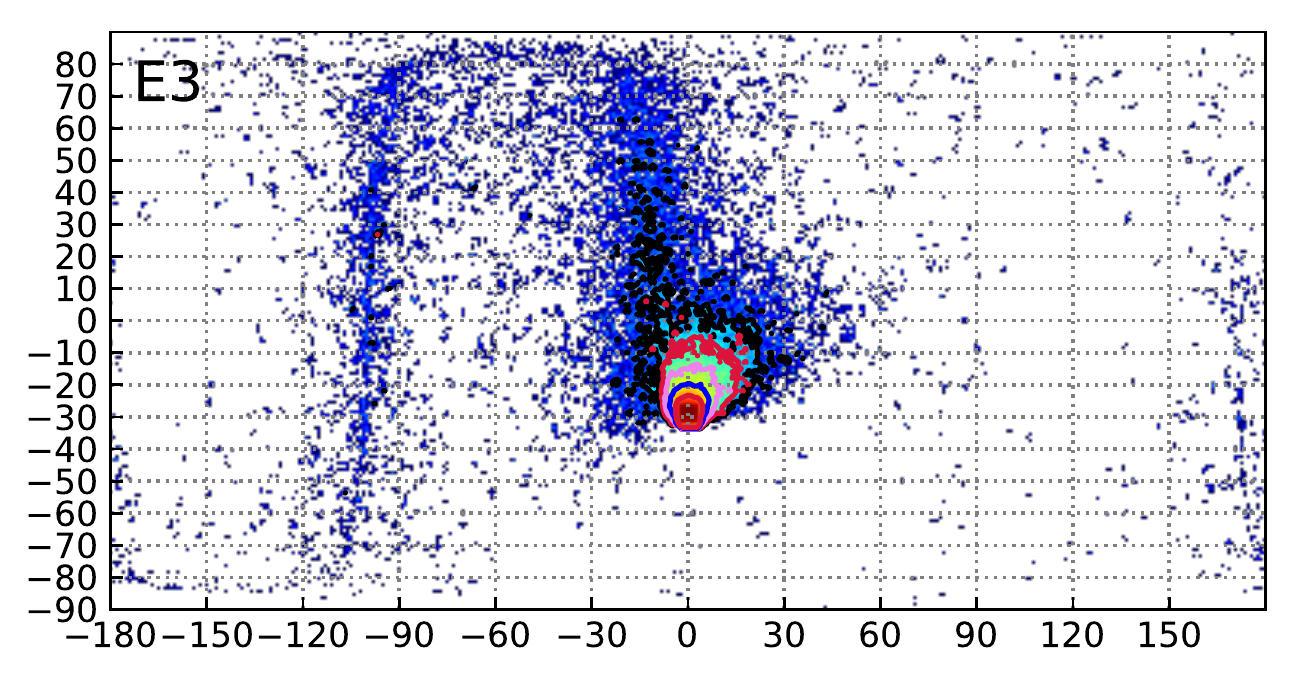}
  \put(0.5,44){\fcolorbox{black}{white}{\bf F3}}
\end{overpic} \\
\begin{overpic}[trim=10 0 8 8, clip, scale=0.66]{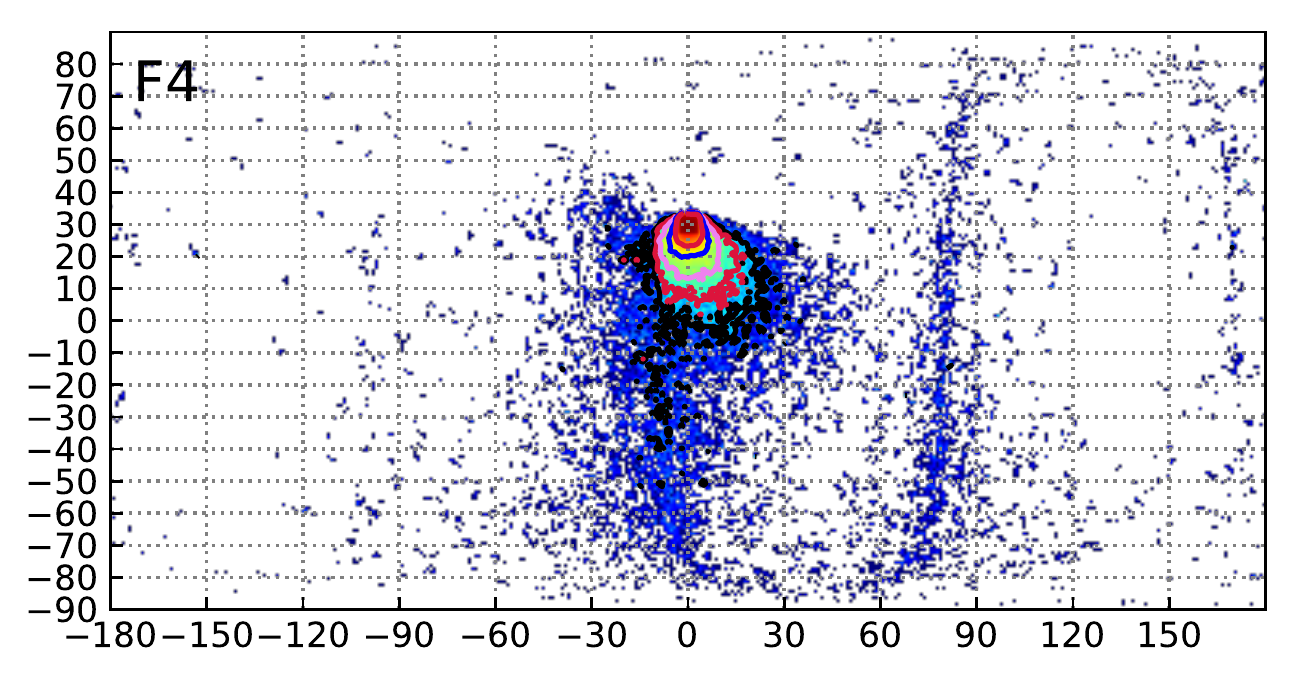}
  \put(6,47){\fcolorbox{black}{white}{\bf E4}}
\end{overpic}
\begin{overpic}[trim=31 0 8 8, clip, scale=0.66]{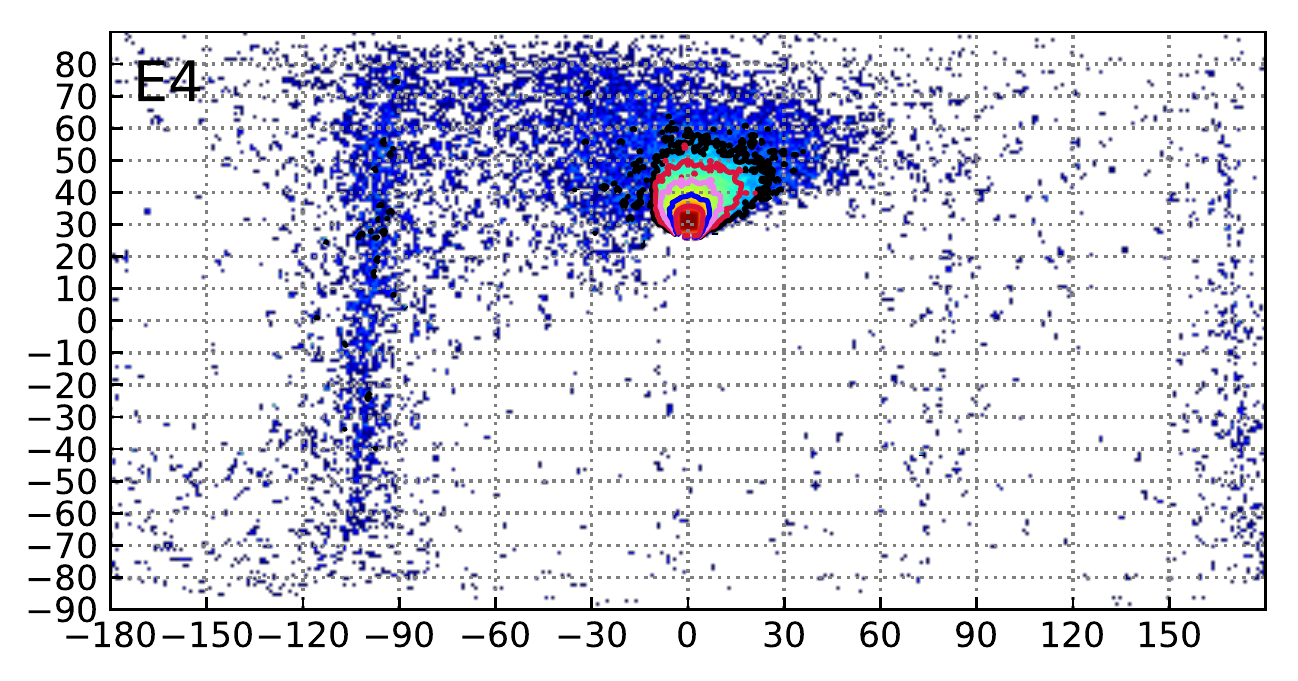}
  \put(0.5,50){\fcolorbox{black}{white}{\bf F4}}
\end{overpic} \\
\caption{
Fluence maps of protons, injected from a power-law with $\gamma=-1.1$, spanning the energy range from 10 to 800 MeV, crossing the \mbox{1 au} sphere, over a time of \mbox{100 hr}. Effects of corotation are removed. Fluence colours are on a logarithmic scale, overlaid by contours, two per decade. Injection regions were $6^\circ \times 6^\circ$. Panels are labelled E1 through F4 according to simulation setup, with dipole tilt angles of 85 degrees, representing a case of significant solar activity.
}\label{fig:maps_EF}
\end{figure*}

In Figure \ref{fig:maps_EF}, we plot fluence maps for protons, with corotation removed, for simulations E1--E4 and F1--F4. These setups are descriptive of periods of high solar activity, with a dipole tilt angle of $85^\circ$ and injection regions centered at latitudes of $\pm 30^\circ$. The injection regions are far from the HCS in latitude, but in longitude the shortest distances to the HCS are less than $2^\circ$. As described in \cite{Dalla2013}, particle drifts depend on latitude, and as particles are injected at higher latitudes, and additionally experience HCS drift to near-polar latitudes, the gradient and curvature drifts cause significant spread of particles over a wide range of longitudes.

Asymmetric drifts in the vicinity of the injection region, such as those seen in figures \ref{fig:corot} and \ref{fig:maps_CD}, are even more clearly visible, with panels E1, E2, F1, and F2 resulting in bi-directional drifts and panels E3, E4, F3, and F4 showcasing drifts where the HCS drift and the gradient and curvature drifts are not in opposition. Due to the strong inclination of the HCS, the simulation sets E and F are qualitatively more like each other than previous comparisons between A+ and A-- IMF configurations, as the truncation of gradient and curvature drift patterns is nearly vertical.

Both IMF configurations presented in Figure \ref{fig:maps_EF} show the capability of the HCS to cause protons to drift and propagate to a wide range of heliolatitudes and heliolongitudes. Due to the large dipole tilt angle and the resulting large HCS inclination, the protons which have experienced significant HCS drift appear to result in longitudinally periodic increases of particle fluence.

\subsection{Energy spectrograms} \label{subsec:energylongitude}
Particle drifts are strongly energy-dependent \citep{Dalla2013}. In order to examine the spread of particles as a function of energy, we produced energy spectrograms of \mbox{1 au} particle crossings versus longitude. We binned particles between energies of \mbox{10 MeV} and \mbox{800 MeV} so that the injection power-law of $\gamma=-1.1$ generated equal amounts of particles into each of the eight energy bins. We defined a bin width of \mbox{$3^\circ$} in longitude and collected all crossings regardless of latitude. We maintained a fixed colour-intensity relation for all spectrograms. For these spectrograms, we did not remove corotation.

\begin{figure*}[!ht]
\centering 
\begin{overpic}[trim=0 20 70 0, clip, scale=0.5]{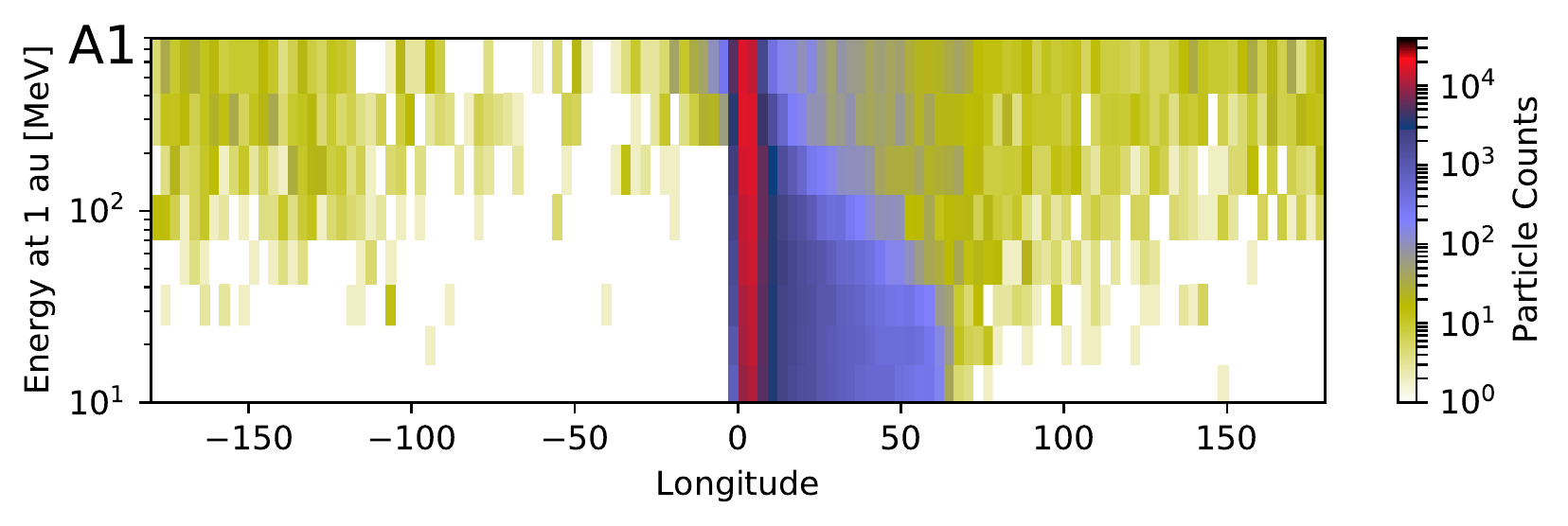}
  \put(27,33){\bf A+ IMF, with corotation}
  \put(5,29){\fcolorbox{black}{white}{\bf A1}}
\end{overpic}
\begin{overpic}[trim=20 20 0 0, clip, scale=0.5]{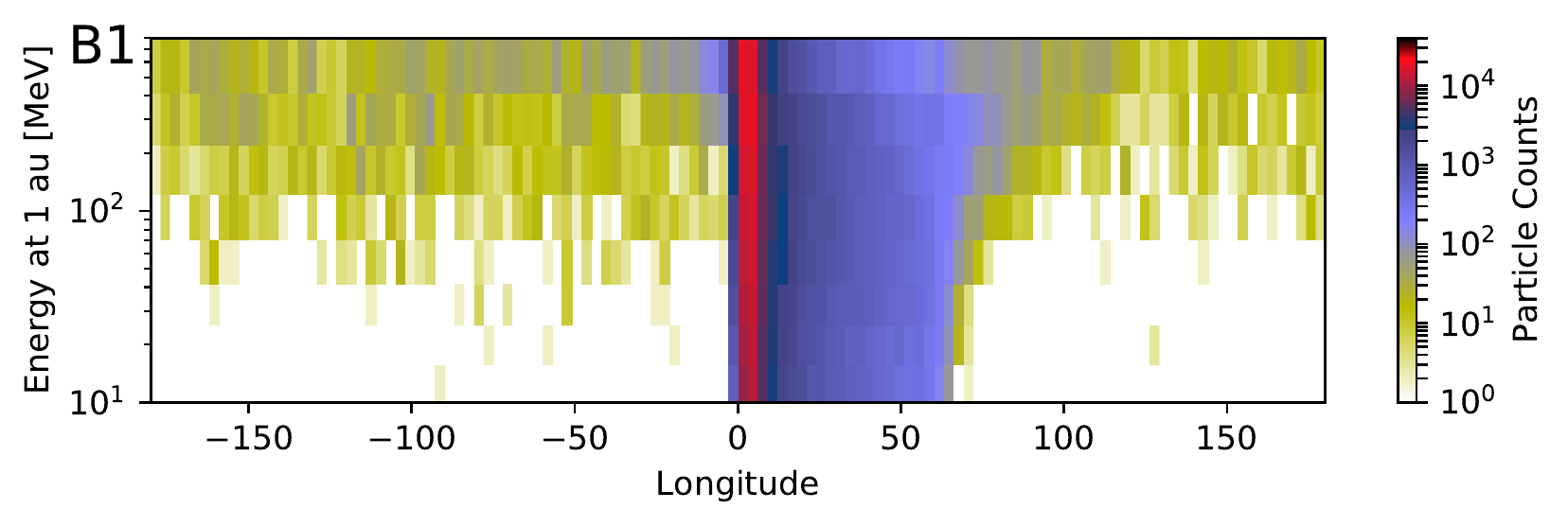}
  \put(19,29){\bf A-- IMF, with corotation}
  \put(0,25){\fcolorbox{black}{white}{\bf B1}}
\end{overpic} \\
\begin{overpic}[trim=0 20 70 0, clip, scale=0.5]{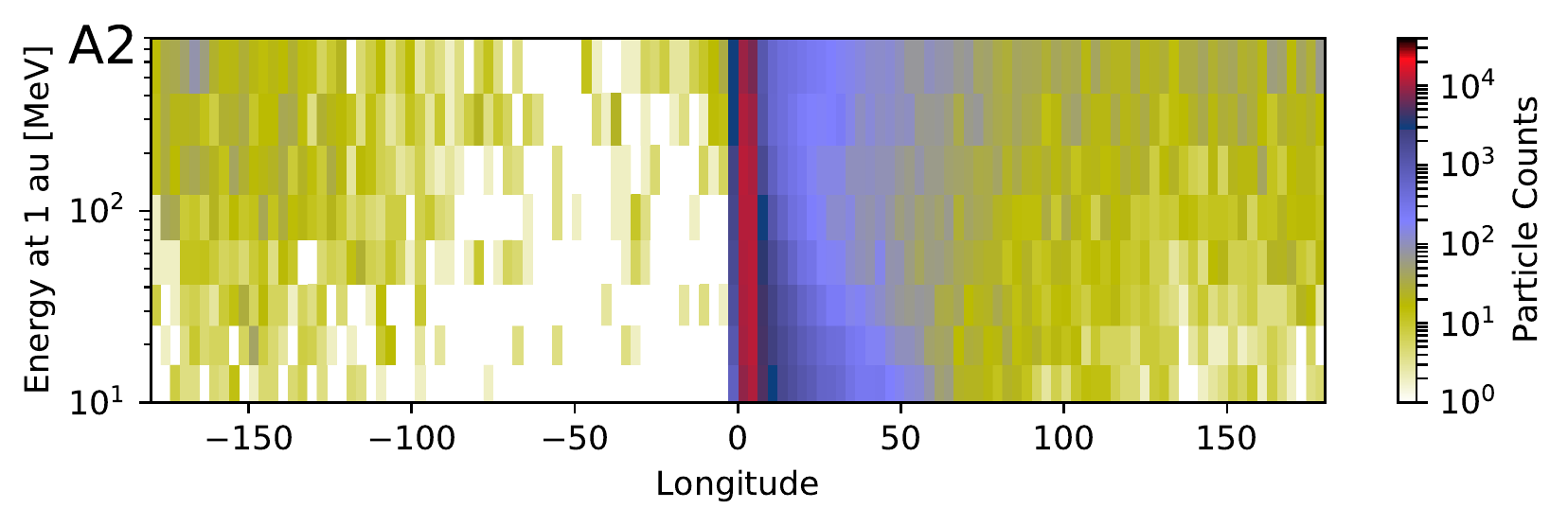}
  \put(5,29){\fcolorbox{black}{white}{\bf A2}}
\end{overpic}
\begin{overpic}[trim=20 20 0 0, clip, scale=0.5]{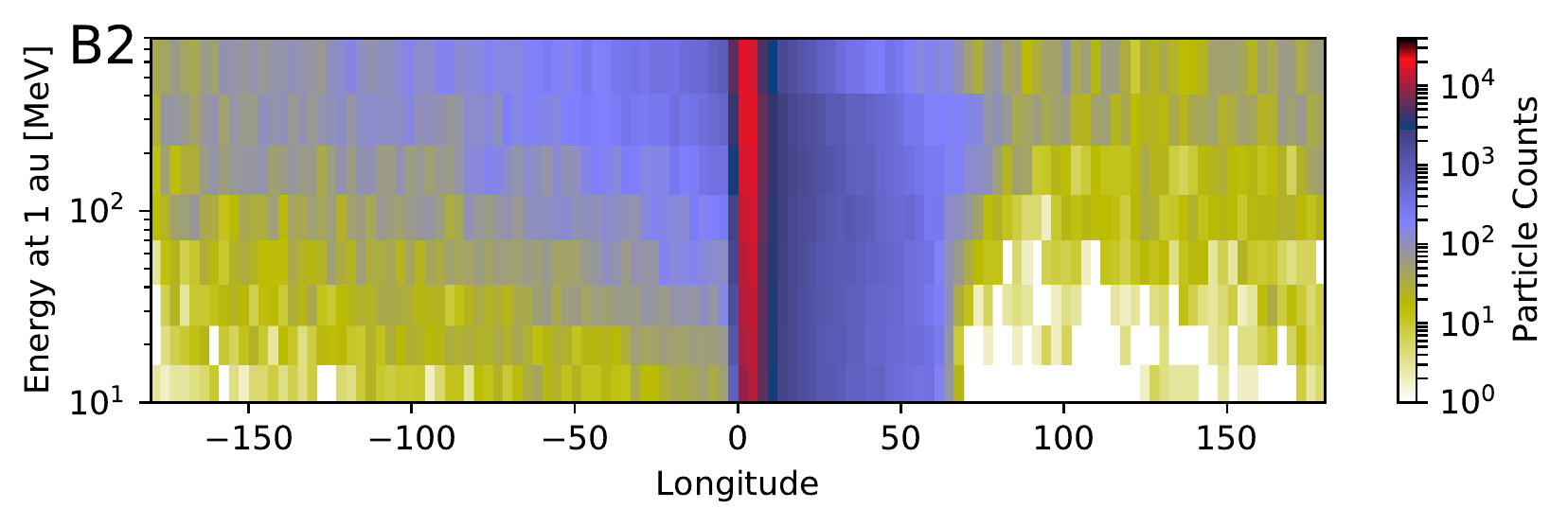}
  \put(0,25){\fcolorbox{black}{white}{\bf B2}}
\end{overpic} \\
\begin{overpic}[trim=0 20 70 0, clip, scale=0.5]{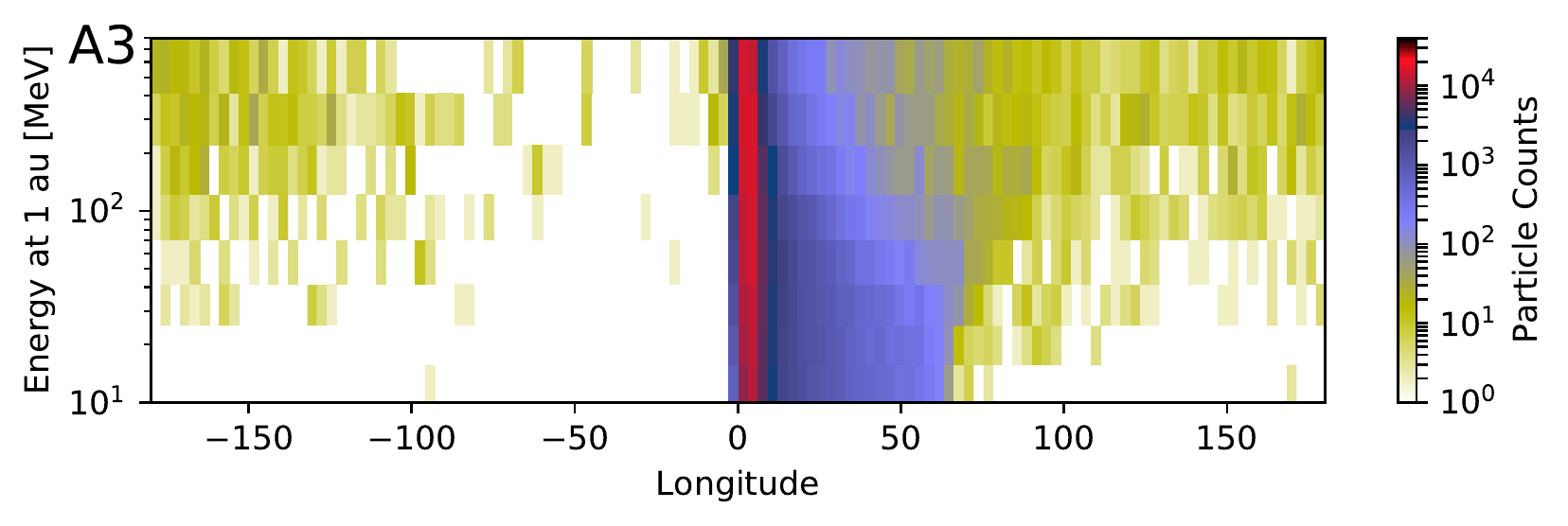}
  \put(5,29){\fcolorbox{black}{white}{\bf A3}}
\end{overpic}
\begin{overpic}[trim=20 20 0 0, clip, scale=0.5]{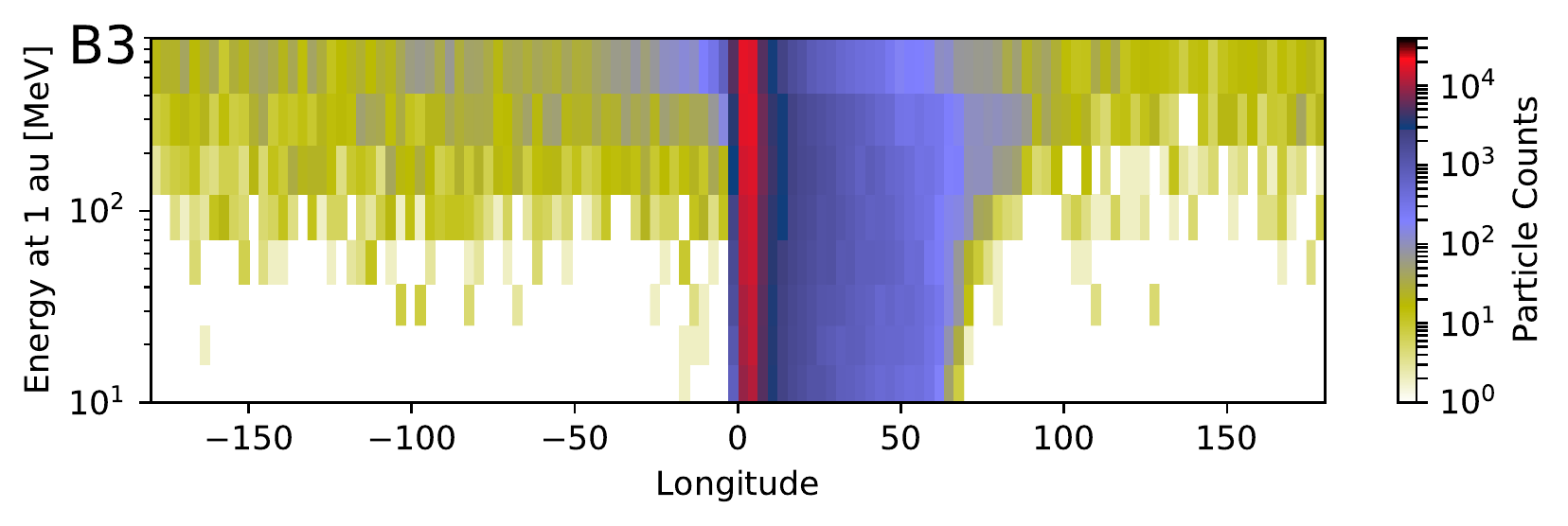}
  \put(0,25){\fcolorbox{black}{white}{\bf B3}}
\end{overpic} \\

\begin{overpic}[trim=0 20 70 0, clip, scale=0.5]{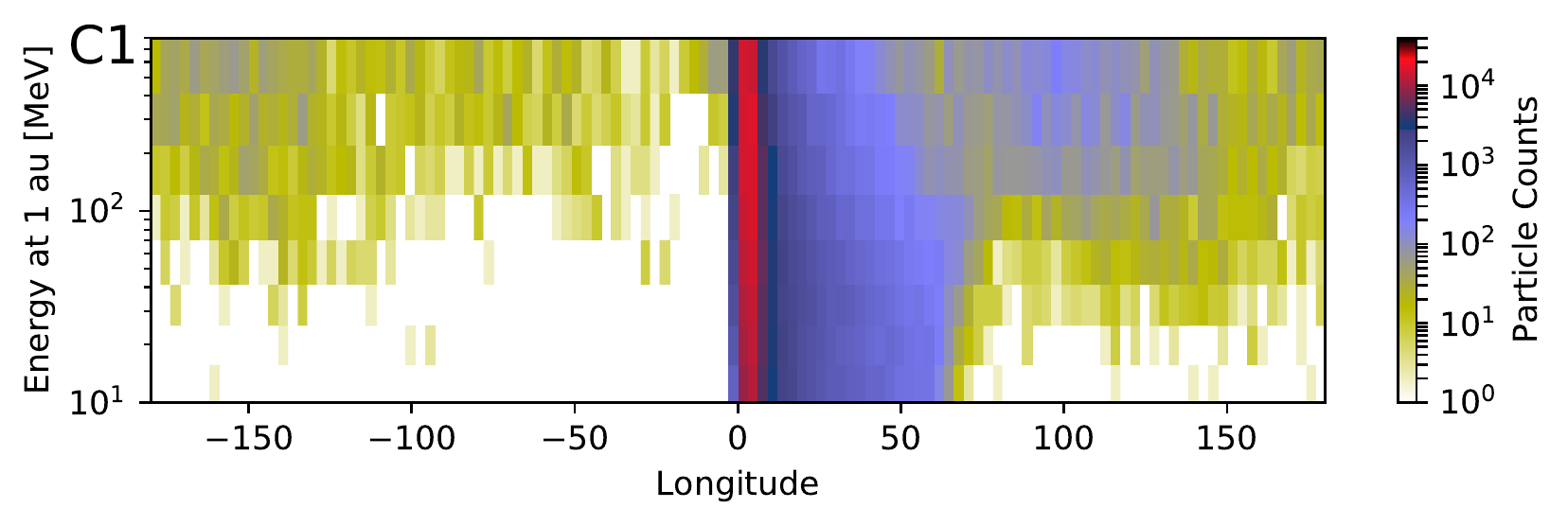}
  \put(5,29){\fcolorbox{black}{white}{\bf C1}}
\end{overpic}
\begin{overpic}[trim=20 20 0 0, clip, scale=0.5]{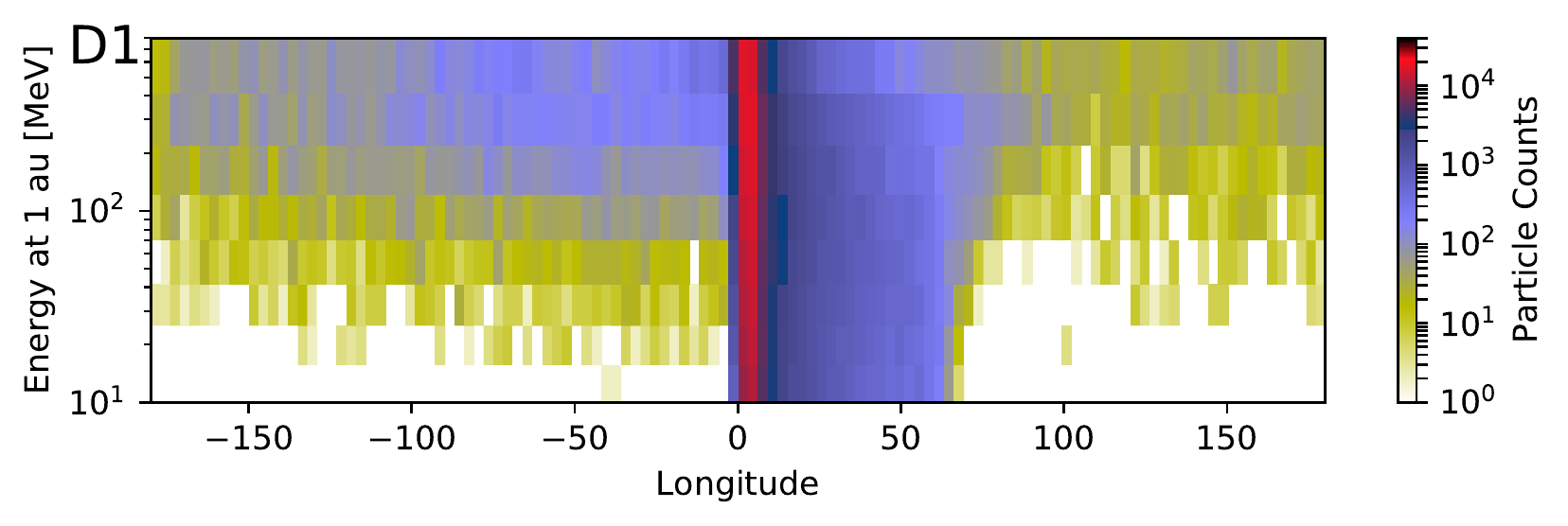}
  \put(0,25){\fcolorbox{black}{white}{\bf D1}}
\end{overpic} \\
\begin{overpic}[trim=0 20 70 0, clip, scale=0.5]{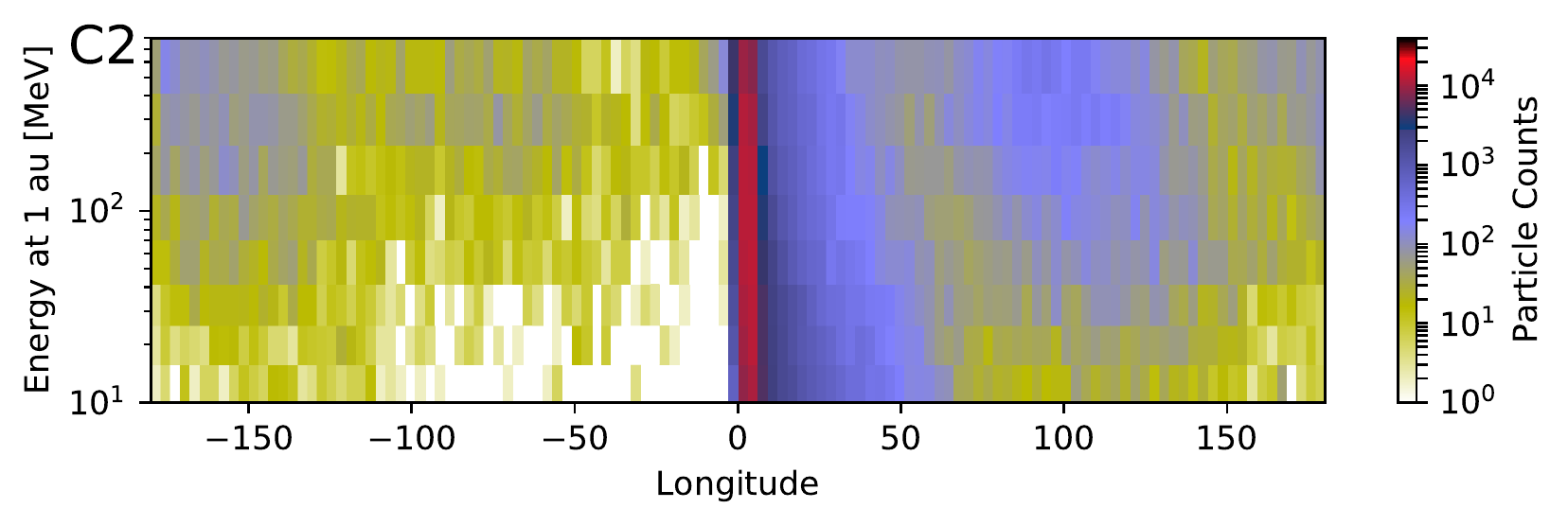}
  \put(5,29){\fcolorbox{black}{white}{\bf C2}}
\end{overpic}
\begin{overpic}[trim=20 20 0 0, clip, scale=0.5]{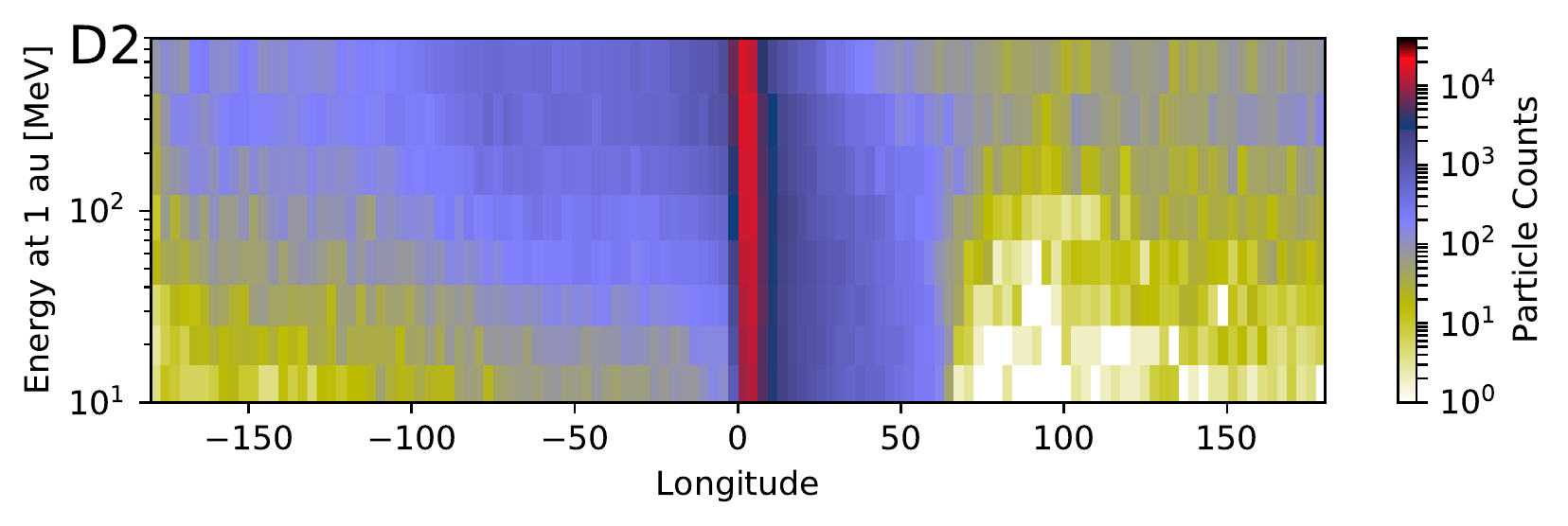}
  \put(0,25){\fcolorbox{black}{white}{\bf D2}}
\end{overpic} \\
\begin{overpic}[trim=0 20 70 0, clip, scale=0.5]{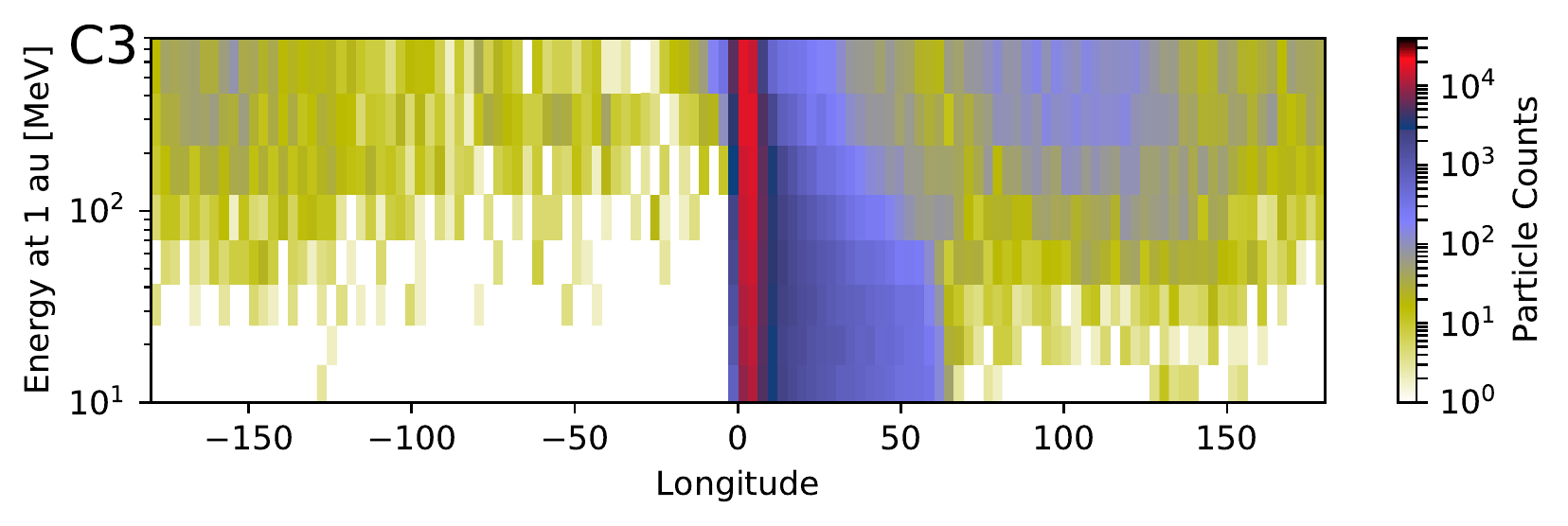}
  \put(5,29){\fcolorbox{black}{white}{\bf C3}}
\end{overpic}
\begin{overpic}[trim=20 20 0 0, clip, scale=0.5]{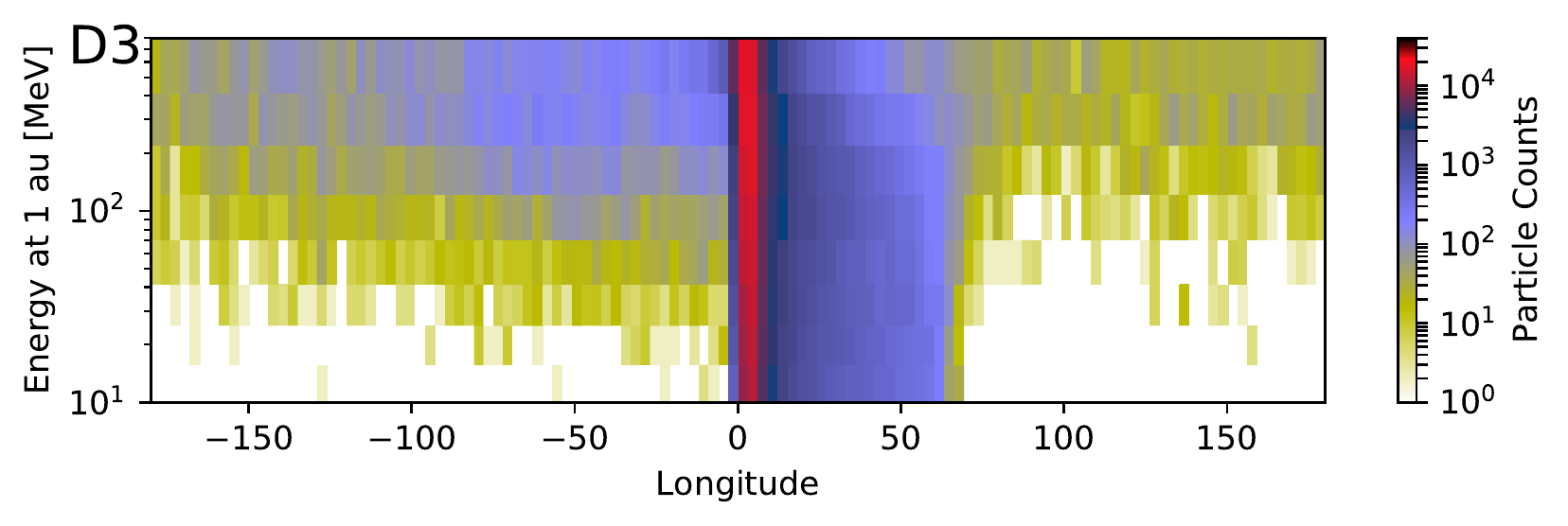}
  \put(0,25){\fcolorbox{black}{white}{\bf D3}}
\end{overpic} \\
\begin{overpic}[trim=0 20 70 0, clip, scale=0.5]{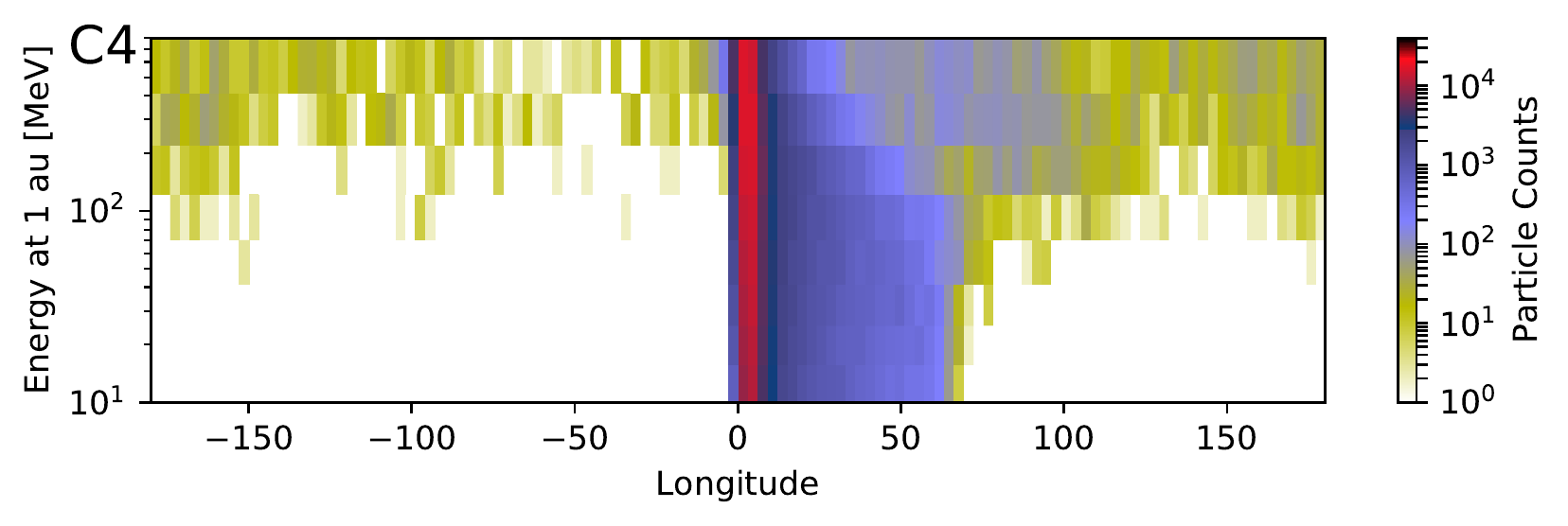}
  \put(5,29){\fcolorbox{black}{white}{\bf C4}}
\end{overpic}
\begin{overpic}[trim=20 20 0 0, clip, scale=0.5]{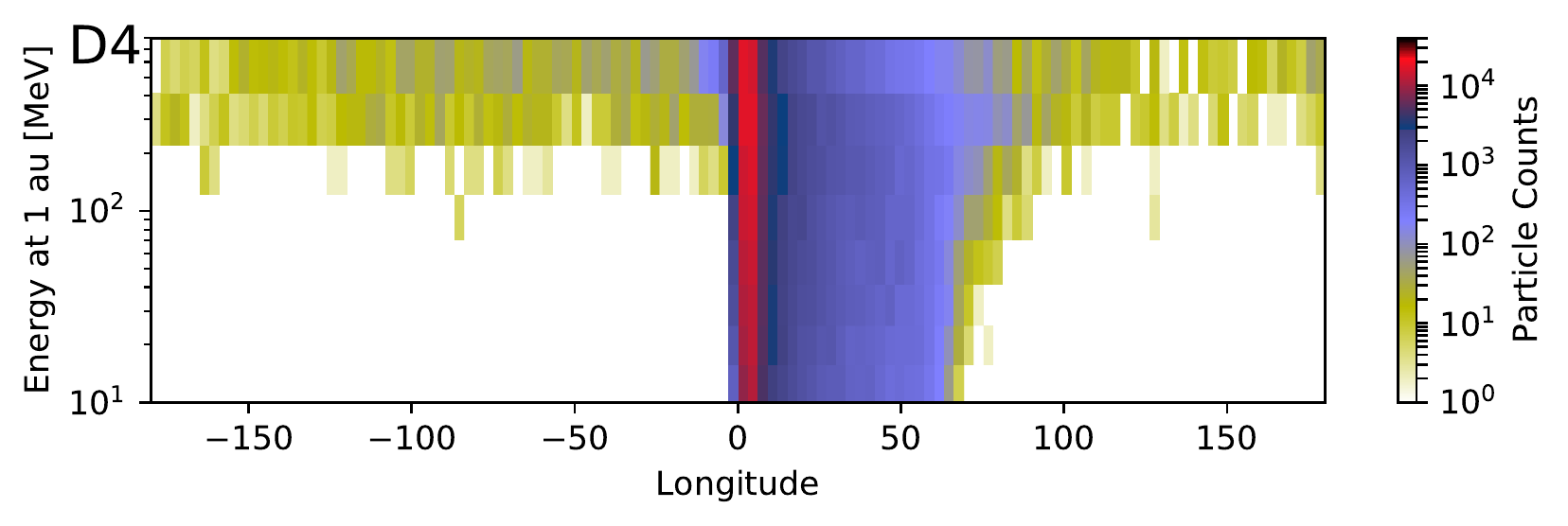}
  \put(0,25){\fcolorbox{black}{white}{\bf D4}}
\end{overpic} \\
\begin{overpic}[trim=0 0 70 0, clip, scale=0.5]{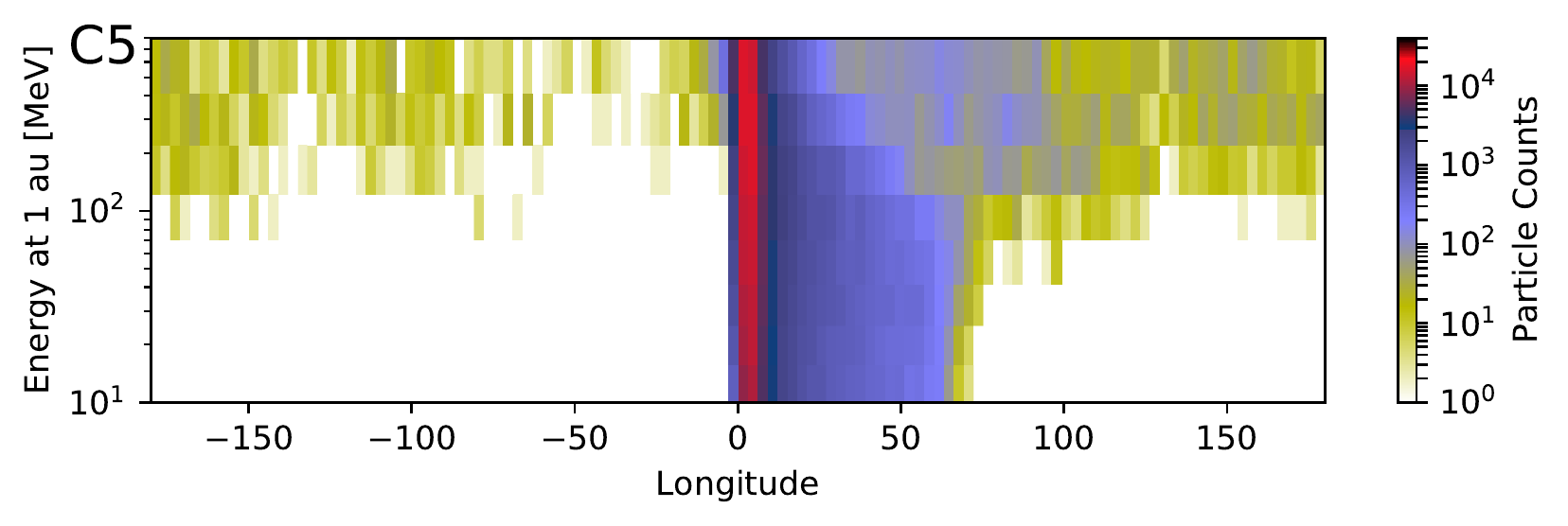}
  \put(5,33){\fcolorbox{black}{white}{\bf C5}}
\end{overpic}
\begin{overpic}[trim=20 0 0 0, clip, scale=0.5]{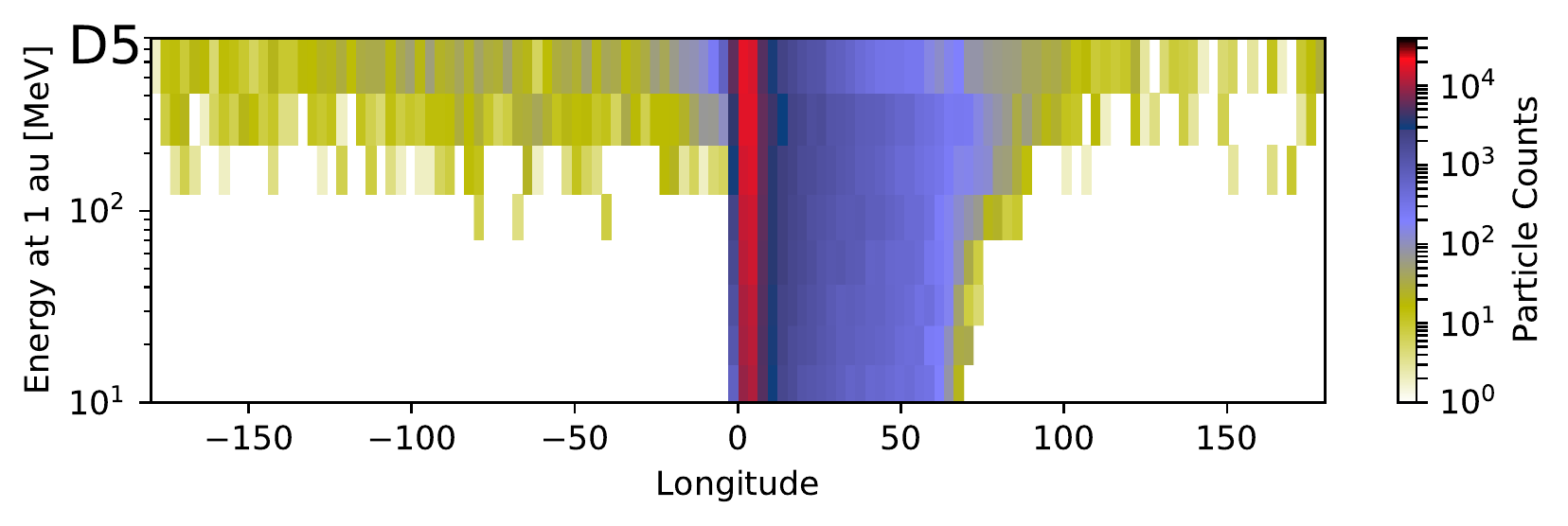}
  \put(0,29.5){\fcolorbox{black}{white}{\bf D5}}
\end{overpic} \\
\caption{Energy spectrograms of protons crossing the \mbox{1 au} sphere, as a function of longitude and energy, with a logarithmic colour scale. Effects of corotation were not removed. Panels are labelled A1 through D5 according to simulation setup. The colour scaling is logarithmic with fixed bounds.
}\label{fig:energyspectrograms}
\end{figure*}

Figure \ref{fig:energyspectrograms} shows the energy spectrograms generated for simulations A1 through D5. First, we discuss the signal due to corotation. In all panels, we see at all energies a strong (red) signal associated with the injection event. Extending westward from this line is continued fluence (in blue) associated with the well-connected field lines as they drift westward due to corotation. For the A+ IMF configuration (sets A and C), we find this fluence to be strong at low energies and weakened at high energies, and for the A-- configuration (sets B and D), the drift effect is mostly uniform in energy with only a small weakening at high energies. 

We suggest that this is due to a combination of two effects. First, particles of all energies scatter and isotropize within the flux tube connected to the injection region, but during the early phase of this process the radial group velocity of low-energy particles is smaller than for high-energy particles, which propagate rapidly into the outer heliosphere. Thus, low-energy particles are more likely to be found close to \mbox{1 au} than high-energy particles, and low-energy particles continue to create more fluence over the \mbox{1 au} sphere as the flux tube co-rotates westward. Secondly, we suggest that the A+ IMF configuration helps high-energy particles to drift rapidly away from the injection region, diluting the counts at high energies within the corotation band. This happens to a smaller extent in the A-- IMF configuration. In addition, A+ and A-- configurations differ in the radial component of HCS drift velocities, which is further investigated in section \ref{subsec:radial}.

Access to longitudes outside the corotation band is mostly facilitated by the HCS drift. In panels A2, B2, C2, and D2 of Figure \ref{fig:energyspectrograms}, particles of all energies are able to drift in excess of $180^\circ$ due to the HCS intersecting the injection region. In simulations numbered 1 and 3, with weaker access to the HCS, we see a requirement of $E\gtrsim 30\,\mathrm{MeV}$ for significant HCS drift. In simulations numbered 4 and 5, with injection regions placed far from the HCS, efficient transport along the HCS is seen only for $E\gtrsim 100\,\mathrm{MeV}$. This is explained by protons with higher energies being able to drift greater distances in latitude, due to the energy-dependence of curvature and gradient drifts, and reach the HCS, unlike particles with lower energies.


In panels C1--C3, we see HCS-associated fluence enhancements at $\sim 90^\circ$ and $\sim 180^\circ$ west of the injection region, coinciding with regions of large HCS inclination found in Figure \ref{fig:maps_CD}. These spectrograms show this to be a true enhancement instead of mere greater spatial spread of particles. The enhancement appears tilted, with enhancements at higher energies apparent at more eastern longitudes and enhancements at lower energies apparent at more western longitudes. This is suggested to be due to corotation, in the same manner as the main corotation-associated fluence enhancement was seen at more western longitudes for low-energy particles and closer to the injection longitude for high-energy particles.

The enhancements seen in panels C1--C3 at regions of large HCS inclination are not found in panels D1--D3. This is in agreement with our proposed explanation for them, as the A-- IMF configuration used in simulations C1--C3 does not drive particles to the HCS, and thus, there is no preference for particles to travel rapidly away from the longitudes associated with small HCS inclination. 

In panels C4, C5, D4, and D5, with the injection region significantly removed from the HCS, we see that only very high energy particles are capable of scattering to the HCS and experiencing HCS drift. The particles that do, are however able to travel to a wide range of longitudes due to their high speed. In agreement with previous panels, simulations C4 and C5 show periodic inclination-associated enhancements in fluences, whereas simulations D4 and D5 do not.

\begin{figure*}[!ht]
\centering 
\begin{overpic}[trim=0 20 70 0, clip, scale=0.5]{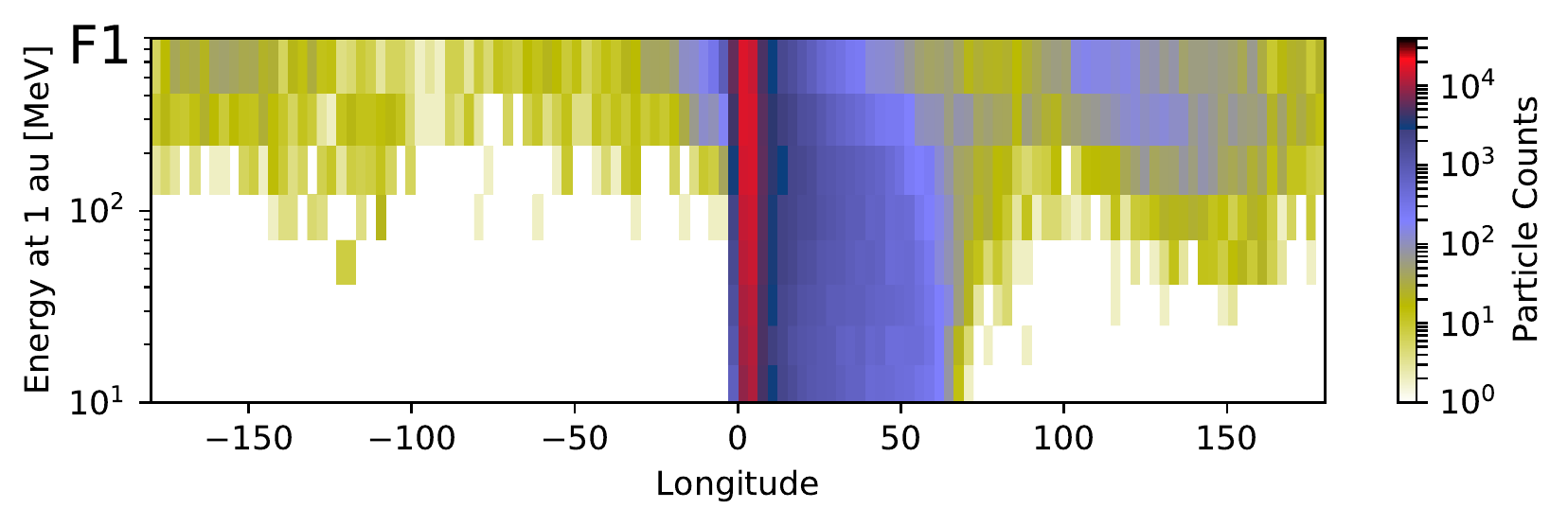}
  \put(27,33){\bf A+ IMF, with corotation}
  \put(5,29){\fcolorbox{black}{white}{\bf E1}}
\end{overpic}
\begin{overpic}[trim=20 20 0 0, clip, scale=0.5]{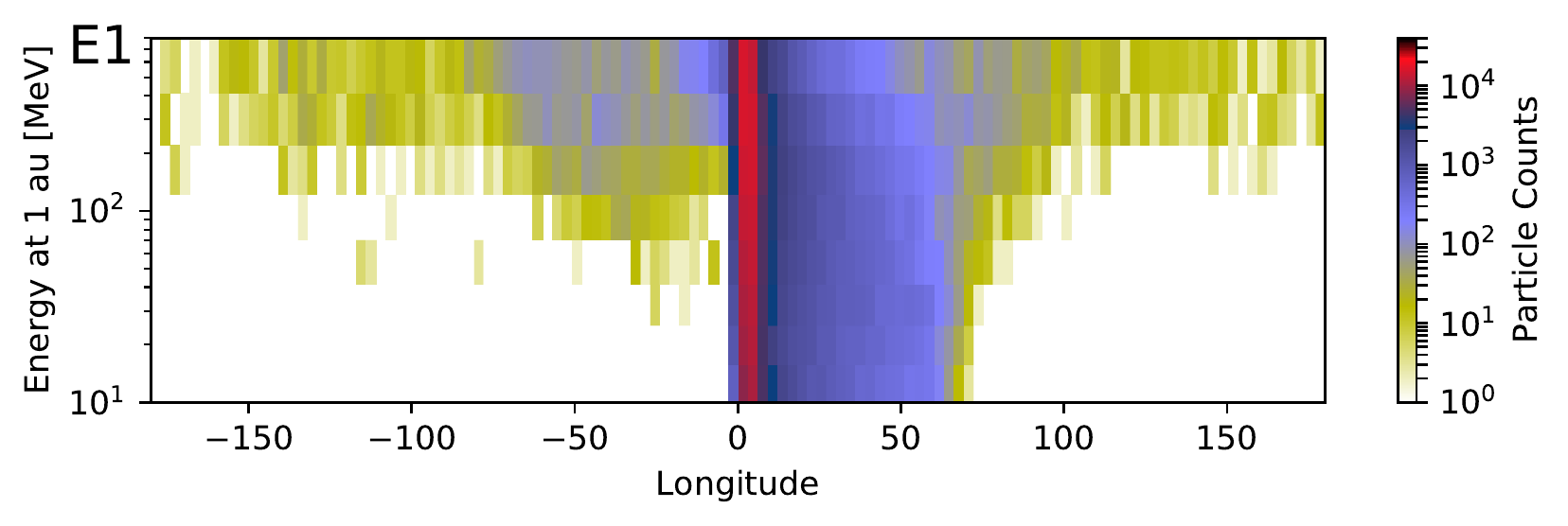}
  \put(19,29){\bf A-- IMF, with corotation}
  \put(0,25){\fcolorbox{black}{white}{\bf F1}}
\end{overpic} \\
\begin{overpic}[trim=0 20 70 0, clip, scale=0.5]{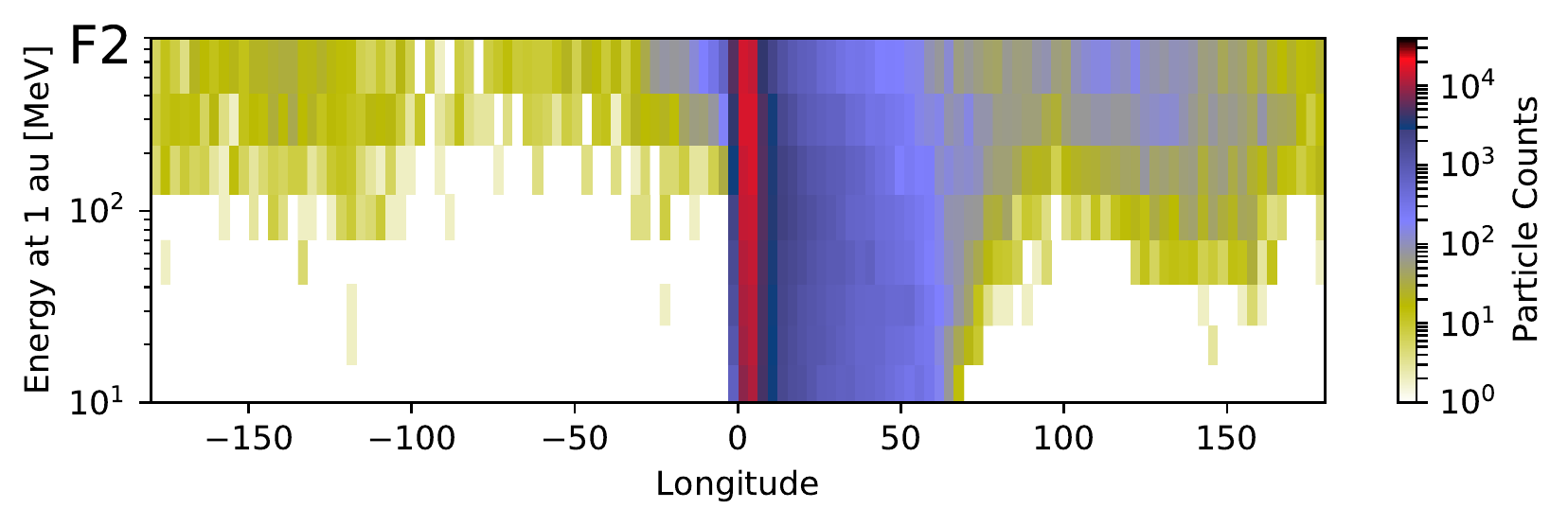}
  \put(5,29){\fcolorbox{black}{white}{\bf E2}}
\end{overpic}
\begin{overpic}[trim=20 20 0 0, clip, scale=0.5]{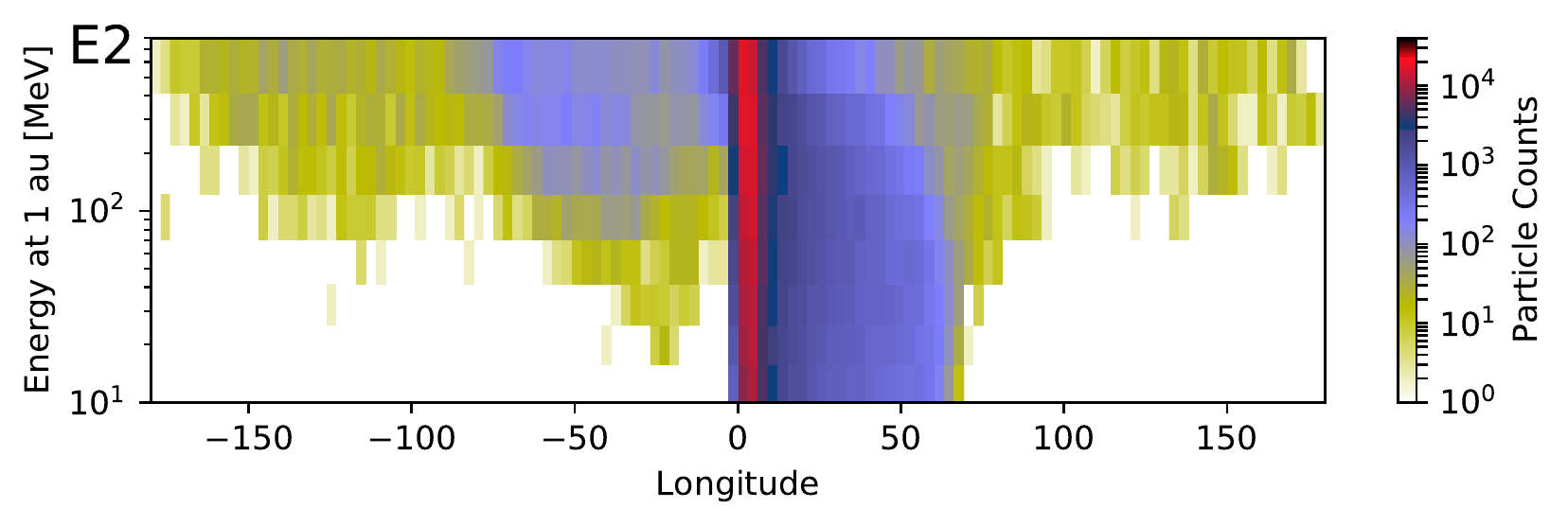}
  \put(0,25){\fcolorbox{black}{white}{\bf F2}}
\end{overpic} \\
\begin{overpic}[trim=0 20 70 0, clip, scale=0.5]{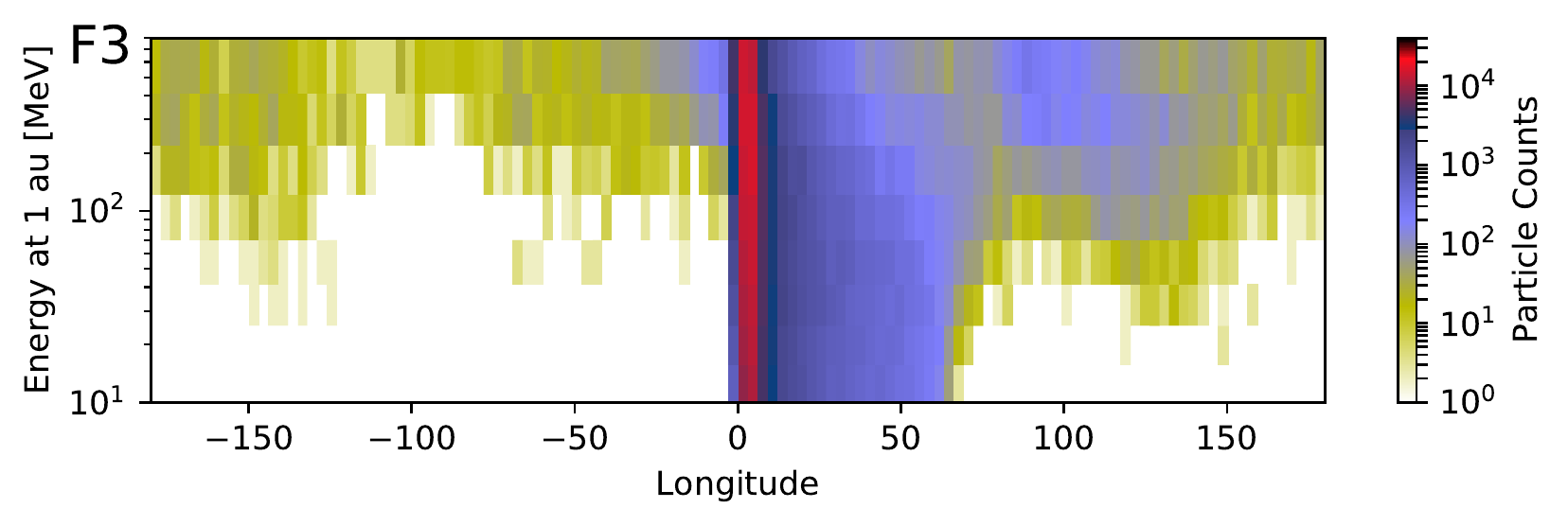}
  \put(5,29){\fcolorbox{black}{white}{\bf E3}}
\end{overpic}
\begin{overpic}[trim=20 20 0 0, clip, scale=0.5]{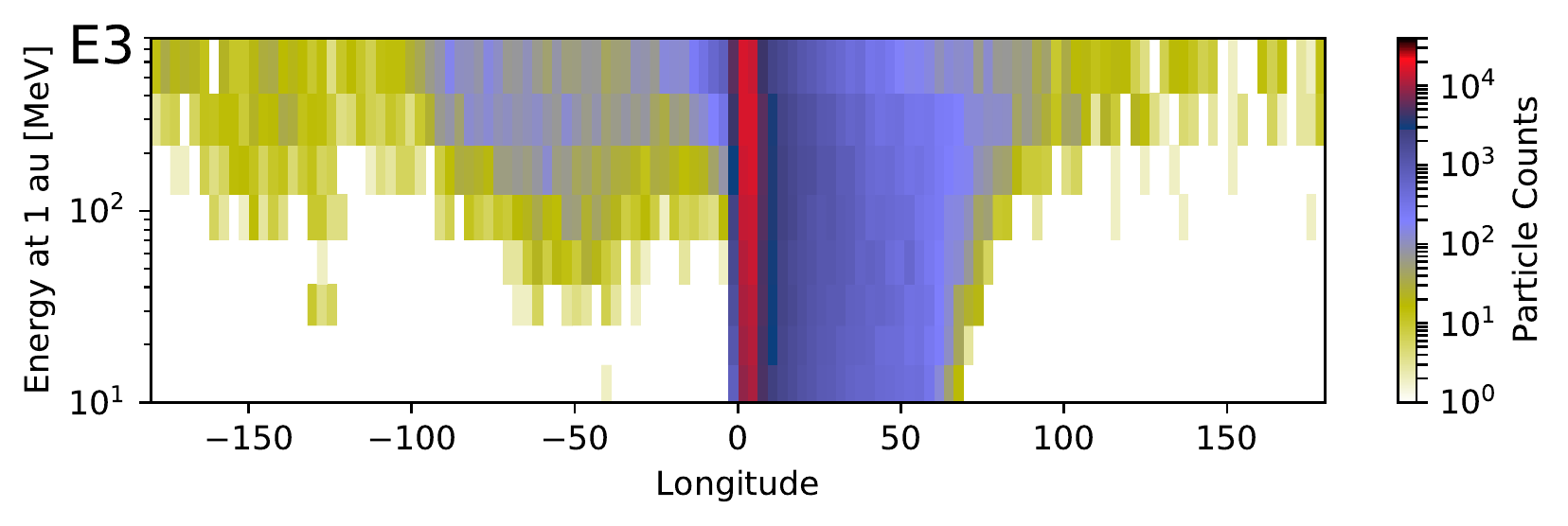}
  \put(0,25){\fcolorbox{black}{white}{\bf F3}}
\end{overpic} \\
\begin{overpic}[trim=0 0 70 0, clip, scale=0.5]{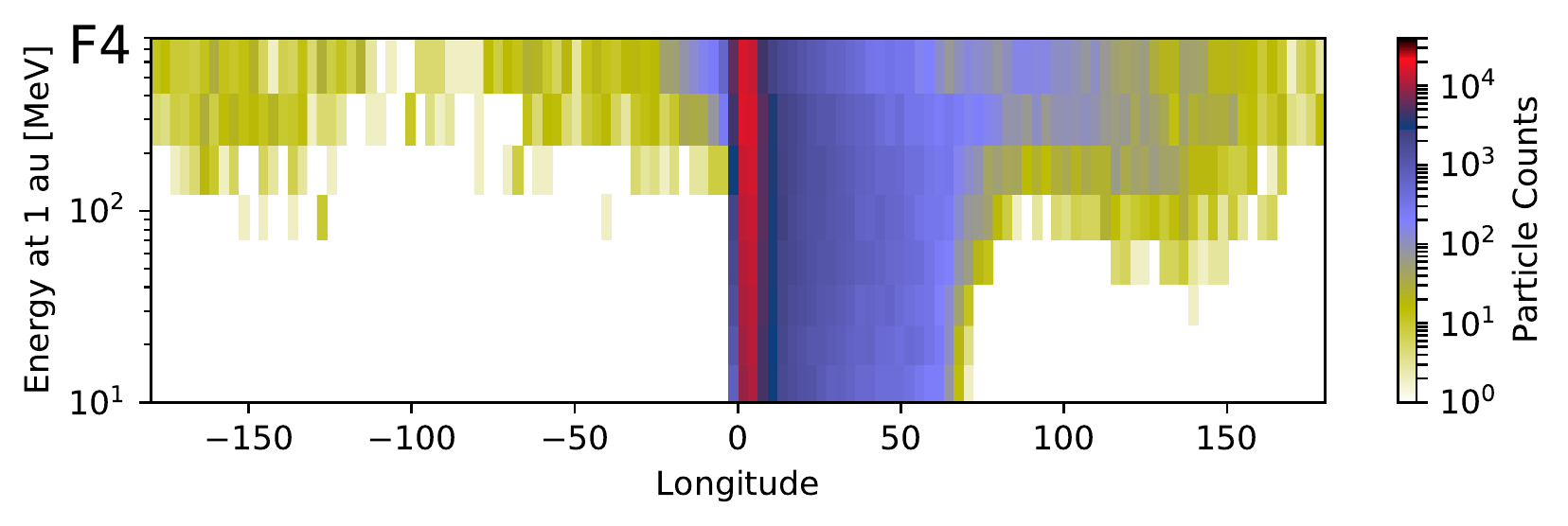}
  \put(5,33){\fcolorbox{black}{white}{\bf E4}}
\end{overpic}
\begin{overpic}[trim=20 0 0 0, clip, scale=0.5]{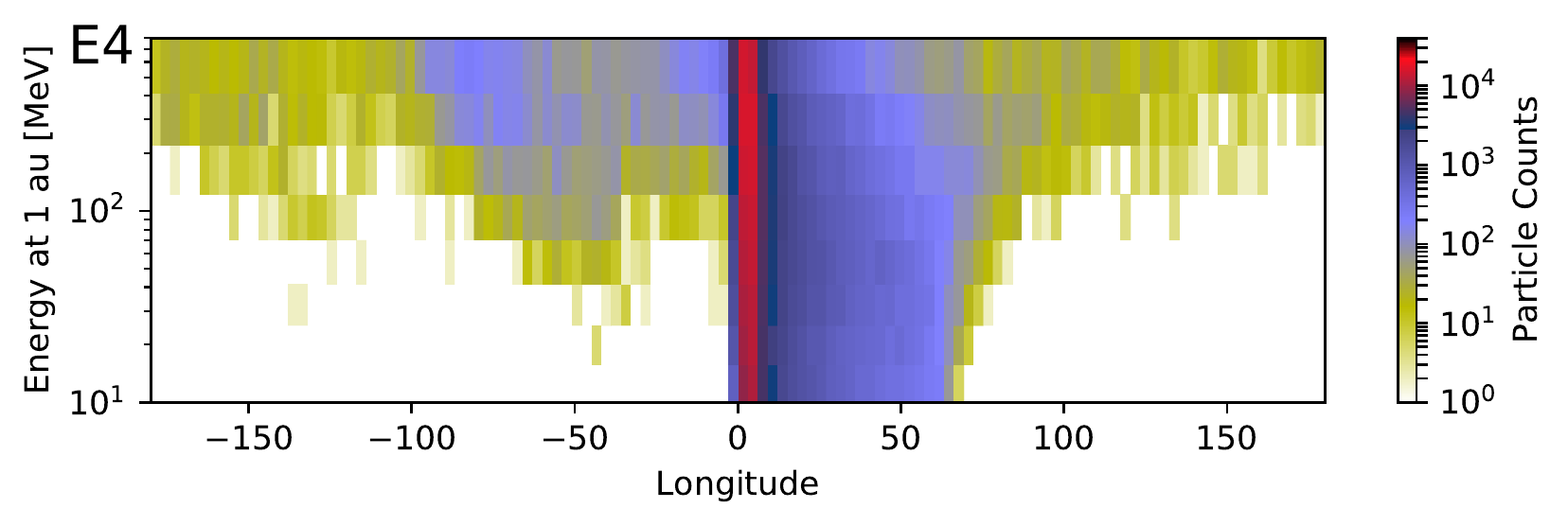}
  \put(0,29.5){\fcolorbox{black}{white}{\bf F4}}
\end{overpic} \\
\caption{Energy spectrograms of protons crossing the \mbox{1 au} sphere, as a function of longitude and energy, with a logarithmic colour scale. Effects of corotation were not removed. Panels are labelled E1 through F4 according to simulation setup. The colour scaling is logarithmic with fixed bounds.
}\label{fig:energyspectrograms_EF}
\end{figure*}

In Figure \ref{fig:energyspectrograms_EF}, we show the energy spectrograms generated for simulations E1 through F4. Due to the large HCS inclination and the fact that the HCS does not intersect the injection region, particles must experience longitudinal gradient and/or curvature drift in order to gain access to the HCS. We see that at this distance, only protons of \mbox{$\gtrsim$ 30 MeV} are efficiently transported to the HCS and then to a wide range of longitudes.

Similar to what was seen in Figure \ref{fig:energyspectrograms}, Figure \ref{fig:energyspectrograms_EF} shows a corotation-associated spread to the west of the injection region. This spread decreases somewhat with increasing energy due to crossings being gathered at \mbox{1 au}, and the fact that fast particles tend to propagate towards the outer portions of the heliosphere. We note that both simulation setups E and F result in periodic enhancements of particle fluence, located at regions of large HCS inclination. In these runs, however, this is likely a projection effect resulting from the $85^\circ$ dipole tilt angle, and not as much due to interplay of gradient, curvature and HCS drifts.

Additionally we note that the lack of gradient and curvature drift truncation allows for runs E1--E4 to show a high-energy particle extension to longitudes east of the injection region. By returning to Figure \ref{fig:maps_EF}, we can see that this drifting particle population is detected both directly due to injection (panels E1 and E2) and also due to particles crossing the HCS and drifting eastward in longitude after the crossing (panels E3 and E4).

\subsection{Fluence spectra} \label{subsec:spectra}
\begin{figure*}[!htb]
\centering 
\includegraphics[trim=0 0 0 0, clip, scale=0.47]{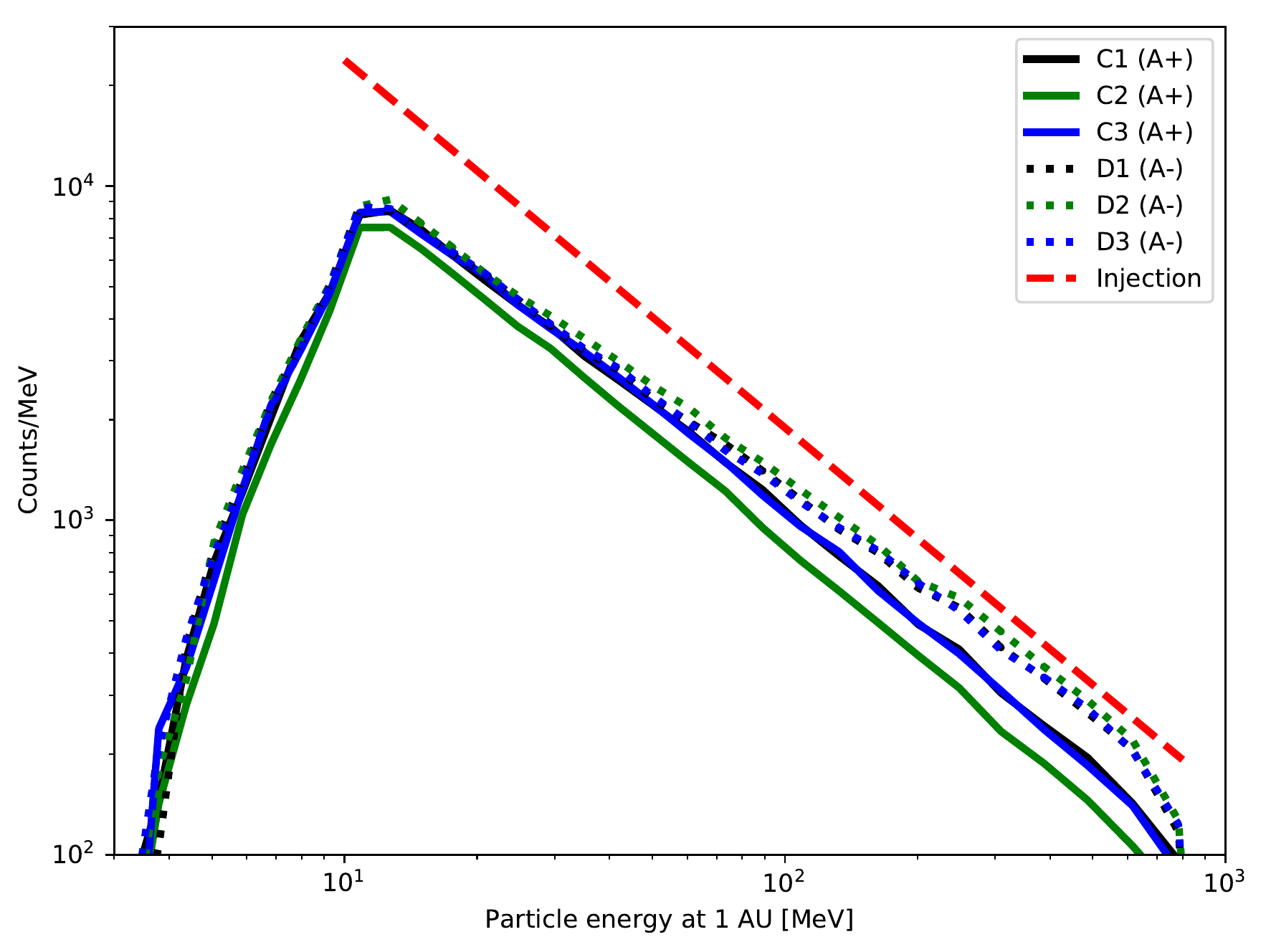}
\includegraphics[trim=0 0 0 0, clip, scale=0.47]{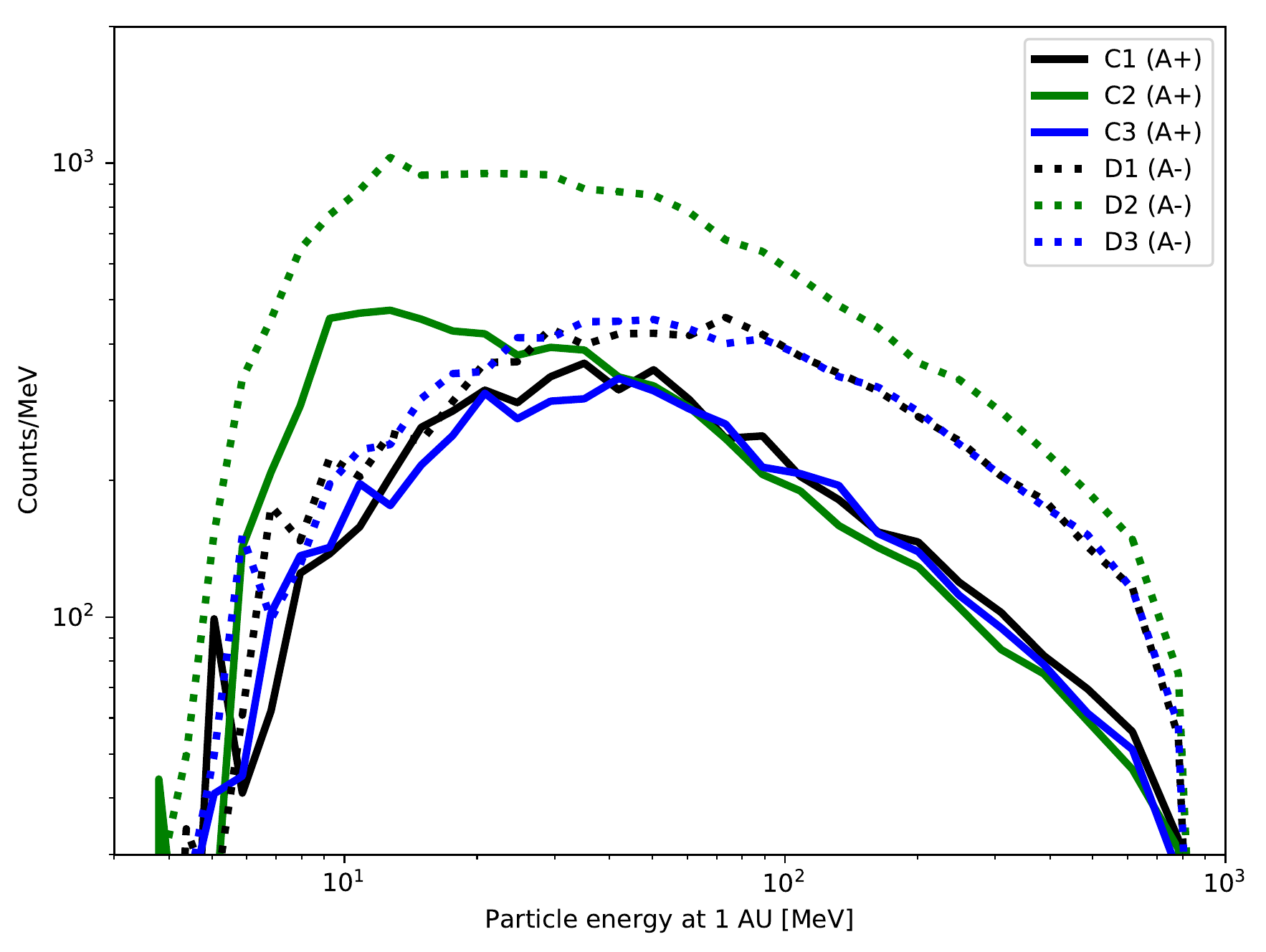}
\caption{
Fluence spectra for proton crossings over the  \mbox{1 au} sphere, in units MeV$^{-1}$ from simulations C1, C2, C3, D1, D2, and D3. The left panel shows the spectra gathered across all crossings. In the right panel the crossings close to the centrally connected field lines (accounting for removal of corotation) have been excluded. In both panels, we see similar results for simulations C1 and C3, and for simulations D1 and D3, with spectra mostly overlapping each other. Simulations C2 and D2 (green curves) have the highest fluences in the right panel due to efficient escape from the injection region to HCS drift.}\label{fig:energy_histo}
\end{figure*}

Particles propagating through interplanetary space experience deceleration, and, as previously mentioned, drifts are energy-dependent. In Figure \ref{fig:energy_histo}, we plot fluence spectra at \mbox{1 au} in units of counts MeV$^{-1}$ for simulations C1--C3 and D1--D3. In the left panel we show spectra gathered over all latitudes and longitudes, along with a power law representing the injected spectrum, and in the right panel we plot spectra gathered over all field lines not connected to areas within $2^\circ$ of the injection region. Although no particles were injected below \mbox{10 MeV}, the \mbox{$<10$ MeV} portion of the spectrum gets populated through deceleration. The left panel shows that although spectra appear mostly similar for all simulations, C2 has slightly lower fluence than C1 or C3, and D2 greater than D1 or D3. The curves for C1 and C3 overlap almost completely, as do the curves for simulations D1 and D3. Overall, set D shows slightly greater fluences, especially at high energies. Thus, the A-- IMF configuration provides higher \mbox{1 au} fluences than the A+ IMF, especially when injecting particles close to the HCS.

The right panel shows how, at energies \mbox{$\gtrsim 20$ MeV}, fluences outside well-connected field lines are significantly lower for simulation set C (A+ IMF) than for set D (A-- IMF). For simulations C2 and D2 (in green), efficient access to the HCS leads to an increase in proton crossings outside the well-connected region. This effect is enhanced particularly at low energies, although for simulation set D, continues to the highest energies. Again, we see that simulations C1 and C3 provide similar results with each other, as do simulations D1 and D3. We discuss the reason behind these fluence differences in section \ref{subsec:radial}.

\subsection{Radial profiles} \label{subsec:radial}
We now discuss why an A-- IMF leads to increased \mbox{1 au} crossings, especially when associated with efficient HCS drift, as seen both in figures \ref{fig:energyspectrograms} and \ref{fig:energy_histo}. First, we must consider the direction of the current sheet drift, as explained in \cite{Burger1985}. It can be calculated to be parallel to the current sheet and perpendicular to the magnetic field at each location along the HCS. As the magnetic field curvature follows the Parker spiral, this results in the HCS drift velocity having a radial component. This component, for an A+ IMF configuration with a mostly westward HCS drift, is away from the Sun, and for an A-- IMF HCS drift it is towards the Sun.

Of all protons propagating within the inner heliosphere, any which end up experiencing some HCS drift will be transported either towards the Sun or outwards toward the heliopause according to the IMF configuration, resulting in a net statistical effect for the total proton population. Thus, for an A-- IMF configuration, particles tend to remain closer to the Sun, and for an A+ IMF configuration, particles are pushed further away from the Sun.
For the A-- IMF configuration, this facilitates more crossings over the \mbox{1 au} sphere than for the A+ configuration as the proton populations isotropize and particles flow back and forth. 

\begin{figure}[!htb]
\centering 
\includegraphics[trim=0 0 0 0, clip, scale=0.47]{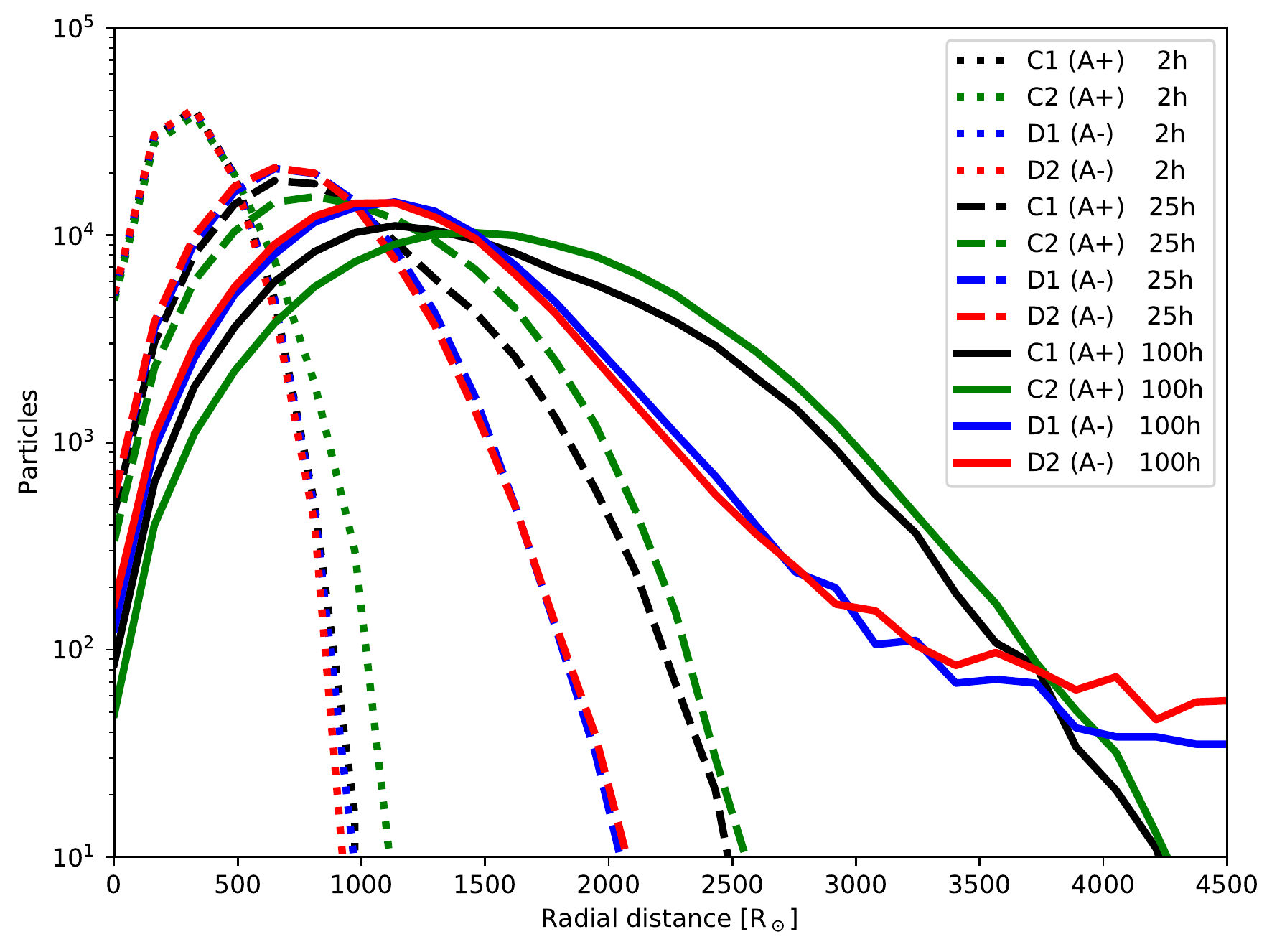}
\caption{
Radial distribution of particles after 2, 25, and 100 hours of propagation for simulations C1, C2, D1, and D2. After 25 hours, the simulations with A+ IMF (C1, C2) show particle populations extending to greater heliocentric mean distances. As particle populations isotropize, this leads to less crossings across the \mbox{1 au} sphere. 
}\label{fig:radial_time}
\end{figure}

This radial preference is shown in Figure \ref{fig:radial_time}, showcasing simulations C1, C2, D1, and D2 after 2, 25, and 100 hours of simulation. Already at the first time step of 2 hours, simulation C2 with an A+ IMF configuration and injection at the current sheet shows particles extending to greater radial distances than the other simulations. At later time steps, the radial distributions of particles show an increasingly strong effect, with the A+ IMF configuration resulting in more particles at large radial distances than the A-- configuration. Simulations C2 and D2 with injection at the current sheet show slightly stronger preferences for this radial effect than simulations C1 and D1, with off-sheet injection. At distances beyond \mbox{$\sim 3500 R_\odot$}, simulation set D overtakes simulation set C, likely due to set C causing more deceleration of high-energy particles, as presented below in section \ref{subsec:deceleration}.

\subsection{Deceleration} \label{subsec:deceleration}
\begin{figure}[!htb]
\centering 
\includegraphics[trim=0 0 0 0, clip, scale=0.47]{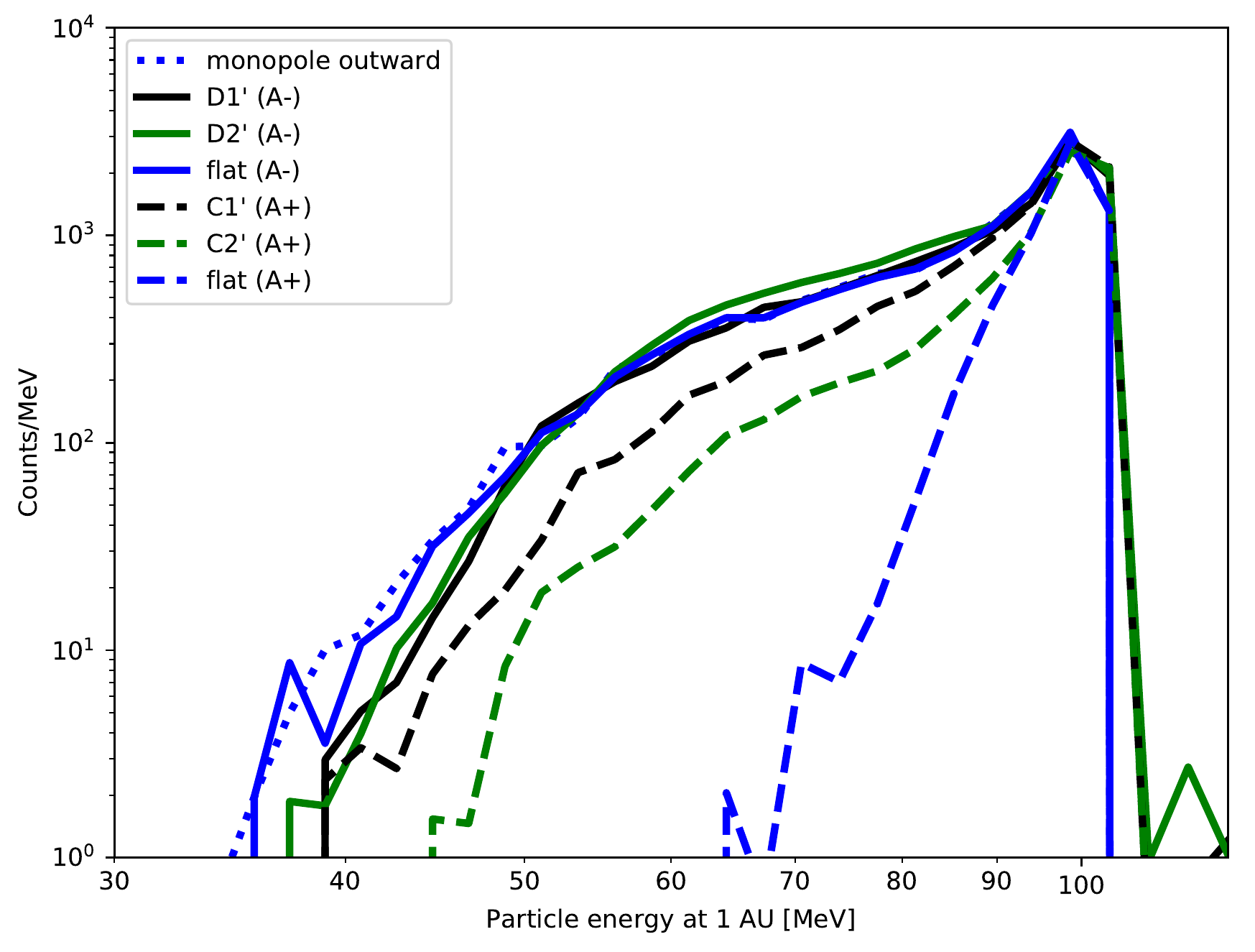}
\caption{
Fluence spectra for proton crossings over the  \mbox{1 au} sphere, in counts MeV$^{-1}$. We used simulation setups C1, C2, D1, and D2, instead injecting $10^4$ \mbox{100 MeV} monoenergetic protons. For comparison, we present three spectra for flat HCS setups and an unipolar field, replicating select results from Figure 13 of Paper I. Results C1' and C2' allow for less deceleration than D1' and D2', but still more than the flat A+ comparison case due to the sheet waviness.
}\label{fig:energy_histo2}
\end{figure}

We also investigate the deceleration of particles by replicating simulations C1, C2, D1, and D2, replacing the power-law injection with a monoenergetic \mbox{100 MeV} population. These simulations are designated C1', C2', D1', and D2'. Fluence spectra for these simulations, along with comparisons values from Figure 13 in Paper I, are plotted in Figure \ref{fig:energy_histo2}. In Paper I, the unipolar and A-- IMF configurations allowed for large latitudinal drift of particles, resulting in strong deceleration, whereas the A+ IMF prevented latitudinal drifts and allowed more particles to retain their energy. In the new simulations, with a wavy HCS, we see that again the A-- IMF (simulations D1' and D2') leads to large deceleration, but also the A+ IMF (simulations C1' and C2') shows much more deceleration than the flat HCS A+ case. This may be due to particles at regions of large HCS inclinations being only weakly bound to the HCS.

\section{Conclusions} \label{sec:conclusions}

In this paper we have presented, for the first time, an analysis of the role of a wavy HCS on the 3D propagation of SEPs.
This was made possible by the use of the test particle approach, coupled with a method for fitting an analytical wavy HCS to coronal source surface maps (SSMs), allowing use of photospheric measurements to be extrapolated into interplanetary space.
 
We have shown that drift along the HCS can distribute SEP protons over a wide range of heliographic longitudes efficiently, in the same direction as corotation (westward) for an A+ IMF configuration, and opposite to corotation (eastward) for an A-- IMF configuration. The degree to which longitudinal transport via the HCS takes place depends on the location of the injection region with respect to the current sheet and on the interplanetary transport conditions. In our simulations, carried out in weak scattering conditions ($\lambda=1\,\mathrm{au}$), we found that injection regions within $10^{\circ}$ of the HCS gave rise to SEP propagation along the current sheet, while the effect was much less for an injection region $20^{\circ}$ above or below the HCS. The HCS-associated transport also depends on the particle energy, as discussed further below.

For a HCS with a tilt angle of around $30^{\circ}-40^{\circ}$ (our simulations A--D), significant qualitative differences in the spatial distributions of SEPs at \mbox{1 au} are observed for different polarity configurations. For A+ IMF configurations, gradient and curvature drifts at locations away from the HCS push protons towards it, tending to concentrate particles near the current sheet. For A-- IMF configurations, the trend is for protons to be pushed away from the HCS, resulting in a larger spread in latitude and less concentration at the current sheet.

For highly inclined HCS situations (tilt angle $85^{\circ}$, our simulations E--F) the differences between results for A+ and A-- IMF configurations are less marked, apart from the opposite directions of HCS drift, due to the near-vertical orientation of the HCS.

Compared to the flat HCS case of \mbox{Paper I}, we find that the introduction of a wavy HCS provides a more varied range of SEP propagation patterns, due to the fact that energetic particle drifts are larger at higher latitudes. For the case of a wavy HCS, particles are able to escape the vicinity of the current sheet more efficiently than for the flat HCS. Especially for A+ IMF configurations, particle escape is related to the shape of the HCS, leading to increased particle fluences in regions of large HCS inclination. This escape from the HCS in regions of large inclination also allows for particles to cross the HCS, in contrast with the results for a flat HCS where crossings, especially in the A+ IMF configuration, were minimal.

In our simulations we injected protons in the energy range of \mbox{$10-800$ MeV}. For injection regions that do not intersect the HCS, the initial kinetic energy affects the strength of curvature and gradient drifts, and consequently, how effectively particles are transported to the current sheet, where they experience HCS drift. The higher the kinetic energy, the faster the propagation to the current sheet. This process is also influenced by the level of scattering, with situations with strong scattering providing more time for particles to drift to the HCS while in the inner heliosphere \citep{Marsh2013}. In the low scattering framework used in our simulations, for injection regions not on the HCS but within 10 degrees of it, we found efficient HCS transport for protons $\gtrsim 30$ MeV. This lower limit may become smaller in high scattering conditions or if additional mechanisms for perpendicular transport, such as magnetic field line meandering \citep{Laitinen2016}, are at play. For protons at the high energy end of the SEP range, e.g. those responsible for Ground Level Enhancements (GLEs), the propagation effects described in this paper will be highly relevant.

The interplay of gradient and curvature drifts and the HCS drift was shown to provide asymmetric particle spreads in the vicinity of the injection region, depending on both IMF polarity and HCS inclination, as the HCS truncated the patterns of gradient and curvature drifts. A further effect was found within A+ IMF configurations, where the statistical interplay of drifts caused preferential enhancements of fluence at regions of large HCS inclination. This effect was not apparent in simulations performed within an A-- IMF configuration.

We found that a wavy A+ current sheet, unlike a flat A+ IMF configuration, allows for particles to drift in latitude, if their motion is quasi-parallel to the inclined current sheet. Thus, protons transported in the vicinity of a wavy HCS can experience deceleration effects associated with latitudinal drifts, which would have been suppressed by a flat HCS.

We examined the direction of HCS drifts for A+ and A-- IMF configurations, and found that the A+ configuration causes a statistical mean drift with a radial
component oriented away from the Sun, whereas for an A-- configuration the radial component is oriented Sunward. This results in, for an A-- (A+) IMF configuration, the particle population being maintained statistically closer to (further away from) the Sun. Later on in the simulation, as the population isotropizes, this effect causes an A-- configuration to exhibit greater \mbox{1 au} fluences than an A+ configuration. This effect becomes even more significant at large particle energies, as those particles were readily propagating far from the Sun, and have larger HCS drift velocities.

Based on our results, we find that realistic SEP transport studies must account for the presence of a non-planar HCS and the associated drifts or risk severely restricting their ability to predict particle propagation effects. As the accuracy of modeling SEP transport conditions increases, additional fluence-enhancing or depleting effects, such as those related to HCS inclination, are found. We do note that our method of fitting the HCS shape from SSMs is limited, especially during periods of solar maximum, and the presented HCS model is only a reasonable approximation in the inner heliosphere. \refbold{If reconnection in the solar wind were to cause statistically significant perturbations in electric and magnetic fields at magnitudes relevant to SEP transport calculations, further steps in modelling such effects are necessary}. Numerical 3D simulations of solar wind structures and preceding ICMEs are a promising topic of future study for the field of SEP transport\refbold{, and will no doubt prove necessary in order to correctly model and predict space weather effects due to SEPs}.

In our simulations, we have considered an injection region of small extent, to obtain a first picture of the patterns of particle propagation. In the case of a wider injection, for example at a Coronal Mass Ejection (CME) driven shock, the overall spatial distribution will be the superposition of a large number of patterns similar to the ones we described, and this will be the subject of future study.

Our simulations prove that the dynamics of SEP propagation in the presence of a wavy HCS are complex and highly dependent on current sheet properties, such as the tilt angle, and on the location of the source region with respect to the HCS. We have shown that 3D test-particle simulations are a key tool in order to fully model the dynamics of solar eruptions.

%

\acknowledgements

 The authors wish to thank the Leverhulme Trust for providing funding for this research through grant number RPG-2015-094.
 SD acknowledges support from the UK Science and Technology Facilities Council (STFC) (grant ST/M00760X/1).
 Wilcox Solar Observatory data used in this study was obtained via the web site \url{http://wso.stanford.edu} at 2017:01:12\_05:40:05 PST courtesy of J.T. Hoeksema.
 The Wilcox Solar Observatory is currently supported by NASA. 
 Historical sunspot data was provided by WDC-SILSO, Royal Observatory of Belgium, Brussels.

%



\appendix
\section{Coronal neutral line fitting}\label{appendix:fitting}
In order to define the wavy heliospheric current sheet used in our model, we fit a wavy neutral line to a source surface model of the solar magnetic field at a given heliocentric distance, e.g., $r=2.5\,R_\odot$. To be able to describe configurations that correspond to actual heliospheric conditions, we make use of solar synoptic source surface maps (SSMs, see, e.g., \citealt{Hoeksema1983}). These are produced by a potential field modeling, using photospheric magnetogram data provided by the Wilcox Solar Observatory. A source surface map is produced for each carrington rotation, but these maps cannot be used directly as a source field due to disagreement at Carrington longitudes 0 and 360, and due to limited latitudinal extent of the SSMs. In order to model particle transport throughout the inner heliosphere, the field description should be continuous.

\begin{figure*}[p!]
\centering 
\includegraphics[trim=0 10 0 0, clip, width=0.48\textwidth]{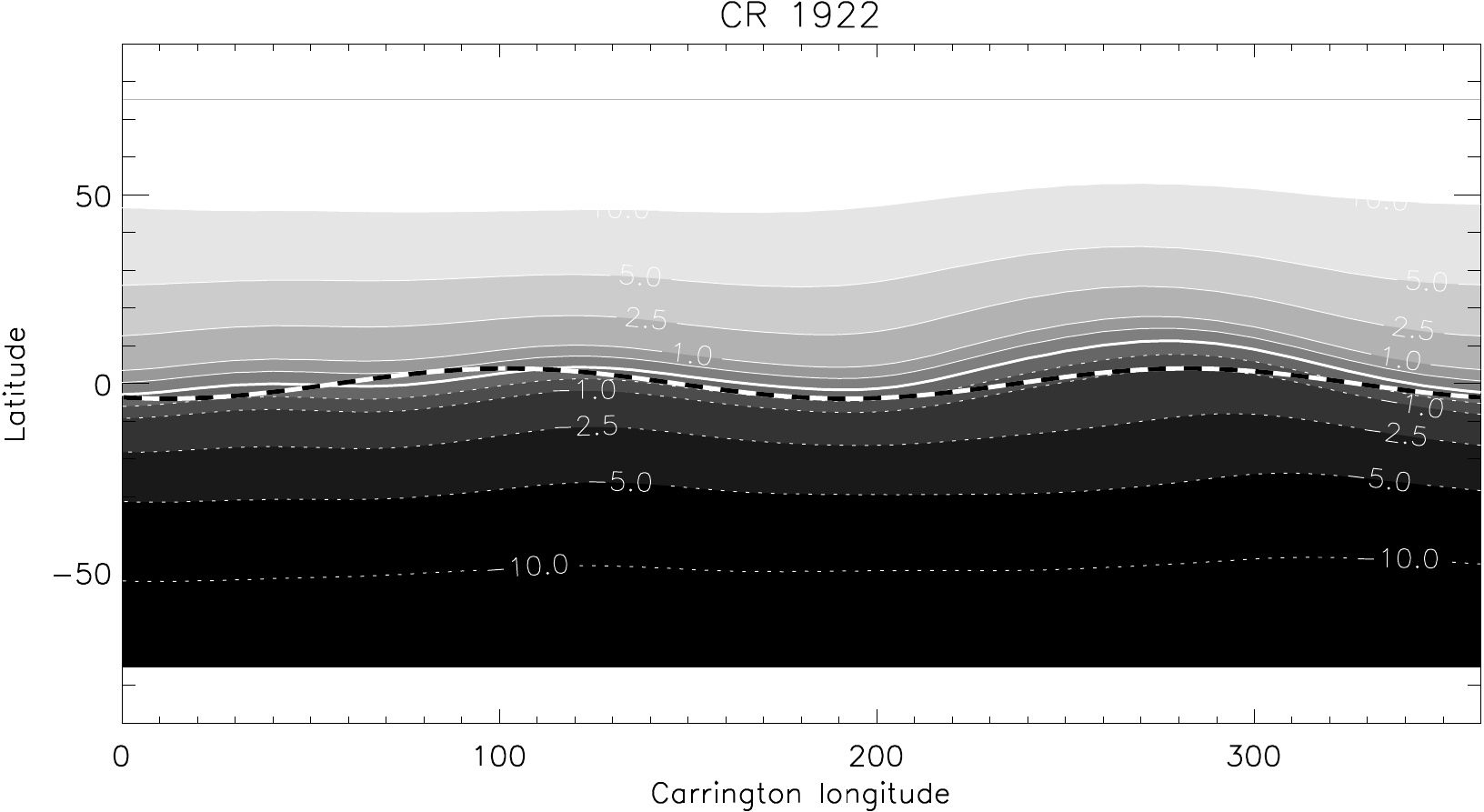} 
\includegraphics[trim=10 10 -10 0, clip, width=0.48\textwidth]{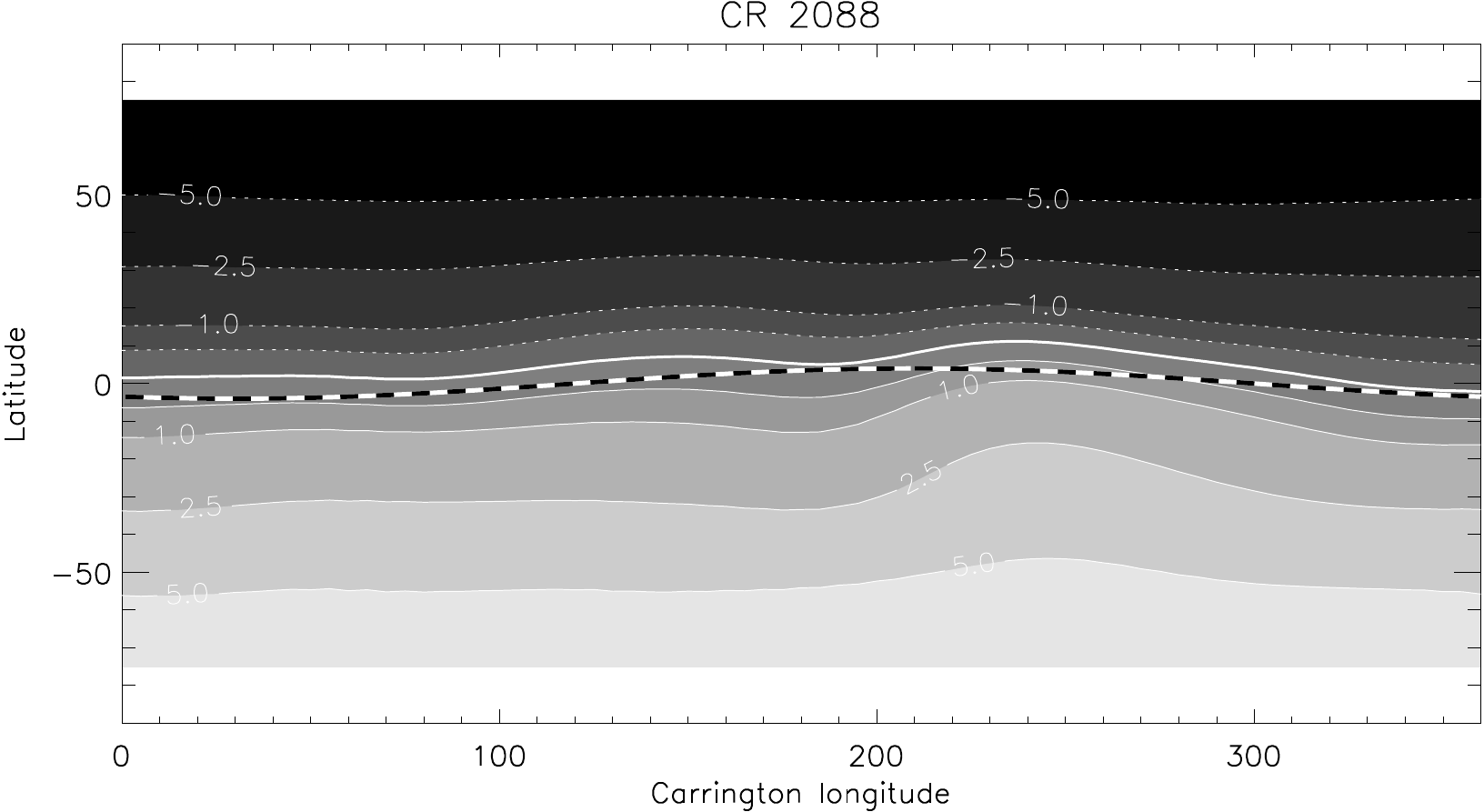} 
\\ \vspace{0.2cm}
\includegraphics[trim=0 10 0 0, clip, width=0.48\textwidth]{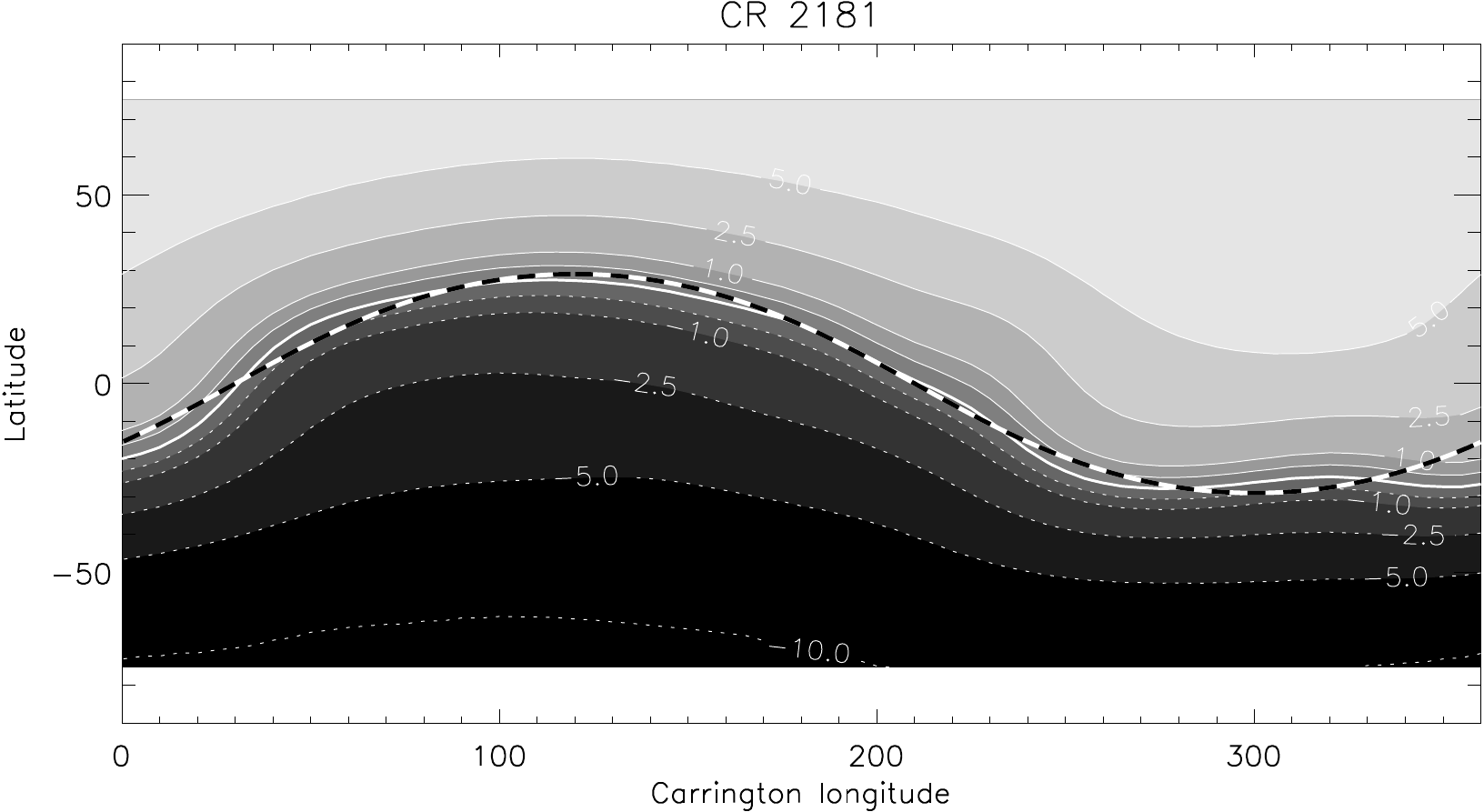} 
\includegraphics[trim=10 10 -10 0, clip, width=0.48\textwidth]{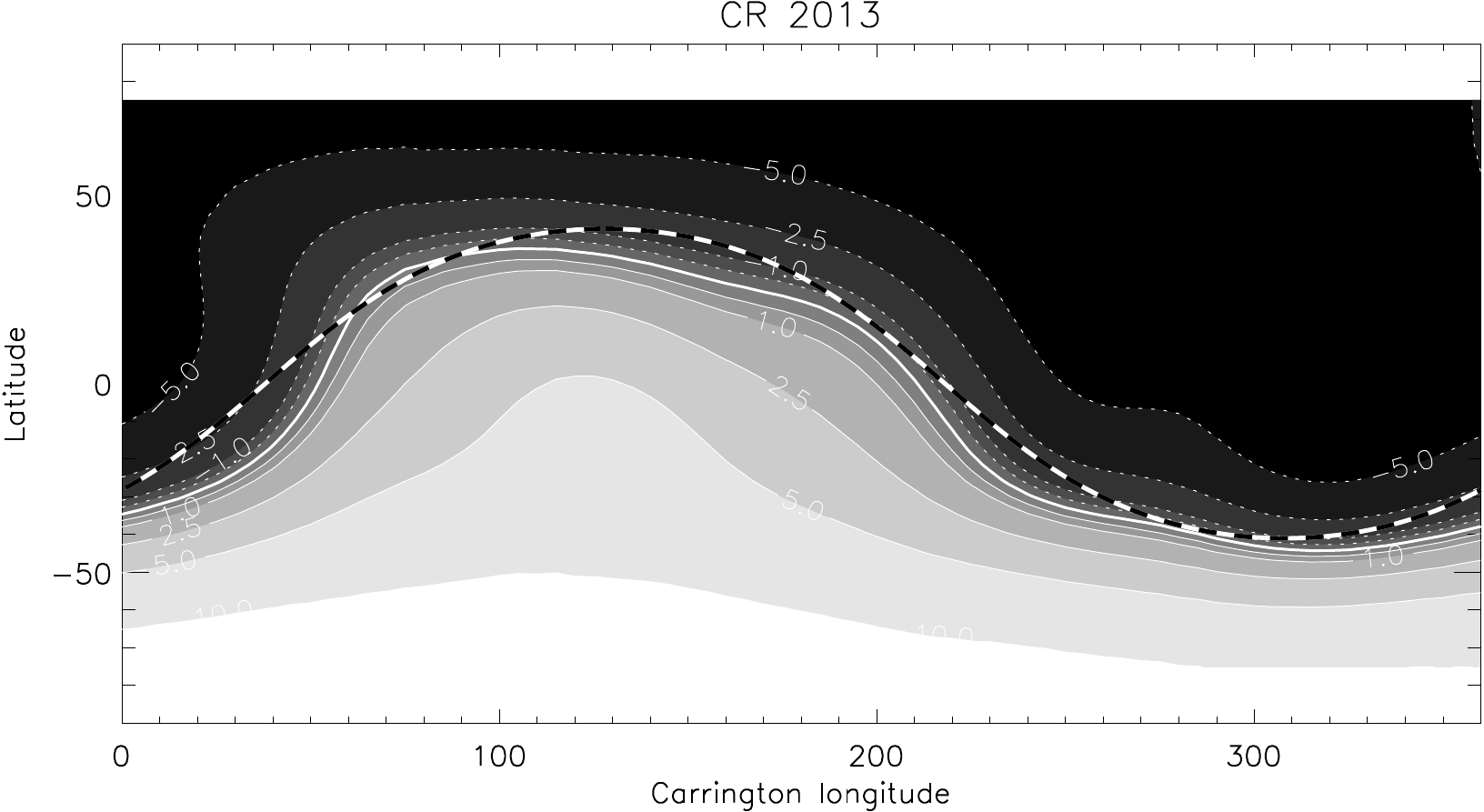} 
\\ \vspace{0.2cm}
\includegraphics[trim=0 10 0 0, clip, width=0.48\textwidth]{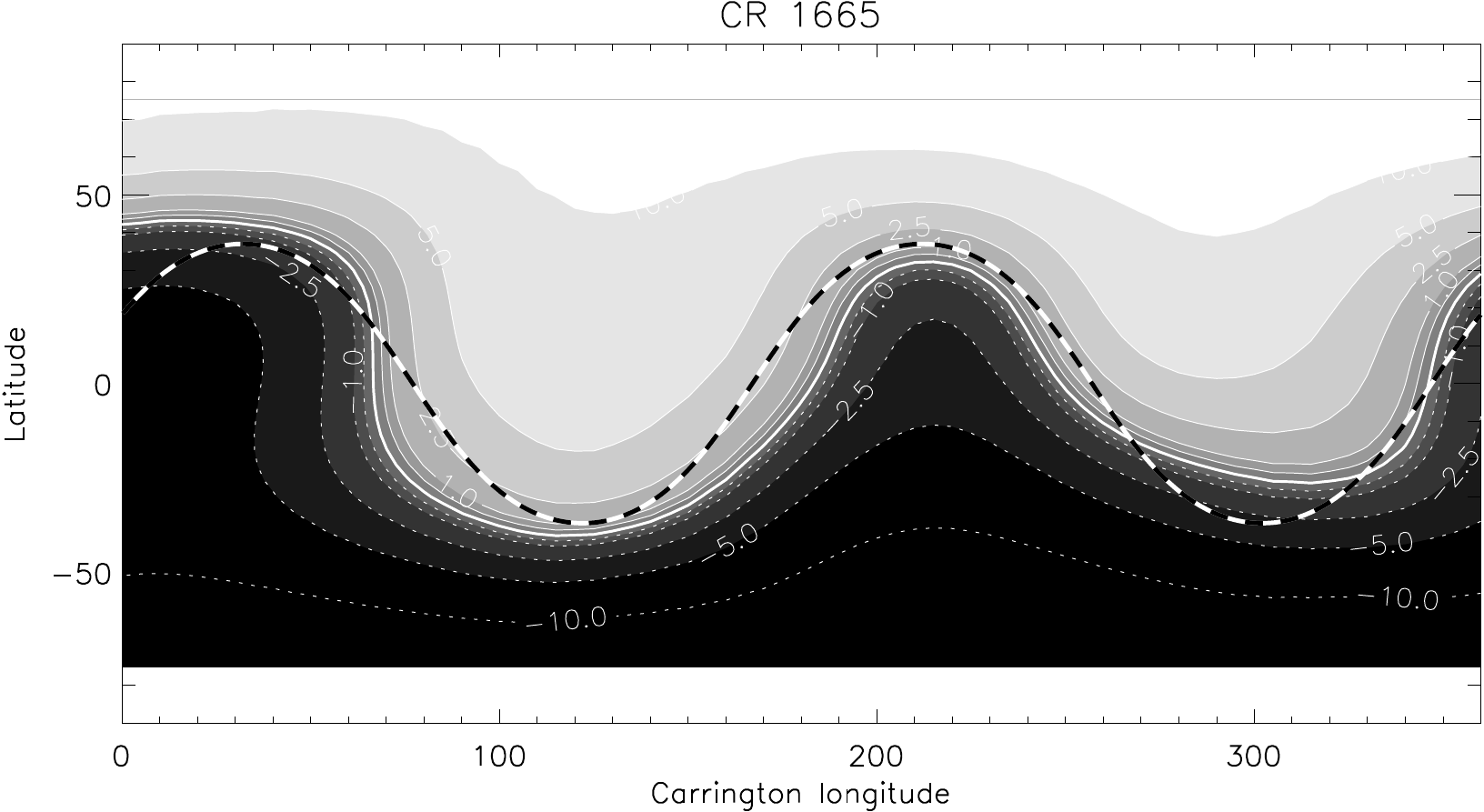} 
\includegraphics[trim=10 10 -10 0, clip, width=0.48\textwidth]{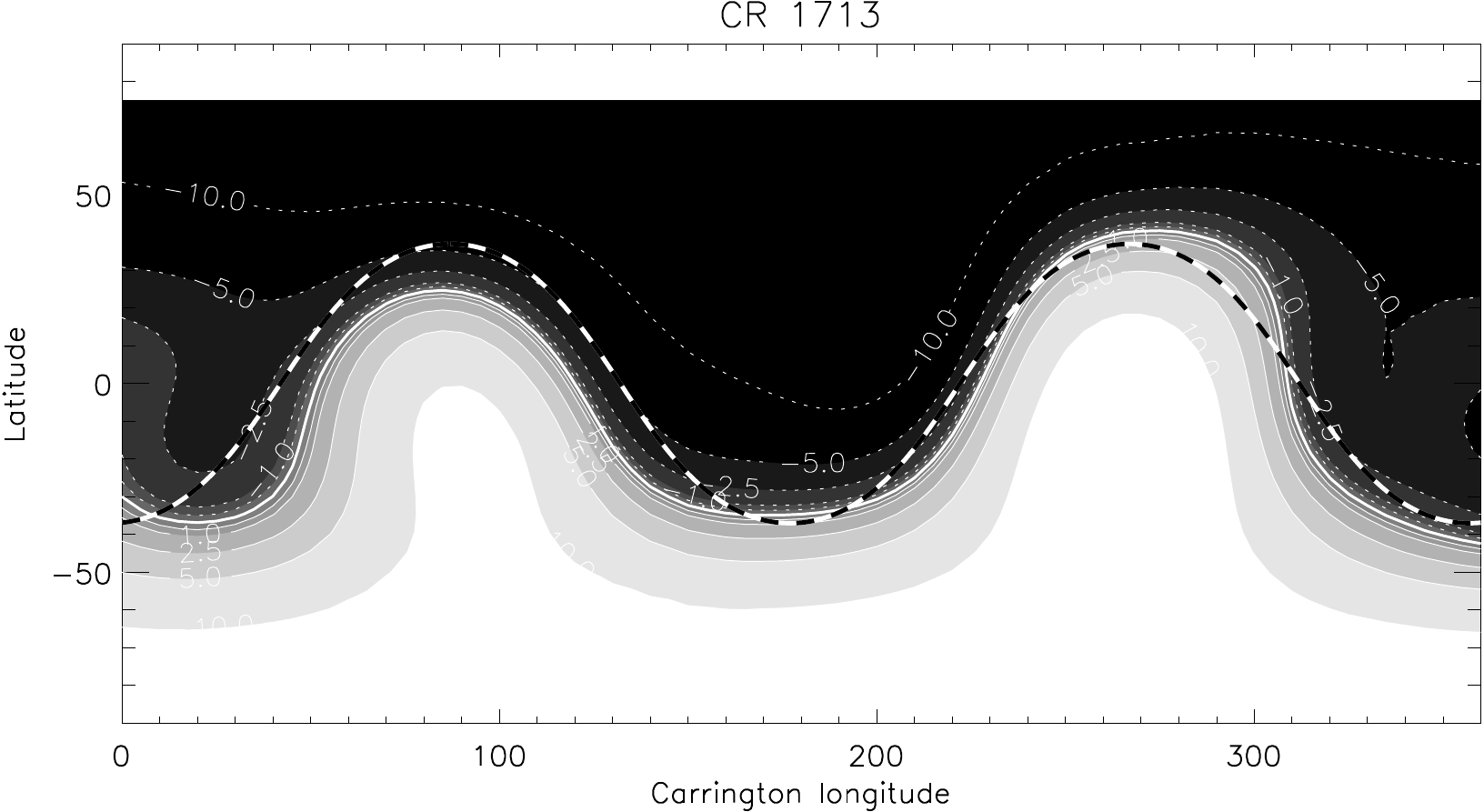} 
\\ \vspace{0.2cm}
\includegraphics[trim=0 0 0 0, clip, width=0.48\textwidth]{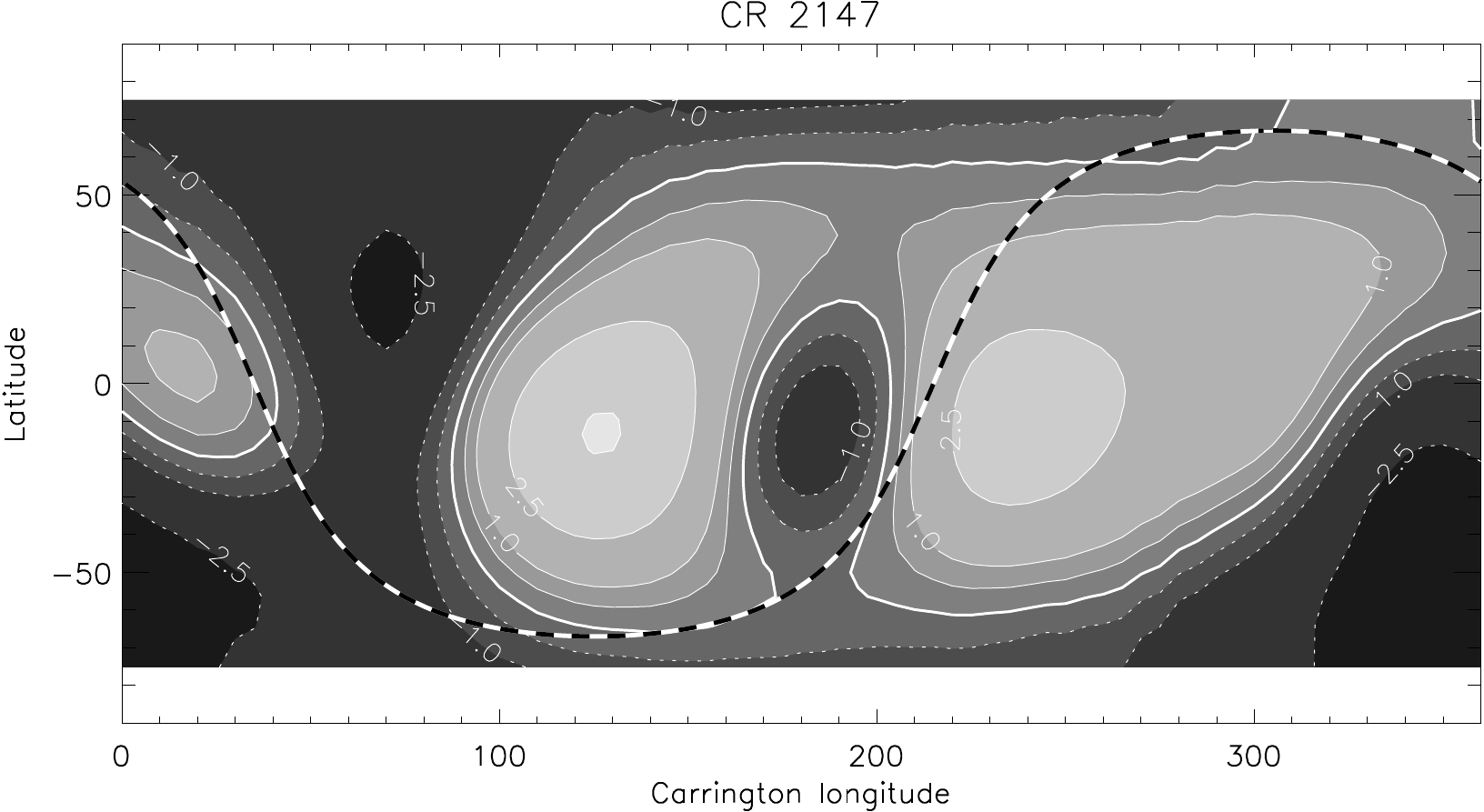} 
\includegraphics[trim=10 0 -10 0, clip, width=0.48\textwidth]{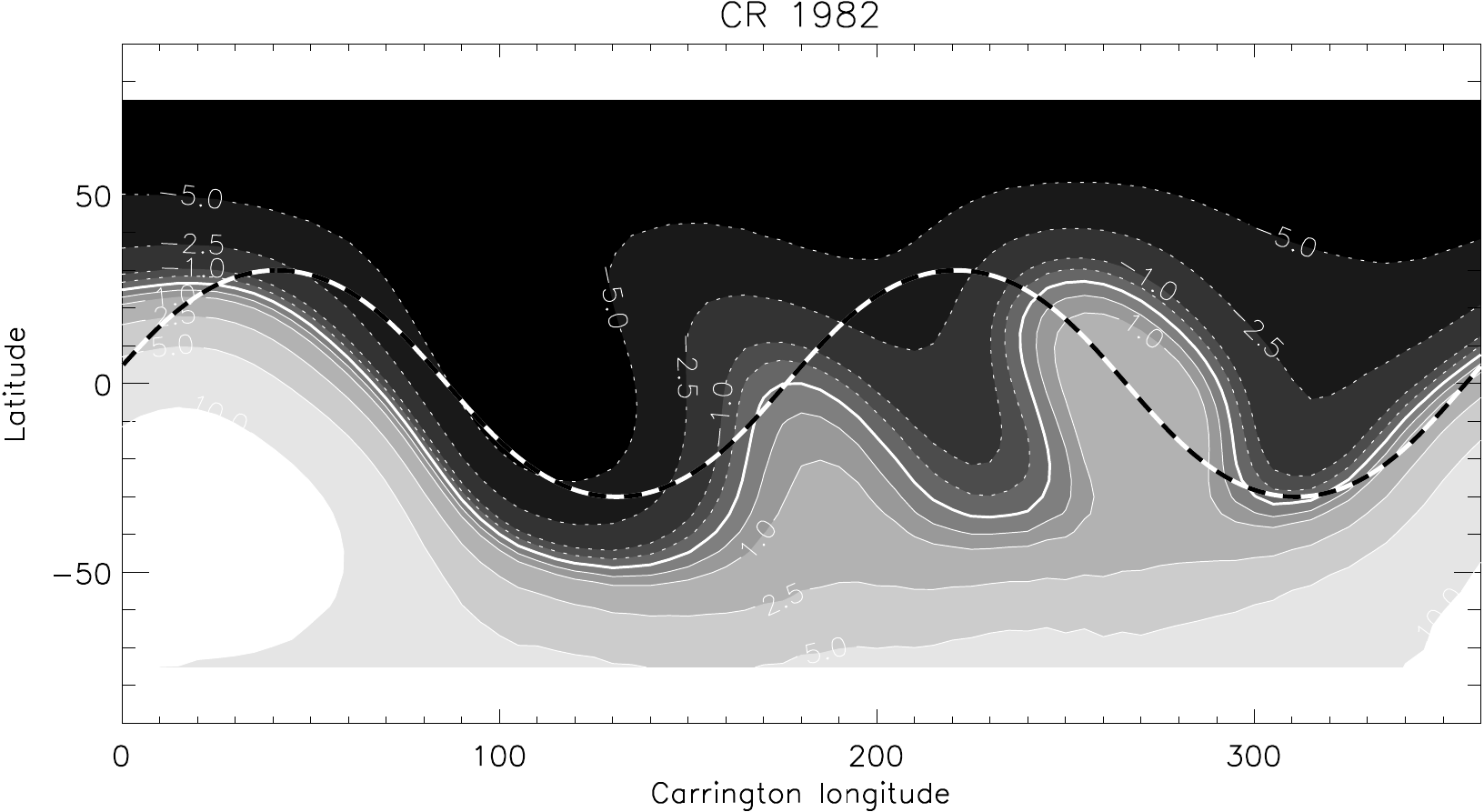} 
\caption{Synoptic source surface maps computed for $r=2.5\,R_\odot$ using photospheric measurements for Carrington rotations 1922, 2088, 2181, 2013, 1665, 1713, 2147, and 1982. Pale regions indicate outward pointing magnetic fields, dark regions inward-pointing magnetic fields, and the boundary neutral line is shown as a thick white curve. Contour values are given in microtesla. The best fit for the wavy neutral line model is shown as a black-and-white dashed curve. Fit parameters are given in Table \ref{tab:HCSfit}. Solar syniptic data is provided by the Wilcox Solar Observatory. The selected plots and fits represent a nearly flat neutral line (CRs 1922 and 2088), single-peak wavy lines (CRs 2181 and 2013), dual-peak amplitude wavy lines (CRs 1665 and 1713), and unsuccesful fits (CRs 2147 and 1982). The source map for CR 2147 is very complicated and unsurprisingly fails to provide a good fit. The source map for CR 1982 appears to have a three-peak structure, but in fact the least bad fit is provided with only two peaks.
}\label{fig:SSMs}
\end{figure*}

Using the radial component of the magnetic field in SSMs, we fit a neutral line along positions of magnetic field reversal. The SSM neutral line is not modeled perfectly by the quasi-sinusoidal neutral line of our model, but in many cases it provides a reasonable and mathematically elegant approximation. In Figure \ref{fig:SSMs}, we show eight sample SSMs, provided by Wilcox Solar Observatory, and the best fits of our wavy neutral line to them. We explored the neutral line parameter space with $n_\mathrm{nl} \in [1,3]$, $\alpha_\mathrm{nl} \in \alpha_0+[-10^\circ , 10^\circ ]$, and $\phi_\mathrm{nl} \in [0^\circ , 360^\circ)$, where $\alpha_0$ is the average of maximum latitude reached in each hemisphere by the SSM neutral line. For each entry in the parameter space, we calculate angular distances between our model neutral line and the SSM neutral line at 1 degree intervals, and we choose the best approximation using a least squares fit. SSMs can be calculated for various heliocentric heights, but in this work we have used maps for a source surface height of $r=2.5\,R_\odot$. 

In Table \ref{tab:HCSfit}, we list the Carrington rotations we considered along with the inferred IMF polarity configuration. We then list the parameters $n_\mathrm{nl}$, $\alpha_\mathrm{nl}$, and $\phi_\mathrm{nl}$ of the best fits to each of the CR neutral lines. We also list the least squares fit quality value $\Sigma \lambda^2$, and an average sunspot number for that CR, calculated from daily sunspot numbers provided by the WDC-SILSO, Royal Observatory of Belgium, Brussels. 

\begin{table*}[!hb]
\caption{Wavy neutral line fit parameters for Carrington rotations 1922, 2088, 2181, 2013, 1665, 1713, 2147, and 1982. The IMF column states the mean polarity according to cosmic ray notation. The rightmost column has an average sunspot number for that CR.}
\label{tab:HCSfit}
\centering
\begin{tabular}{l l r r r r r}
\hline
CR & IMF  & Peak count $n_\mathrm{nl}$ & Tilt angle $\alpha_\mathrm{nl}$ & Longitudinal offset $\phi_\mathrm{nl}$ & Fit quality $\Sigma \lambda^2$ & Avg. sunspot number \\    
\hline
\hline
   1922 & A+  & 2 & 4  & 147 & 5287  & 18.6  \\
   2088 & A-- & 1 & 4  & 300 & 8541  & 8.3   \\
   2181 & A+  & 1 & 29 & 210 & 2770  & 52.7  \\
   2013 & A-- & 1 & 41 & 218 & 15714 & 64.4  \\
   1665 & A+  & 2 & 37 & 77  & 22199 & 114.3 \\
   1713 & A-- & 2 & 37 & 132 & 20214 & 233.8 \\
   2147 & A+  & 1 & 67 & 35  & 34343 & 144.6 \\ 
   1982 & A-- & 2 & 30 & 86  & 66538 & 209.4 \\ 
\hline
\end{tabular}
\end{table*}

\end{document}